\documentclass[final]{siamart1116}

\usepackage{color}
\usepackage{graphicx}
\usepackage{tikz}
\usetikzlibrary{shapes,arrows}
\usepackage{amssymb,amsmath}
\DeclareMathAlphabet\mathbfcal{OMS}{cmsy}{b}{n}

\title{Optimization methods for in-line holography%
\thanks{
\funding{The work of the second author was supported by a National Science Foundation (NSF) graduate research fellowship. The research of the first, third, and fourth authors was supported by MINECO grants 
MTM2014-56948-C2-1-P (AC, PV) and MTM2013-43671-P (VS).
Part of the computations of this work were performed in EOLO, the HPC of Climate Change of the Moncloa International Campus of Excellence, funded by MECD and MICINN.}}}

\author{A. Carpio%
  \thanks{Departamento de Matem\'atica Aplicada, Universidad Complutense de Madrid, Madrid, 28040, Spain
  (\email{carpio@mat.ucm.es, pervidal@ucm.es}).}%
  \and
  T.G. Dimiduk%
   \thanks{Department of Physics, Harvard University, Cambridge, MA 02138, 
   USA (\email{tdimiduk@physics.harvard.edu}).}
  \and
  V. Selgas 
  \thanks{Departamento de Matem\'aticas, Escuela Polit\'ecnica de Ingenier\'{\i}a, Universidad de Oviedo, Gij\'on 33203, Spain (\email{selgasvirginia@uniovi.es}).}
  \and 
  P. Vidal 
  \footnotemark[2]
}

\headers{Optimization methods for in-line holography}
{A. Carpio, T.G. Dimiduk, V. Selgas, P. Vidal}

\ifpdf
\hypersetup{ 
  pdftitle={Optimization methods for in-line holography} }
\fi

\begin{document}
\maketitle

\begin{abstract}
We present a procedure to reconstruct objects from holograms recorded in in-line holography settings.
Working with one beam of polarized light, the topological derivatives and energies of functionals quantifying hologram deviations yield predictions of the number, location, shape and size of objects with nanometer resolution.
When the permittivity of the objects is unknown, we approximate it by parameter optimization techniques. Iterative procedures combining topological field based geometry corrections and parameter optimization sharpen the initial predictions. 
Additionally, we devise a strategy which exploits the measured holograms to produce  numerical approximations of the full electric field (amplitude and phase) at the  screen where  the hologram 
is recorded. Shape and parameter optimization of functionals  employing such approximations 
of the electric field also yield images of the holographied objects.
\end{abstract}




\section{Introduction}
\label{sec:intro}

\begin{figure} [!h]
\centering
\includegraphics[width=8cm]{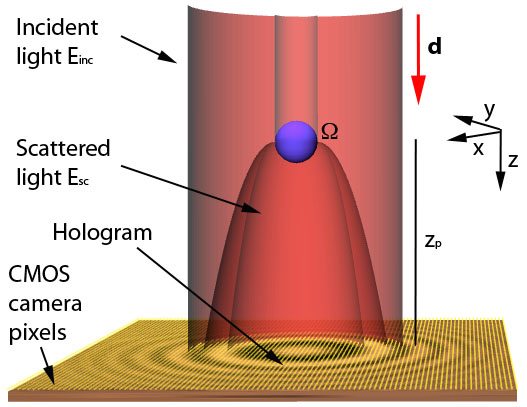}
\caption{Schematic representation of in-line hologram formation.
The incident light beam ${\mathbfcal E}_{inc}$ is scattered by the object  $\Omega$. The resulting interference pattern 
${\mathbfcal I}=|{\mathbfcal E}_{inc}+{\mathbfcal E}_{sc}|^2$ is recorded at a grid of CMOS camera pixels, forming the hologram. We consider here objects diameters in the range $10 \, {\rm nm}$ - $2 \, \mu${\rm m} ($1 \, \mu${\rm m} = $10^{-6} {\rm m}$, $1$ {\rm nm} = $10^{-9} {\rm m}$).  Typical distances $z_p$ to the screen  are about $5 \, \mu${\rm m}. The screen covers an area $10  \, \mu{\rm m} \times 10 \, \mu{\rm m}$ and the CMOS grid step is $0.1 \, \mu${\rm m}.}
\label{fig1}
\end{figure}

Digital holographic microscopy is a three dimensional optical 
imaging technique with high spatial (about 10 nanometers) 
and temporal (microseconds) precision. It furnishes a noninvasive 
approach for high speed 3D imaging of soft matter and live cells
\cite{fung,holocell} which avoids the use of toxic stains and 
fluorescent labels. Numerical processing makes it possible 
to  analyze different object planes without optomechanical 
motion \cite{holosection,holosection2}. It also allows
for postprocessing to remove aberrations and improve
resolution \cite{holoimprovedepth,holoenhanced}. 
However, the adoption of holography as a characterization 
technique has been restricted due to the inherent difficulty of 
recovering  3D structures from the 2D holograms they generate. 
Holograms encode the light field scattered by an object as an 
interference pattern  \cite{holodigital,holo}, see Figure \ref{fig1}.
Successful reconstructions habitually require considerable 
knowledge about the sample being imaged (the approximate 
positions of  particles in the field of view, for example), as well 
as  proficiency in scattering theory.
Advances in digital holography are strongly related to 
progress in the mathematical methods used to decode light 
measurements.

Unlike when working with acoustic waves or microwaves \cite{capatano,guzinaacoustic}, light waves oscillate much to fast for any detector to measure the phase of the wave. This means we can only measure an averaged intensity of the electric field  $|{\mathbfcal E}|^2$. This lost phase information is why you cannot extract three dimensional information from a single camera picture or microscope image. Holography lets us get at this missing phase information by mapping it into intensity patterns though interference. Specifically, by interfering light from a sample ${\mathbfcal E}_{sc}$ with a known plane wave ${\mathbfcal E}_{ref}$, we obtain a holographic interference pattern ${\mathbfcal I}_h = |{\mathbfcal E}_{sc} + {\mathbfcal E}_{ref}|^2$. Because ${\mathbfcal E}_{ref}$ is known, this hologram lets us get at the ${\mathbfcal E}_{sc}$ which is needed to compute information about three dimensional scenes. In our previous work \cite{siims} we demonstrated that if you knew ${\mathbfcal E}_{sc}$ directly for a holographic recording, you could use topological techniques to recover the three dimensional scene. In this work, we demonstrate an improved technique that can work directly with ${\mathbfcal I}_h$, which is physically measurable.

Different holographic settings are possible, we focus here
on in-line holography. The principle of in-line holographic 
microscopy is as follows \cite{inline}. 
An initially collimated laser beam ${\mathbfcal E}_{inc}$ is 
scattered by a particle $\Omega$, see Figure \ref{fig1}. 
The scattered light beam ${\mathbfcal E}_{sc}$ 
interacts with the unscattered beam ${\mathbfcal E}_{inc}$ 
in the focal plane of a microscope objective. The interference 
pattern 
${\mathbfcal I} = | {\mathbfcal E}_{sc} + {\mathbfcal E}_{inc} |^2$
is recorded at a screen, forming the in-line hologram.
Knowing the emitted light and the measured hologram, we aim 
to approximate the geometry of the scatterers (number, location, 
size, shape), as well as their permittivities, when unknown.

When the approximate location in the field of view is known,
objects describable by simple parametrizations, such as  spheres
or rods, can be approximated from in-line holograms by fitting the parameters so that the difference between the synthetic holograms generated by them and measured hologram diminishes \cite{fung,inline,kaz,holotom}.
For spherical particles, the scattered electric field is given explicitly by the Mie solution \cite{fung,inline}. Spheres and rods have also been handled through the discrete dipole approximation (DDA) \cite{holotom}. The employed techniques are computationally intensive,  though this drawback can be diminished while keeping the resolution by resorting to a smaller number of randomly distributed detectors \cite{holotom2}.

To image scatterers without previous knowledge of their geometry, number and approximate location in the field of view, we may address more general optimization problems  in which the design variable is the unknown domain. Different techniques differ in the way the objects are represented and deformed to decrease the value of the cost functional. When the number of scatterers is known, classical shape deformation along adequately chosen vector fields may be used  \cite{masmoudi, zolesio,feijooOberai}. However, this procedure does not  allow for topological changes: the exact number of contours has to be known from the beginning \cite{santosa}. Deformations inspired in level set methods, instead, allow to create and destroy contours during the process \cite{oliver1,oliver2,santosa}. All these methods require an initial guess of the object contours to proceed. Topological derivative techniques provide such guesses without  a priori information  and do not require a specific parametrization of the objects \cite{feijoo}.
They have been combined with both shape derivatives and level sets to approximate objects in a variety of aplications, such as flow studies and electric impedance tomography \cite{caubetfluid,hintermuller}. 
Methods entirely built on topological field based approximations have 
been employed  in acoustics, materials testing and electric impedance tomography \cite{park,ip2012,topologicalenergy,feijoo,guzinaacoustic,samet}, for instance. 
The results may be refined by combining multiple frequencies \cite{park}, many directions \cite{guzinalarge}  or by iteration \cite{ip2008}. They may require less iterations than level sets, though some previous calibration work is needed. Most practical implementations have used incident directions and detectors distributed over wide angle ranges, usually in 2D settings.
When imaging with one light beam, and assuming that the electric field instead of the hologram was known, we showed that topological methods provide approximations of the 3D scatterers \cite{siims}. If the object permittivity is known and the objects size is small enough compared to the wavelength, these approximations may be improved by a combination of topological derivative techniques, blobby molecule fittings and  a forward solver \cite{siims}.
Here we tackle the inversion problem of recovering the 3D scatterers from the recorded hologram in two ways.
A first possibility is to find a strategy to predict numerically the electric field from the recorded hologram and use this prediction as approximate data. This is done here employing  topological methods, gradient optimization  and Gaussian filtering. 
Alternatively, we may seek to optimize the holographic cost functional,
which measures the deviation with respect to the recorded hologram.
Combining topological and descent methods we succeed in
approximating objects and their permittivities with no a priori knowledge, 
other than the ambient permittivity and the measured hologram.
We observe that the initial reconstructions provided by 
both strategies are similar. However, the later one is more straightforward.
Whether the first one may lead to more accurate approximations
when iterated or combined with other techniques is a matter of study since other factors, such as the distance to the detectors and the size of the hologram, play a role too.

Our numerical approximations of the  electric field pave
the way to adapting to this framework techniques based on its knowledge developed for larger wavelengths.  A wide variety of 
methods tracking permittivity variations  have been introduced 
whose applicability in holography settings may be explored \cite{ip2012,capatano,chaumet,eyraud,li,yu,zaeytijd}.
Qualitative techniques such as  linear sampling \cite{capatano,coltonsampling},
as well as factorization and MUSIC methods \cite{music, coltonsampling}, have been analyzed when a much  wider distribution of incident waves and  observation  points are employed. 
In holography settings, we must work with penetrable objects using only one incident wave and intensities measured at observation points located on a limited flat screen, for moderate to large dimensionless wavenumbers.
When the optical properties of objects are unknown, we are able to
infer them in an additional optimization step.

\begin{figure}[h!]
\centering
\hskip -3mm (a) \hskip 3.4cm (b)  \hskip   3.4cm (c)\\
\includegraphics[width=4cm]{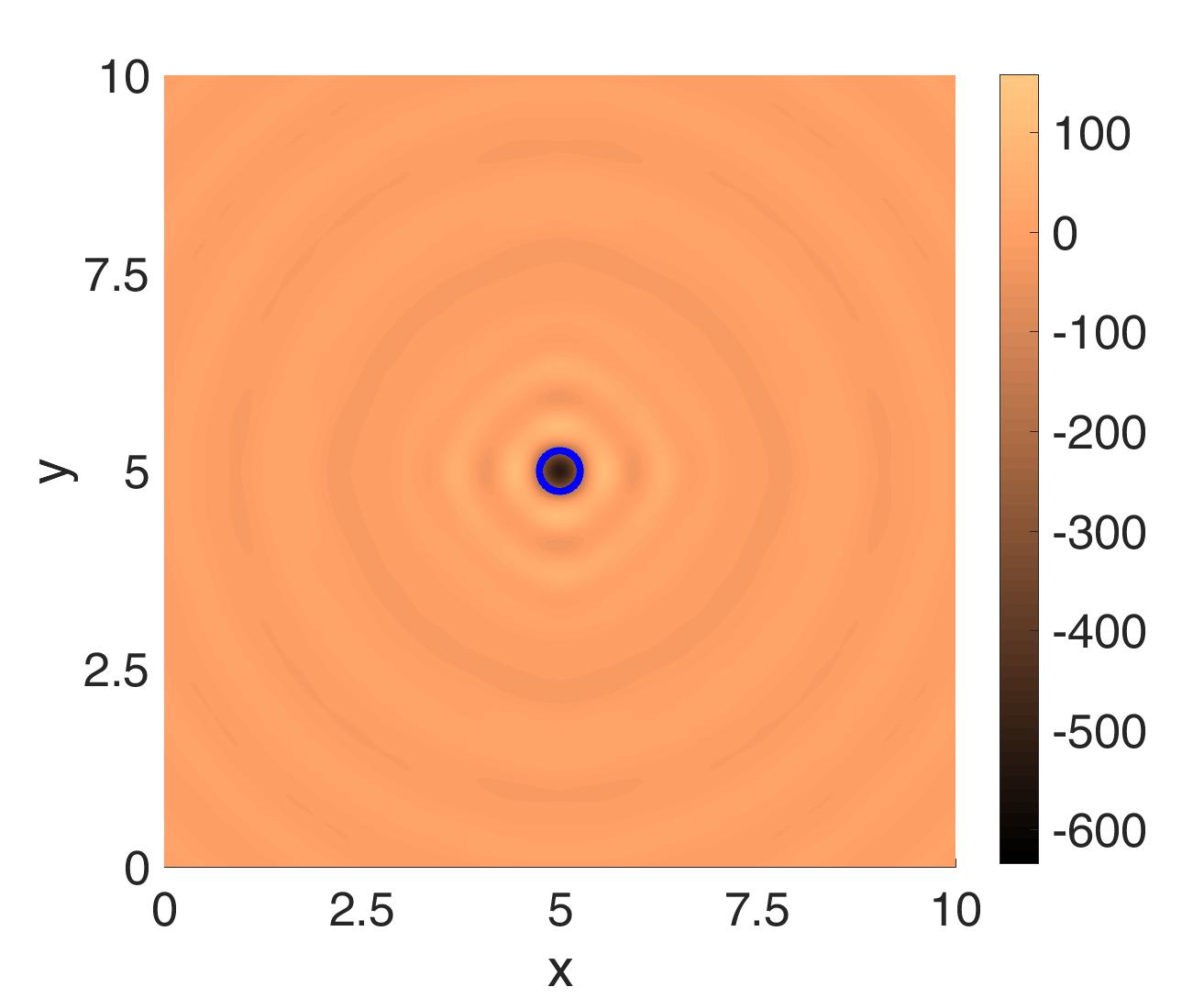} 
\hskip -1mm
\includegraphics[width=4cm]{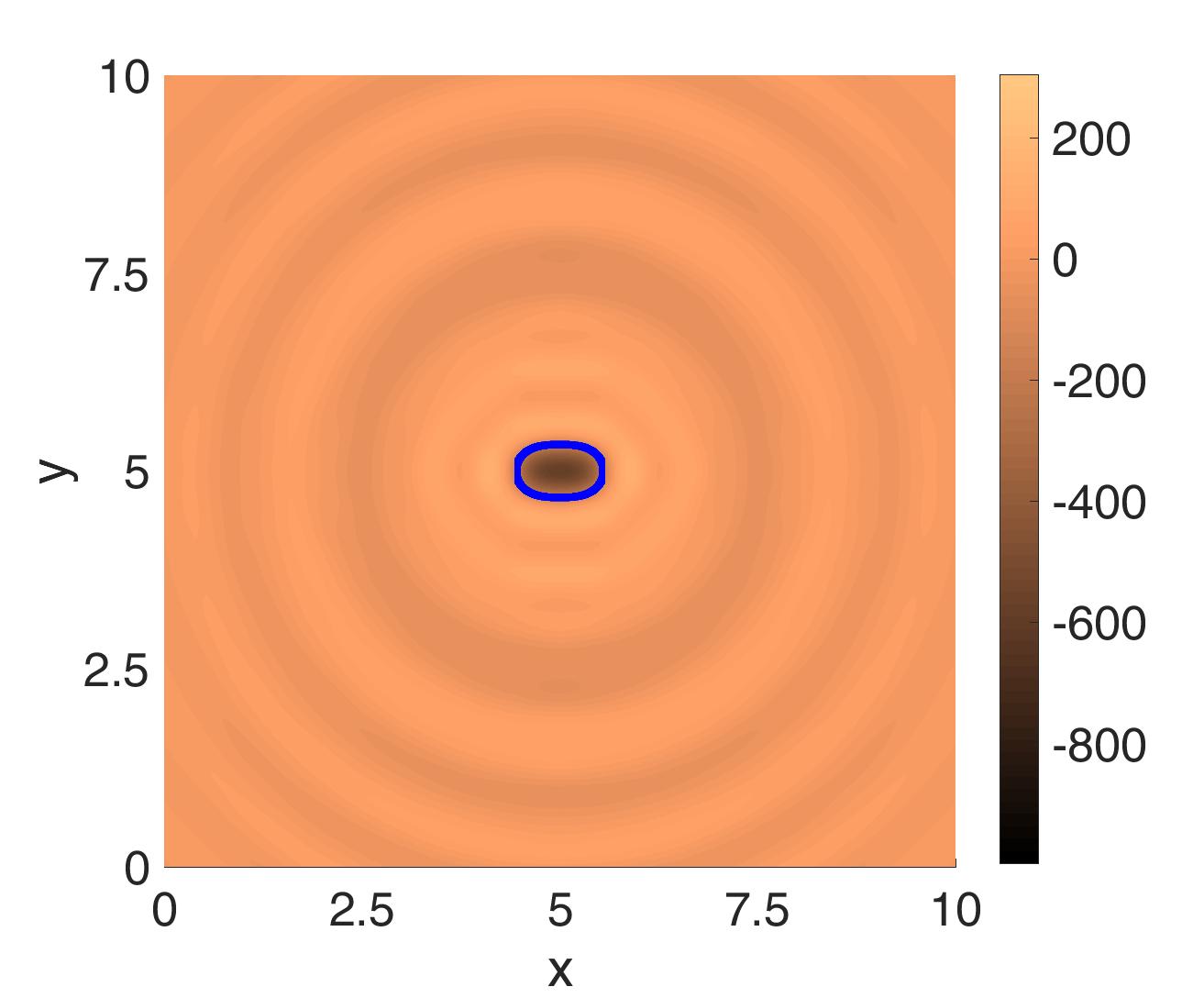} 
\hskip -1mm
\includegraphics[width=4cm]{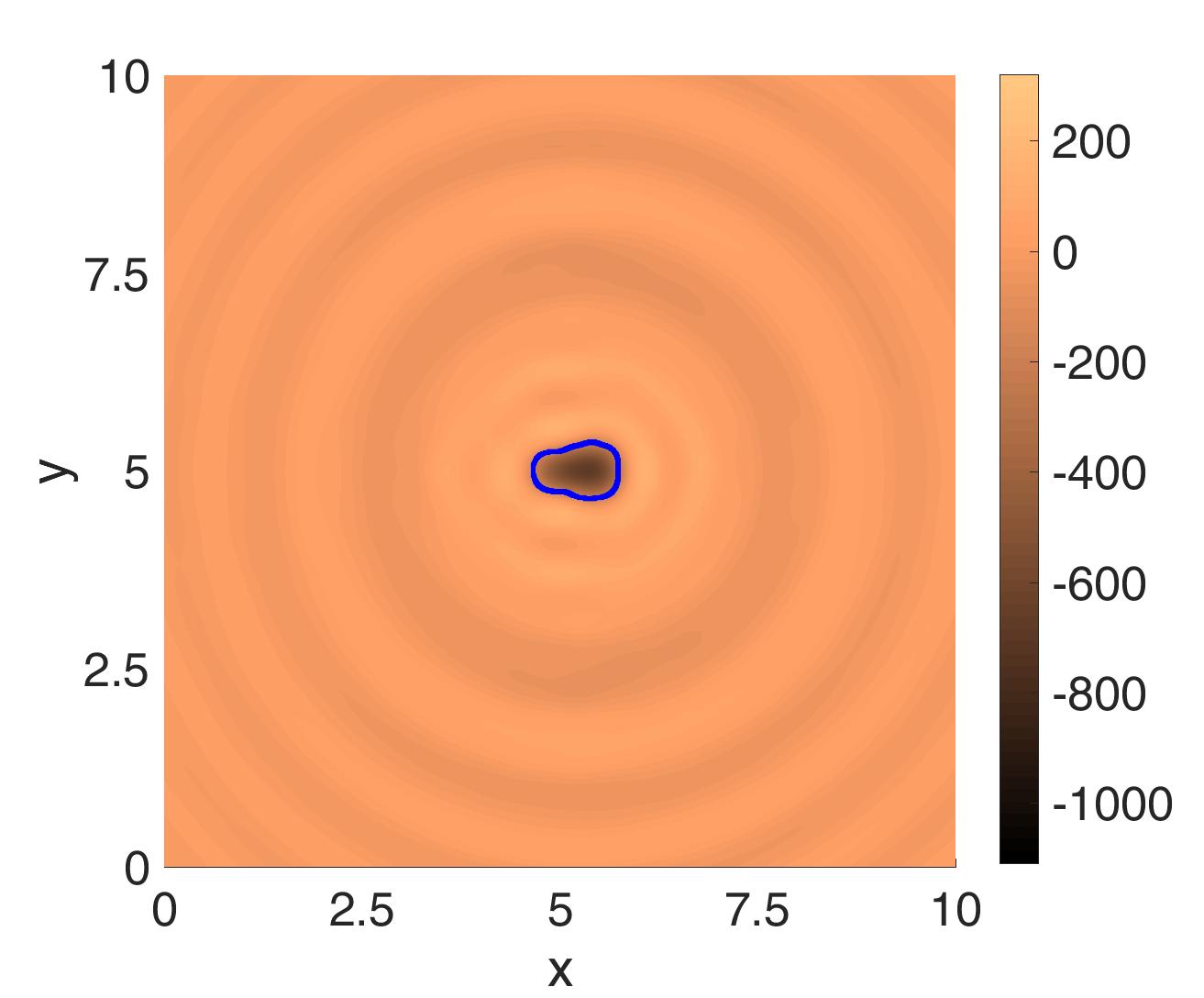}  \\ 
\hskip -3mm (d) \hskip 3.4cm (e)  \hskip   3.4cm (f)\\
\includegraphics[width=4cm]{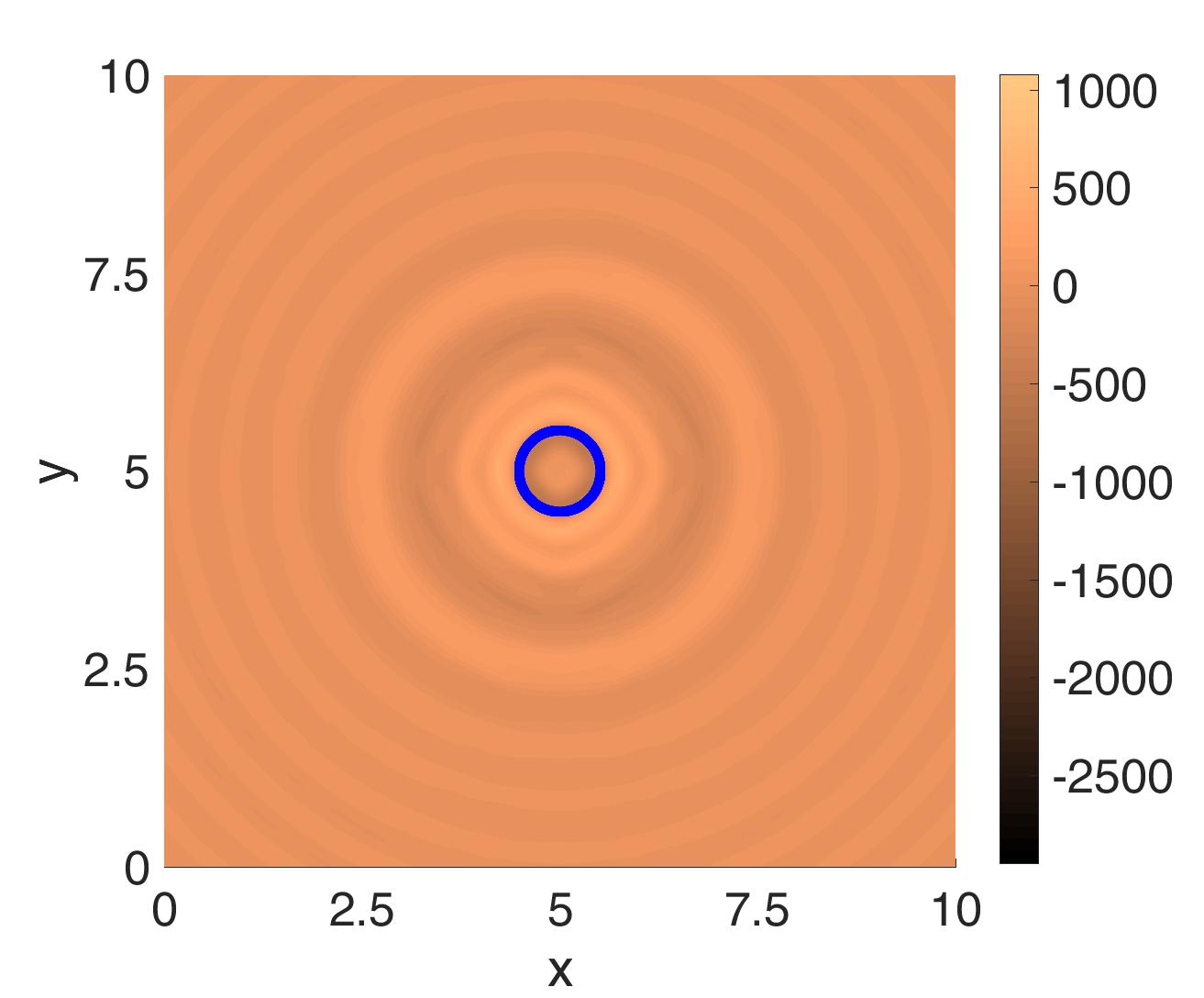} 
\hskip -1mm
\includegraphics[width=4cm]{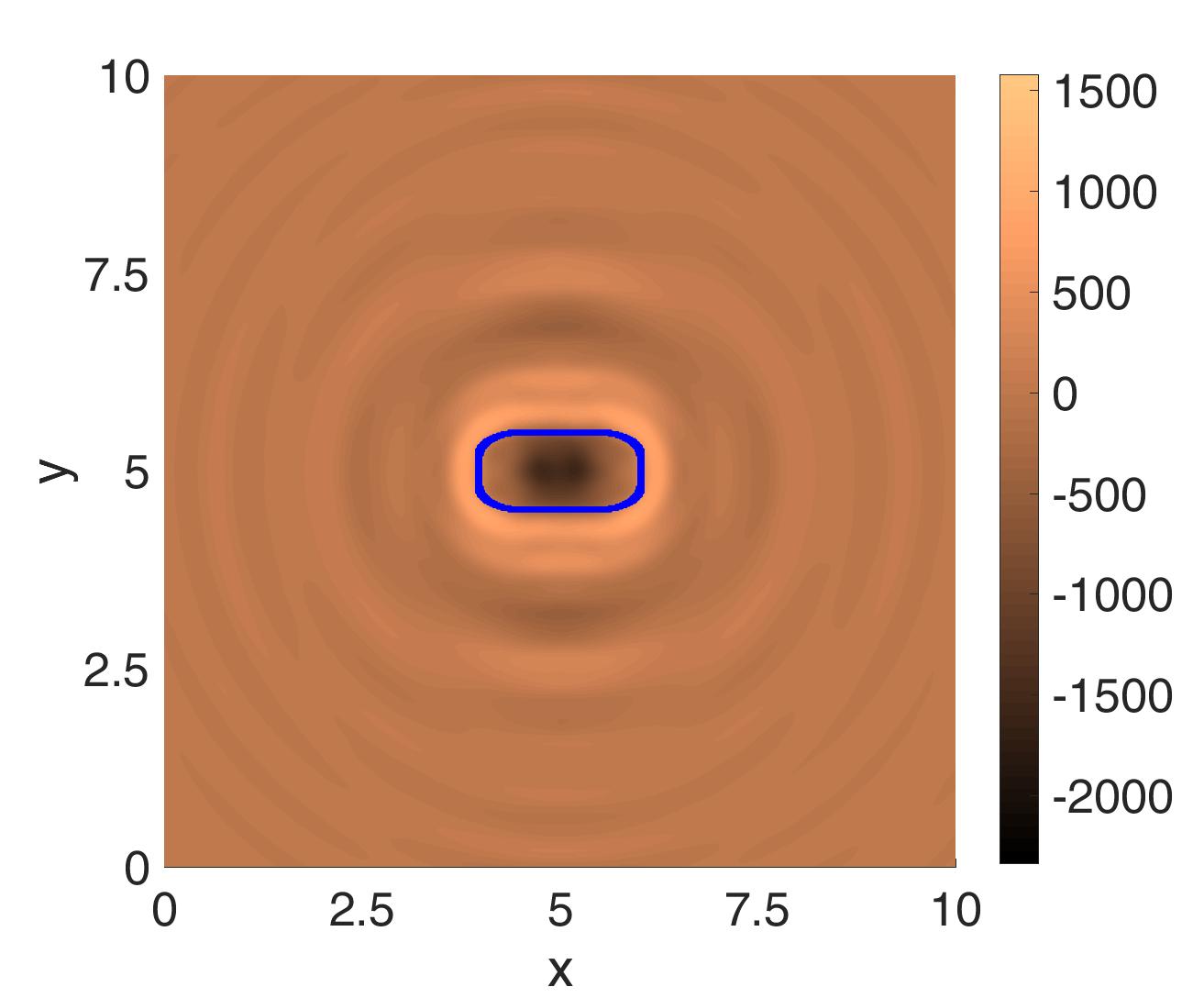} 
\hskip -1mm
\includegraphics[width=4cm]{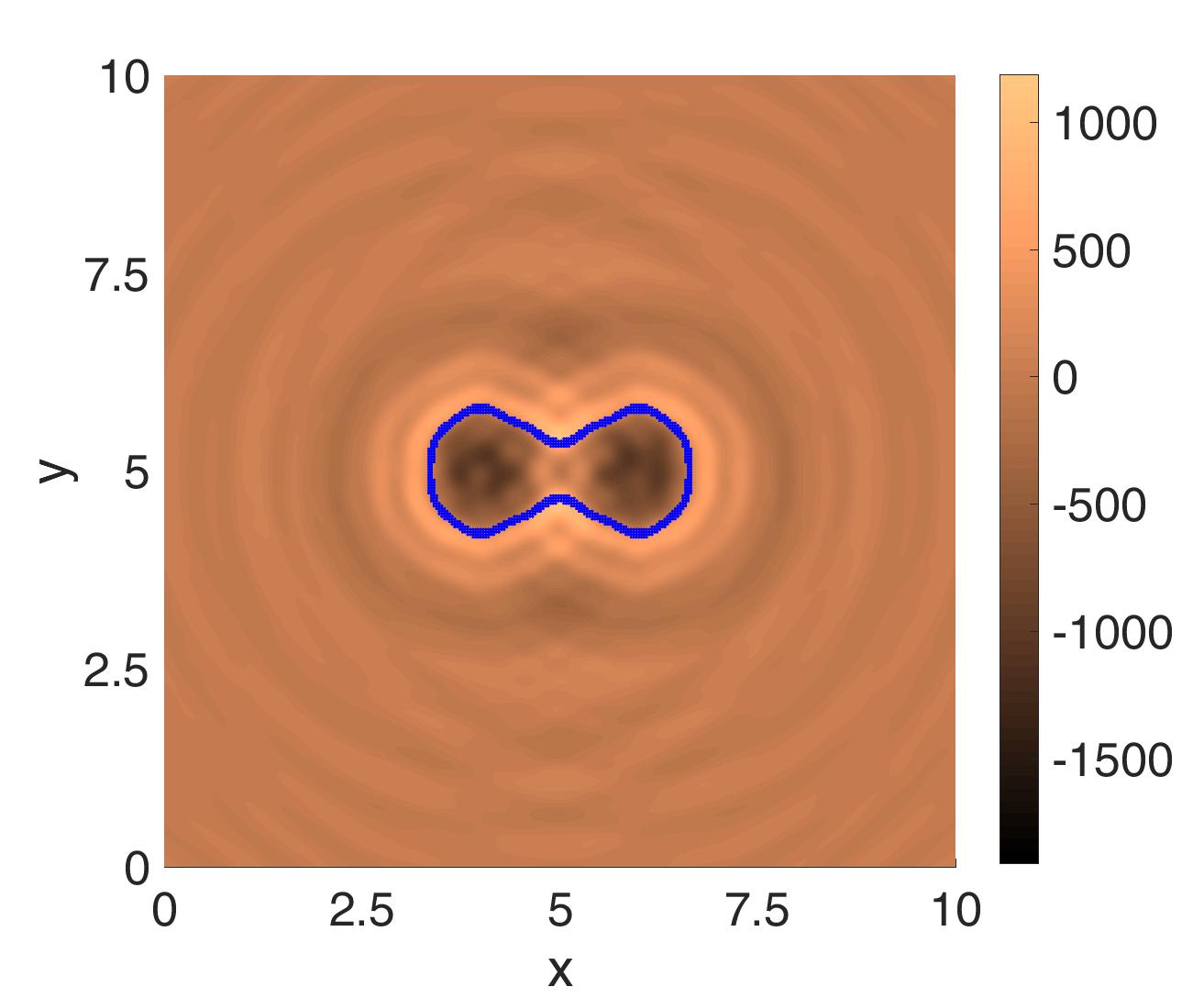}
\caption{Slices $z=5$ of the topological 
derivatives for the holography cost functional
(\ref{costH}), computed evaluating (\ref{dtempty}) with 
(\ref{forwardempty}) and (\ref{adjointemptyH}).
For ${k}_e= 12.6$ and ${k}_i=15.12$ 
(red light of wavelength $660 \, {\rm nm}$):
(a)  Sphere of radius $0.25$ ($250 \, {\rm nm}$, $1 \, {\rm nm}
= 10^{-9} \ {\rm m}$),
(b)  Spherocylinder of maximum radius  $0.25$ 
and length  $1$, oriented along the $x$ axis,
(c) Pear shaped object oriented along the $x$ axis, with 
maximum radius $0.25$.
For ${k}_e= 20.60$ and ${k}_i=24.79$
(violet light of wavelength $405$ {\rm nm}):
(d) Sphere of radius $0.5$.
(e)  Spherocylinder of maximum radius  $0.5$ 
and length  $2$, oriented along the $x$ axis.
(d)  Sand clock shaped object, with maximum radius 
$1$ and length  $4$, oriented along the $x$ axis.
The darkest regions represent the approximate shape. 
The true object borderline is  superimposed.
Objects are placed at the center of a box of size $10$
 in the imaging setting represented in 
Figure \ref{fig1}. The distance to the screen and to the 
emitter is $5$ ($5 \, \mu${\rm m},  $1 \, \mu{\rm m}
= 10^{-6} \ {\rm m}$). }
\label{fig2}
\end{figure}

\begin{figure}[h!]
\centering
\hskip -2mm (a) \hskip 3.2cm (c)  \hskip   3.4cm (e)\\
\includegraphics[width=4cm]{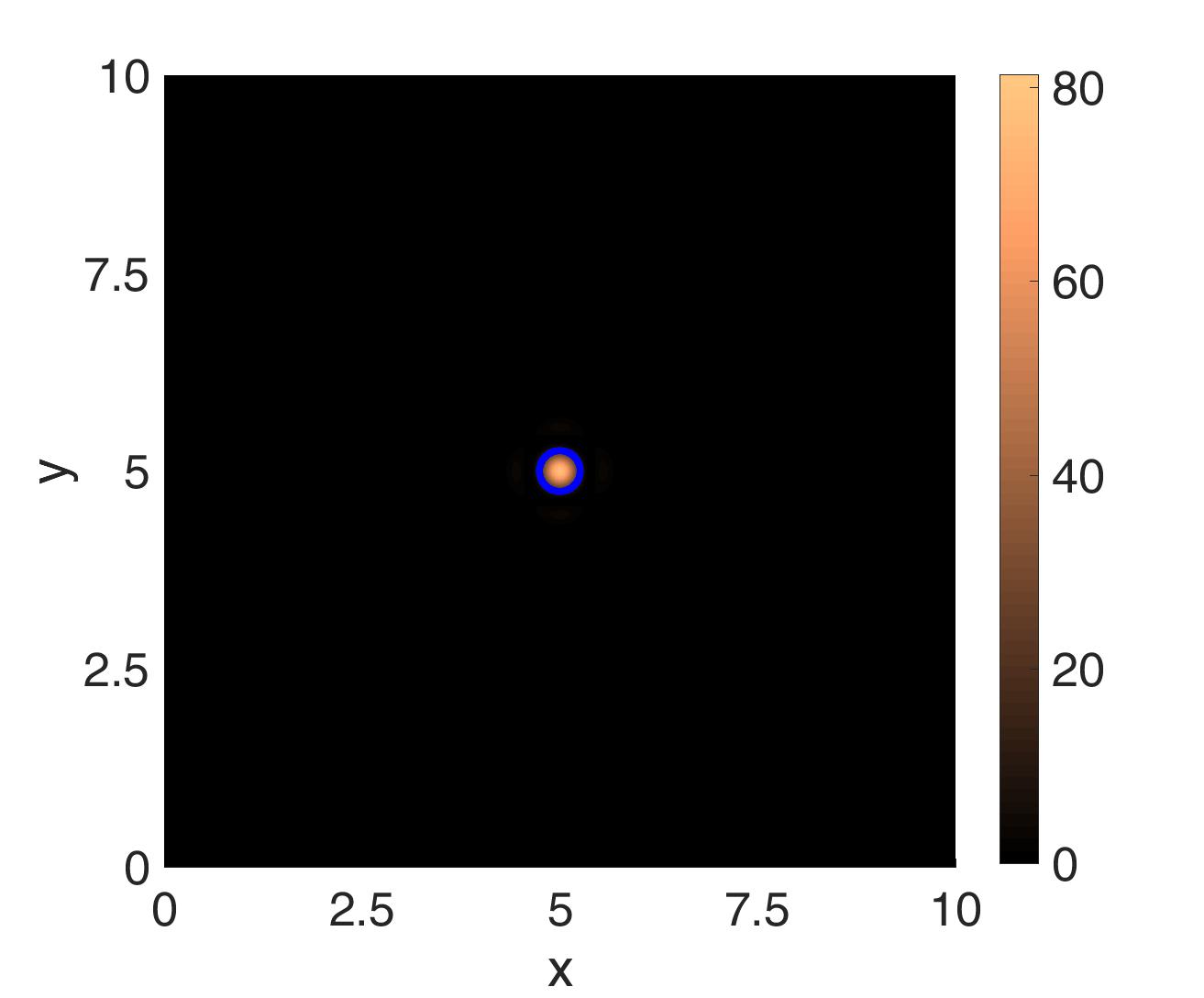} 
\hskip -3mm
\includegraphics[width=4cm]{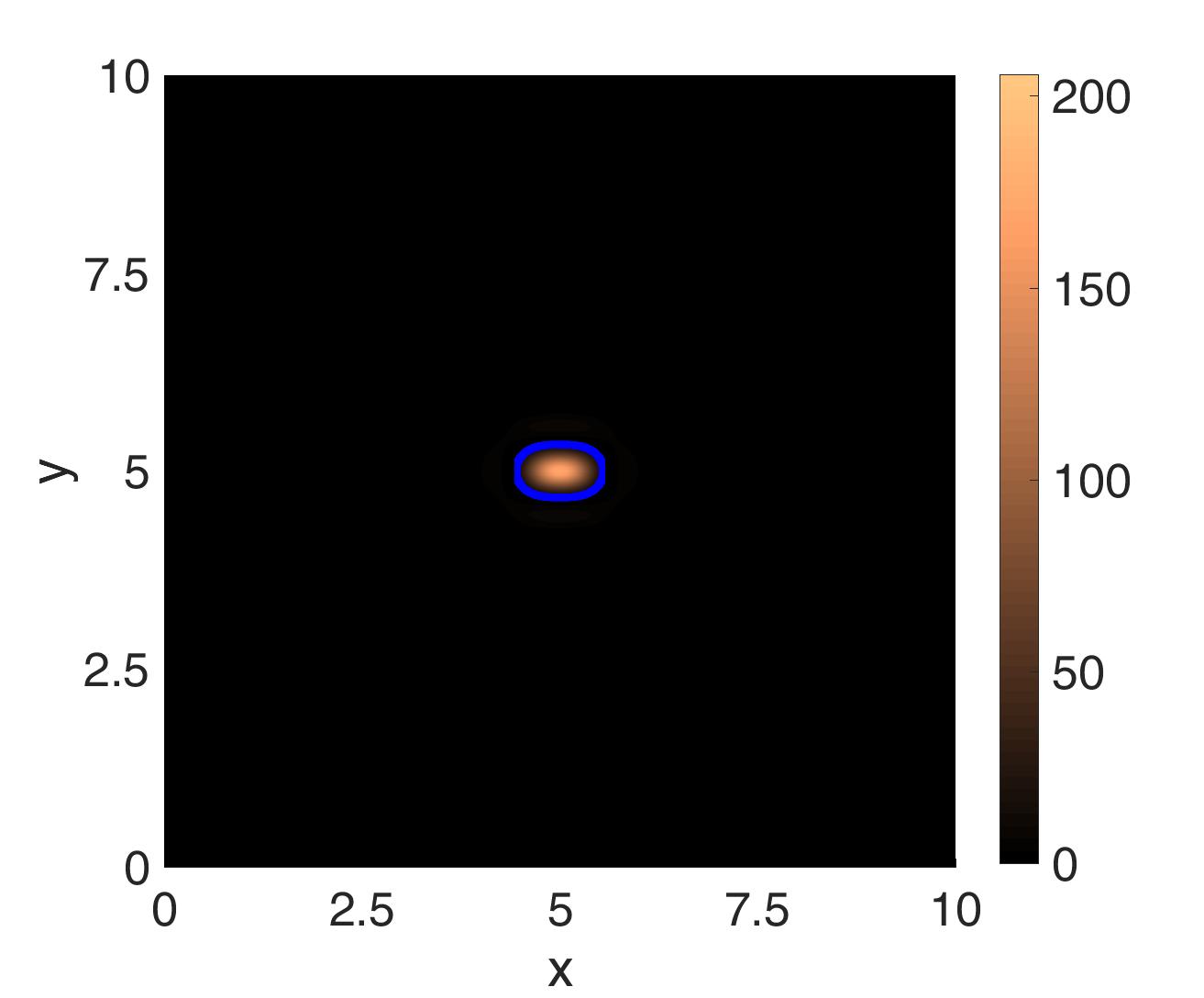} 
\hskip -2mm
\includegraphics[width=4cm]{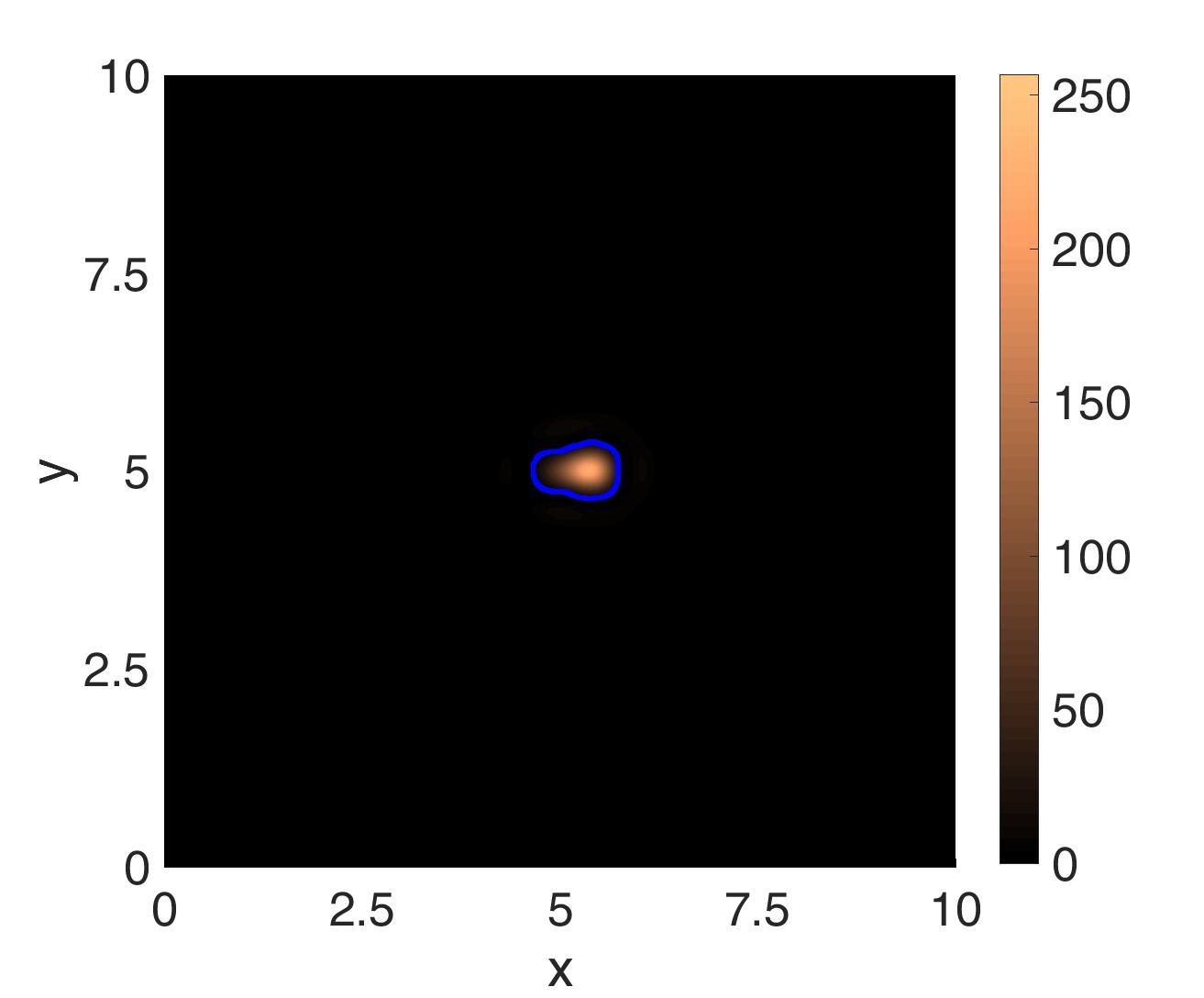} \\
\hskip -2mm (b) \hskip 3.2cm (d)   \hskip   3.4 cm (f)\\
\includegraphics[width=4cm]{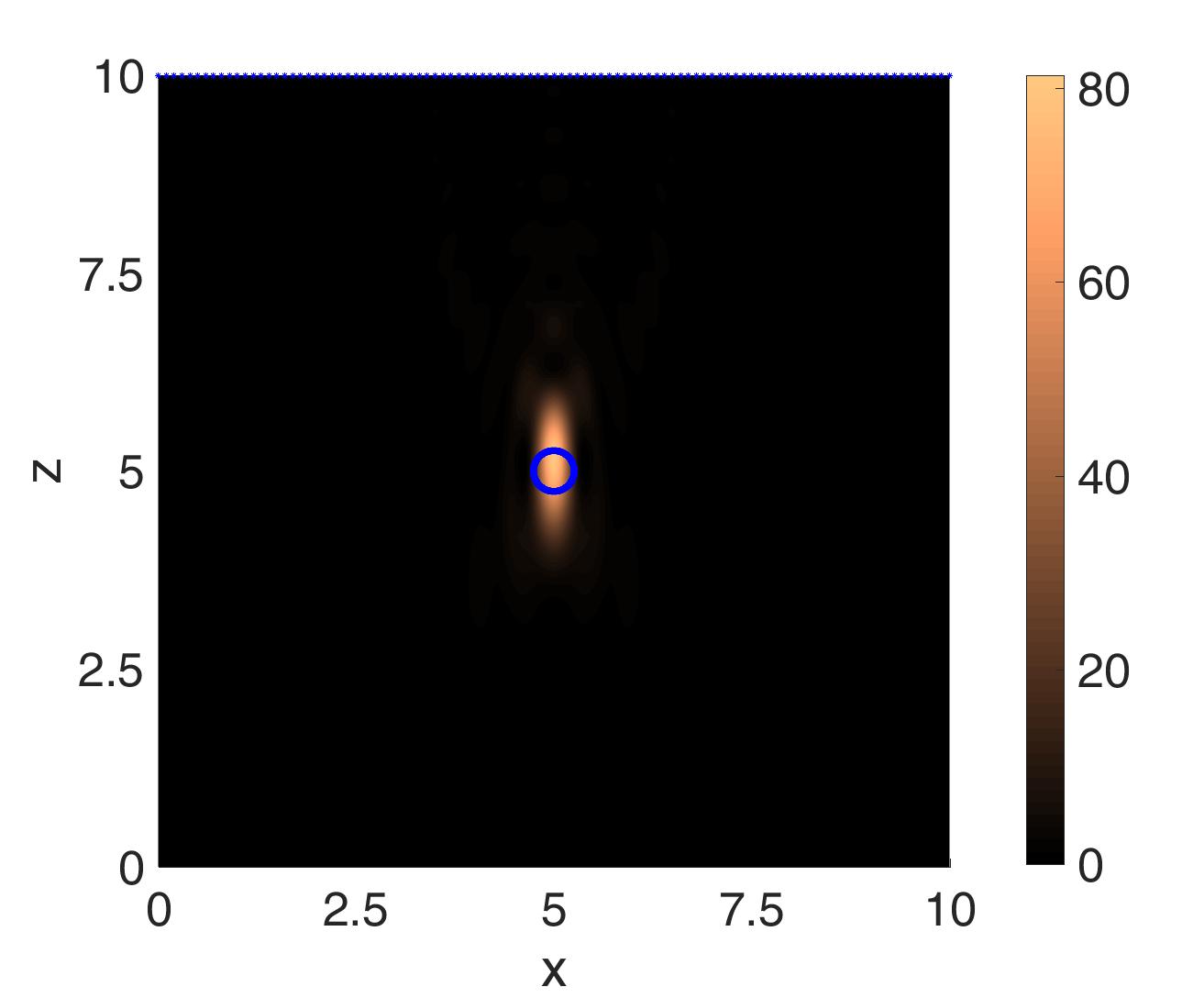} 
\hskip -3mm
\includegraphics[width=4cm]{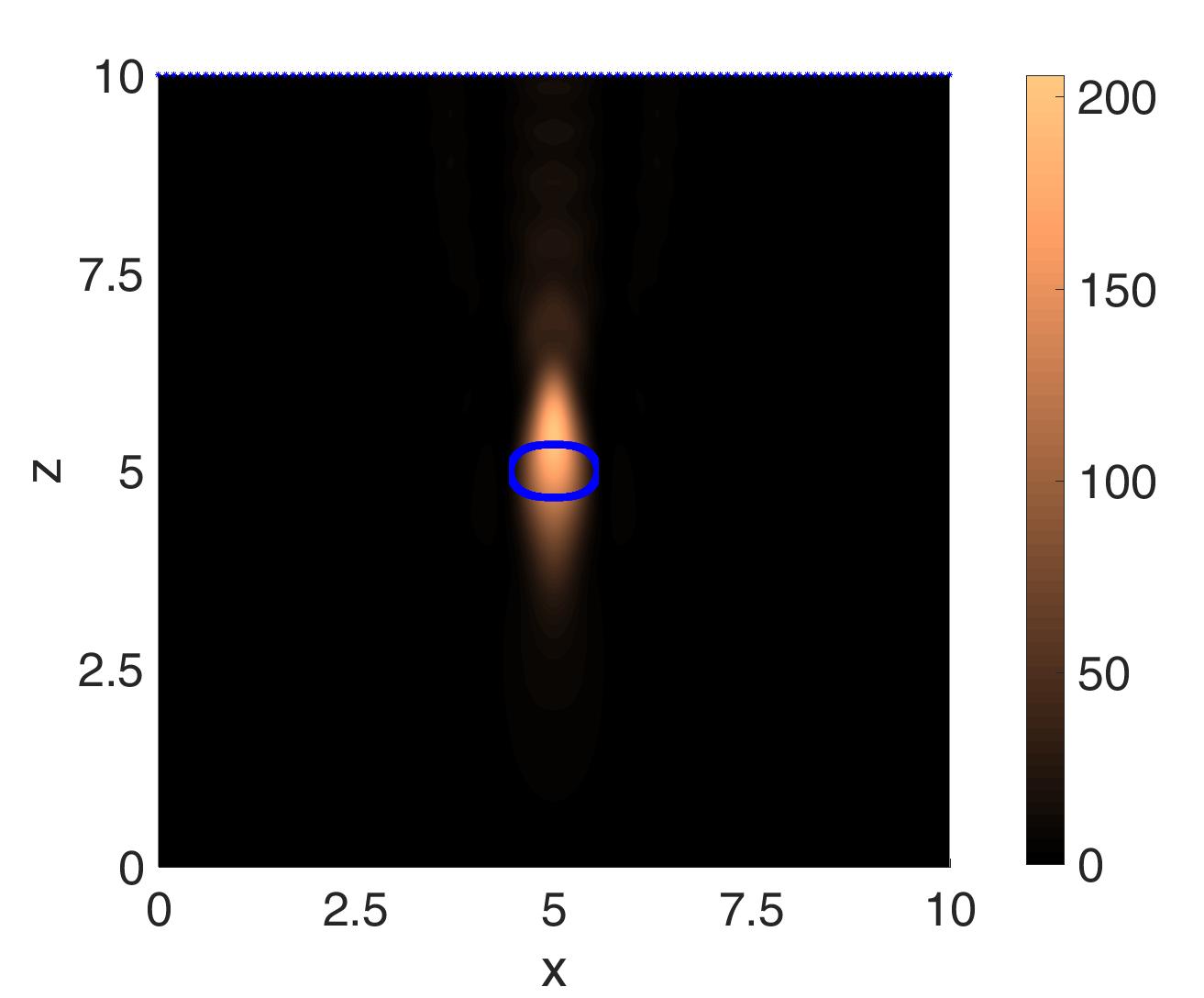} 
\hskip -2mm
\includegraphics[width=4cm]{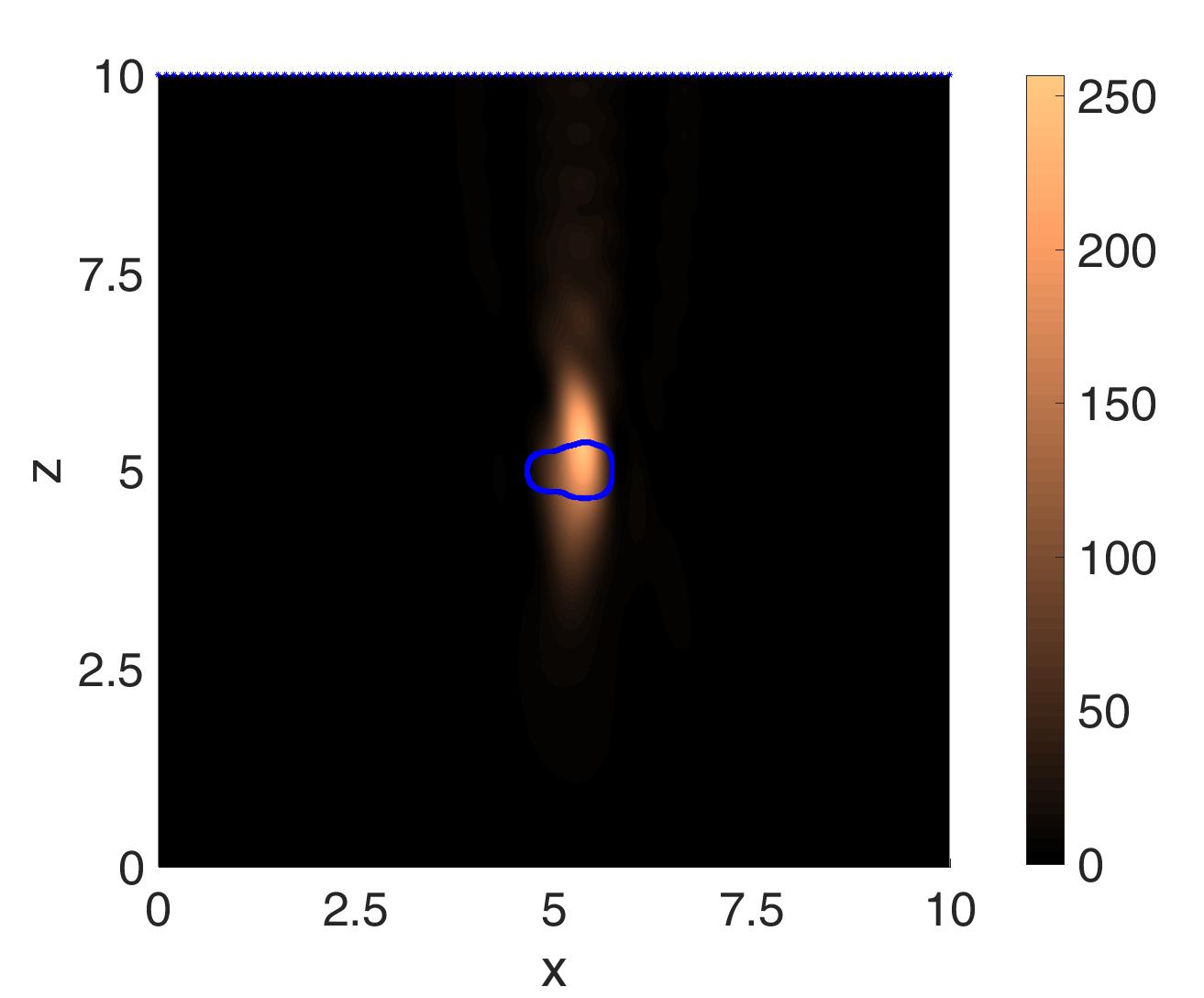}   
\caption{Slices $z=5$ and  $y=5$ of the topological 
energies of the holography cost functional
(\ref{costH}), given by (\ref{etempty}) with 
(\ref{forwardempty}) and (\ref{adjointemptyH})
for the same objects and parameters considered in Fig. \ref{fig2}
(a)-(c).
(a),(b)  Sphere.
(c),(d)  Spherocylinder.
(e),(f)   Pear shaped object. 
The brightest spots represent the approximate
objects. The true borderline  is superimposed.  
Whereas the slices $z=5$ reproduce correct shapes and positions,
the objects appear to be shifted and elongated in the $y=5$ slices.}
\label{fig3}
\end{figure}

\begin{figure}[h!]
\centering
\hskip -3mm (a) \hskip 4cm (b)   \\
\includegraphics[width=4.3cm]{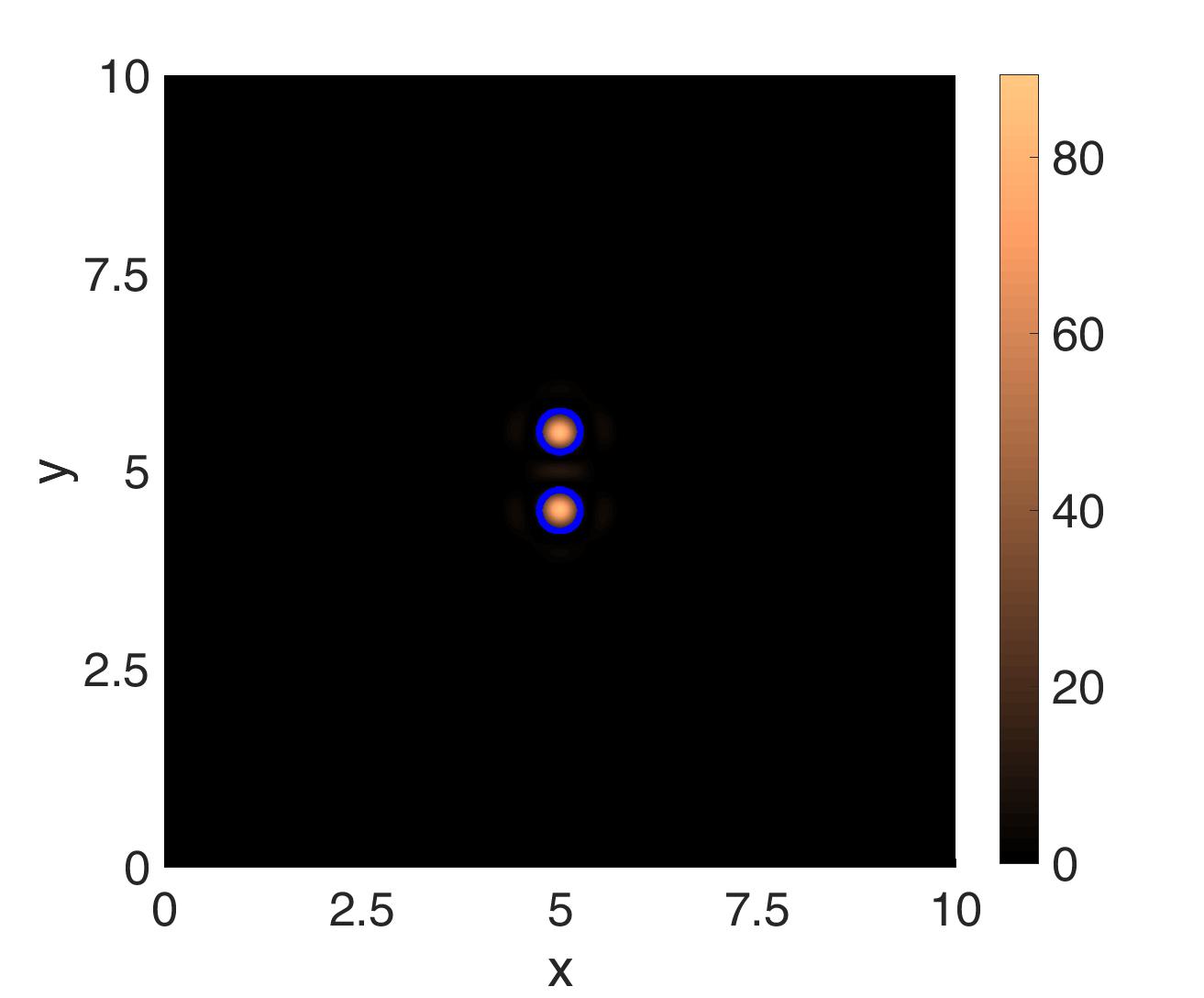}
\includegraphics[width=4.3cm]{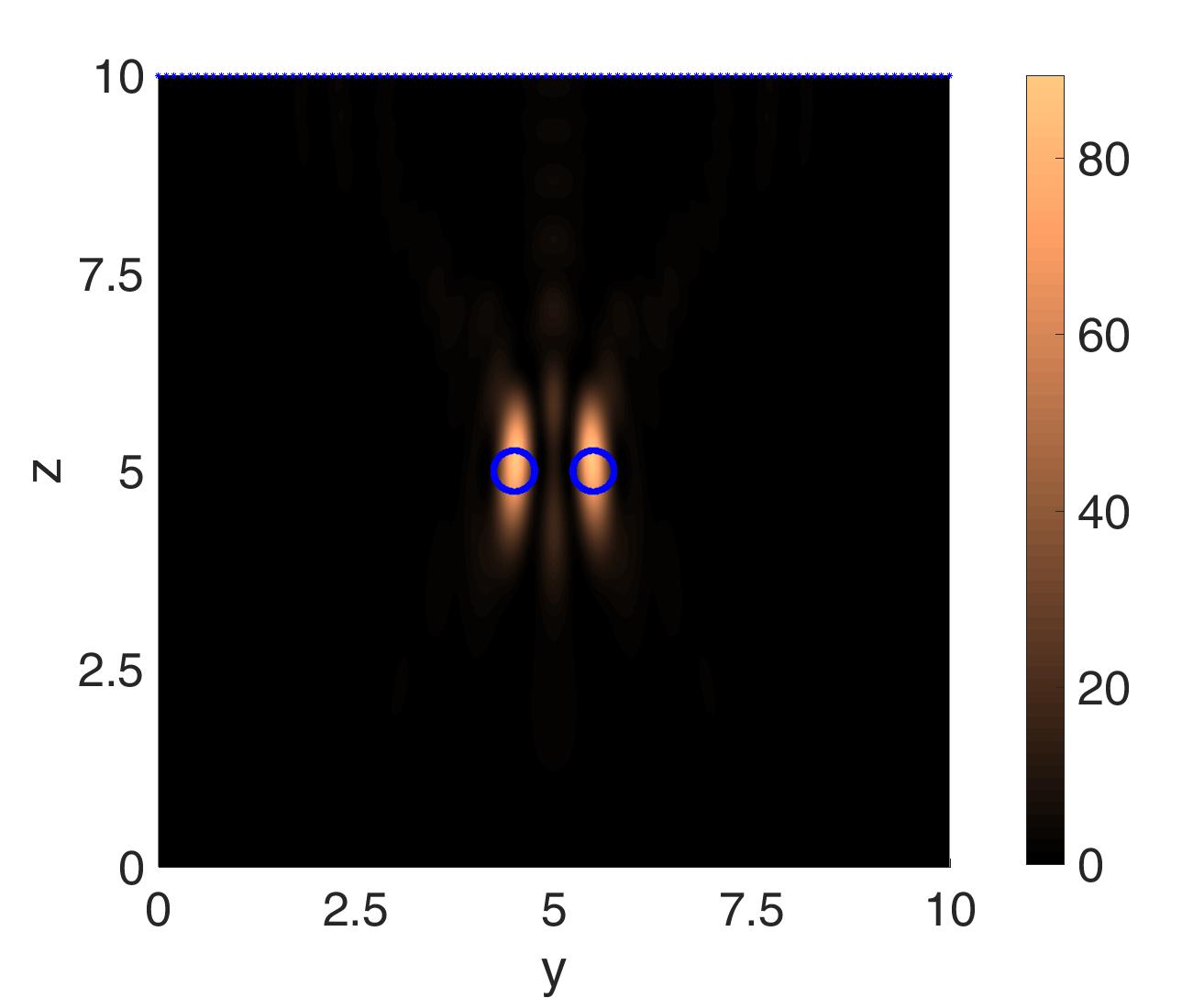} \\
\hskip -3mm (c) \hskip 4cm (d)   \\
\includegraphics[width=4.3cm]{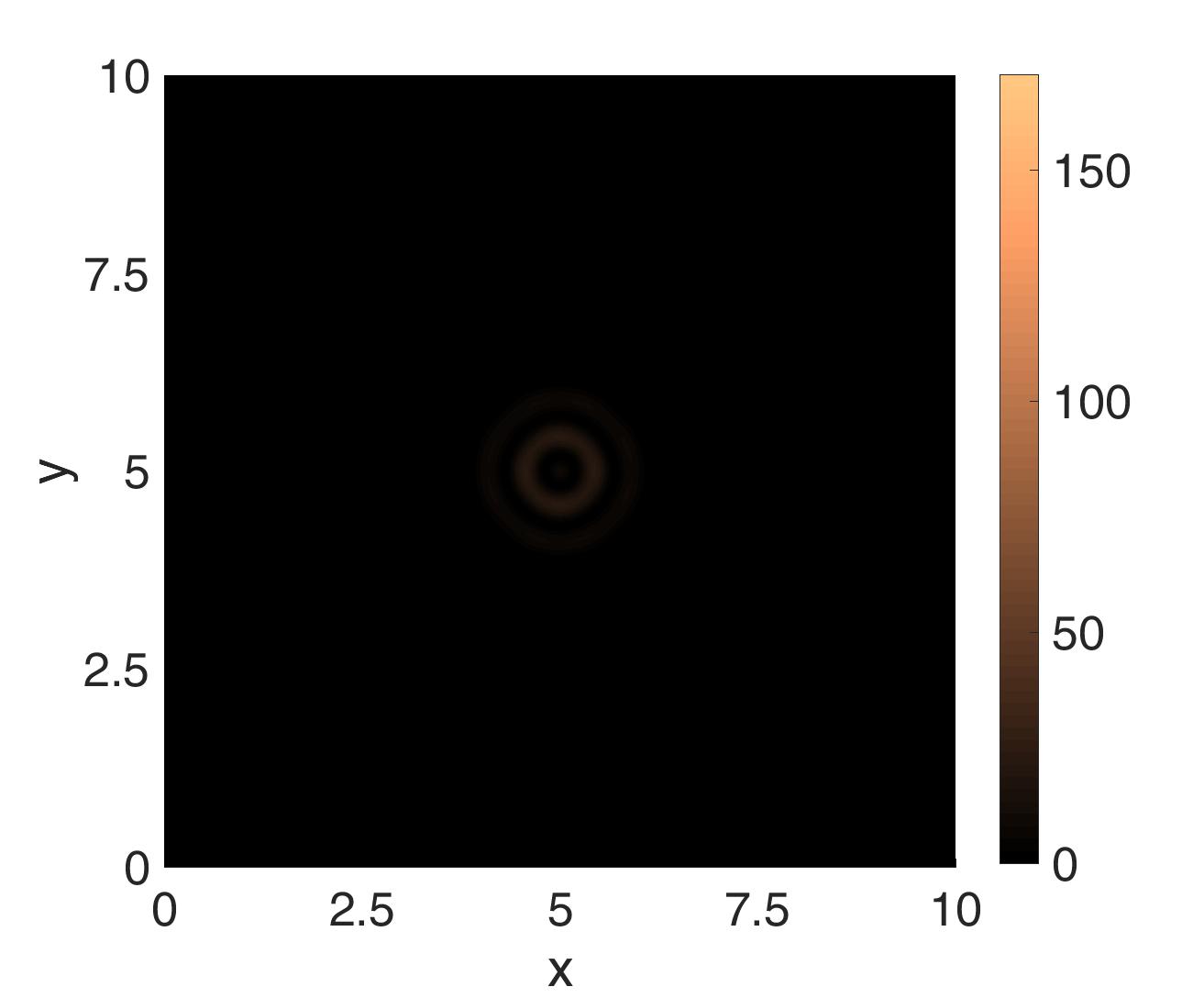}
\includegraphics[width=4.3cm]{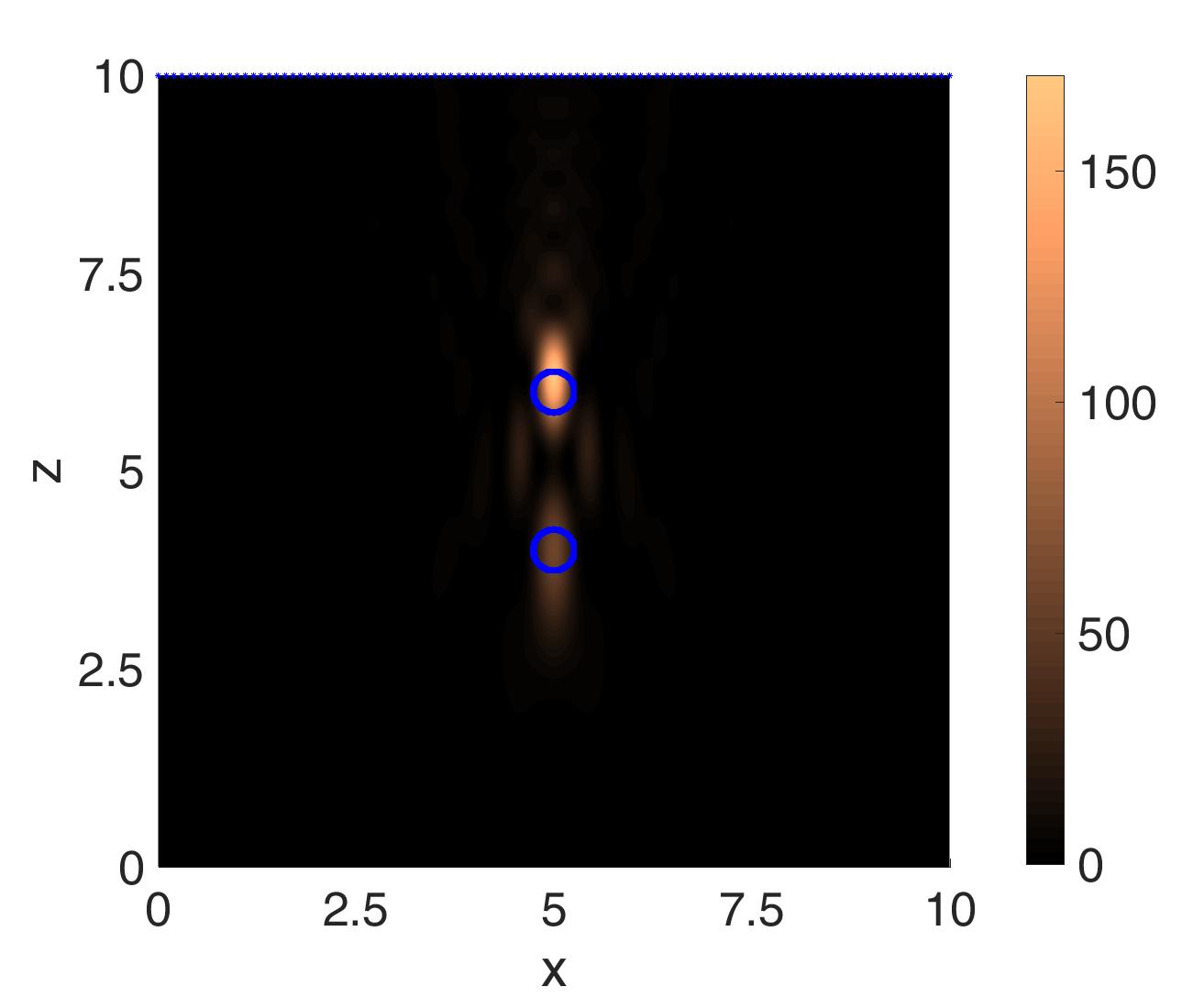} 
\caption{Slices $z=5$ and $x=5$ (resp. $y=5$) of the topological energies 
for the holography cost functional (\ref{costH}), 
computed evaluating (\ref{etempty}) with (\ref{forwardempty}) and 
(\ref{adjointemptyH}) when ${k}_e= 12.6$ and ${k}_i=15.12$.
(a),(b) Two spheres of radius $0.25$ ($250$ {\rm nm}), aligned along the $y$ 
axis and separated a distance $0.5$.
(c),(d) Two spheres of radius $0.25$, aligned along the $z$ axis,
located at $z=4$ and $z=6$. The borderline of the true objects
is superimposed.}
\label{fig4}
\end{figure}

The paper is organized as follows. Section \ref{sec:variational} provides 
a variational  formulation of the inverse scattering problem in holography,  assuming light polarized in the direction $x$ and neglecting the $y$,$z$ components of the electric field.  With pure $x$ polarization, we see $y$ polarization at around a factor of $15$ lower intensity, both for single spheres and sets of two spheres. 
Such observations motivate the assumption. 
We consider here geometrical shapes typically adopted 
by bacteria and viruses, whose sizes are of a similar order or
smaller than the wavelength.  For simple arrangements, the polarization
assumption we make is reasonable. As the geometrical configurations
become more complex, enhanced scattering may require the use
of the full vector Maxwell equations. The performance of the
technique in the vectorial case is to be explored.
Section \ref{sec:tholography} constructs initial approximations of the 
objects exploiting  the topological derivative and energy associated to 
a functional quantifying the deviation between the hologram generated
by any arbitrary shape at the recording screen and the hologram 
generated by  the true object. 
Both are calculated using the incident light beam and an 
explicit adjoint  field, which depends on the recorded hologram.
Topological fields quantify the variations of a shape functional when creating or 
modifying objects  in the  ambient medium. 
Multiple and non convex objects can be handled, down to nanometer 
sizes, see Figures \ref{fig2}-\ref{fig4}. 
As it is often the case in microscopy, working with one incident wave results here in a different resolution in planes orthogonal to the incidence direction and along the incidence direction.
Whereas the shape and location of the objects are correctly represented in orthogonal slices, they may be shifted and elongated in the incidence direction, specially for small sizes. Elongation and loss of axial resolution are also present in traditional holographic reconstruction techniques based on numerical backpropagation \cite{holo}.
For object sizes of the same order or smaller than the light wavelength, Section \ref{sec:shape} explains how to  reduce the offset and determine the number of object components by a topological derivative based iterative scheme.  
Moreover, we suggest strategies to combine these procedures with additional optimization  with respect to the permittivity of the objects to determine it when unknown.   
We illustrate the horizontal and axial resolution of the technique studying the test cases of two objects with different permittivities located on the same $xy$ plane and two objects aligned along the incidence  direction. By iteration, we are able to detect secondary objects with smaller contrast as well as objects screened by another scatterer aligned with them in the incidence direction, reducing in both cases the shift towards the screen in the objects  position. We observe that a good approximation of the scatterers location is essential to improve the predictions of their parameters.
Section \ref{sec:electricholography} explores an alternative initialization procedure which might help to reduce the elongation in the incidence direction, using the hologram to produce approximations of the electric field at the recording screen. The idea is to seek real and imaginary values that minimize the error when compared to the recorded hologram. This is done by a gradient method starting from the explicit field scattered by objects fitted to the peaks of a rough initial topological energy. Gradient optimization is alternated with Gaussian filtering to smooth out local  spikes. 
Section \ref{sec:conclusions} summarizes our conclusions. 
Two final appendices compute the derivatives of the holography cost functional 
with respect to  domains and coefficients and give explicit formulas for forward and adjoint fields in presence of penetrable spheres, which are useful to obtain the formulas for the topological derivatives and to lower the computational complexity in specific situations.

\section{Variational formulation of the inverse holography problem}
\label{sec:variational}

Electromagnetism equations  for light have been used to fit
real holograms to spheres and rods
in references \cite{inline,holotom}, for instance.
Ref. \cite{inline} exploits explicit Mie solutions of Maxwell's 
equations whereas ref. \cite{holotom} relies on discrete dipole 
approximations. For light polarized in the direction $x$ (resp. $y$), 
the component $x$ (resp. $y$) of the electric field is the relevant one.
Since we work with polarized light, we reduce the Maxwell system
to a single scalar wave equation for the relevant component.

In presence of a  set of  particles occupying a domain $\Omega$, 
part of the incident wave is transmitted inside the objects and 
the remaining portion is scattered. 
The evolution of the selected component
\[
{\mathcal E}(\mathbf{x},t):=\left\{\begin{array}{ll}
{\mathcal E}_{inc}(\mathbf{x},t)+{\mathcal E}_{sc}(\mathbf{x},t),&
\quad\mbox{$\mathbf{x}\in\mathbb R^3\setminus\overline{\Omega}$},\ \ t>0,\\
{\mathcal E}_{tr}(\mathbf{x},t),&\quad\mbox{$\mathbf{x}\in\Omega$}, \ \ t>0,
\end{array}
\right.
\]
is governed by the wave equation
\begin{equation}
\varepsilon(\mathbf{x}) {\mathcal E}_{tt}(\mathbf{x},t) -
{\rm div} \left({1\over {\mu(\mathbf{x})}} \nabla {\mathcal E}(\mathbf{x},t)
\right) = 0,
\label{ec:vble}
\end{equation}
where $\varepsilon$ represents the permittivity and $\mu$  the permeability.  
When the incident light is time-harmonic with frequency $\nu$, that is,
$
{\mathcal E}_{inc}(\mathbf{x},t)=\mbox{Re}[e^{-2\pi\imath \nu t} E_{inc}(\mathbf{x})],
$
the  resulting electric wave field $\mathcal E$ is time-harmonic too:  
${\mathcal E}(\mathbf{x},t)=\mbox{Re}[e^{-2\pi\imath\nu t} E(\mathbf{x})]$,
with (complex)  amplitude $E(\mathbf{x})$. 

The electric field $E_{sc}$
scattered by the objects $\Omega$ generating an 
hologram ${\mathcal I}$ satisfies 
\begin{eqnarray}
|E(\mathbf{x}_j)|^2= |E_{sc}(\mathbf{x}_j) + 
E_{inc}(\mathbf{x}_j)|^2 = {\mathcal I}(\mathbf x_j),
\quad j=1,\ldots, N,
\label{inverse}
\end{eqnarray}
at the detectors where the hologram ${\mathcal I}$ is recorded.
Finding the objects whose scattered electric field
satisfies (\ref{inverse}) is the inverse holography problem, which
can be recast as an optimization problem: Find regions 
$\Omega \subset {\mathbb R}^3$  minimizing
\begin{eqnarray}
J(\mathbb R^3 \setminus \overline{\Omega}) = {1\over 2}  \sum_{j=1}^N
| |E(\mathbf{x}_j)|^2- {\mathcal I}(\mathbf{x}_j)|^2, \label{costH}
\end{eqnarray}
where $E=E_{sc}+E_{inc}$ is the (complex) amplitude of the 
total electric field in presence of $\Omega$ and ${\mathcal I}$
is the measured hologram.
The true objects are a global minimum at which functional (\ref{costH})
vanishes. 

This formulation assumes the parameters characterizing the
optical properties of the objects known. 
When unknown, we may consider (\ref{costH}) a functional  
depending on two variables 
$J(\mathbb R^3 \setminus \overline{\Omega},k_i)$ and seek  
regions $\Omega \subset {\mathbb R}^3$ and 
parameter functions $k_i : \Omega \rightarrow {\mathbb R}^+$
minimizing $J(\mathbb R^3 \setminus \overline{\Omega},k_i)$ 
in presence of objects $\Omega$
with wavenumber $k_i$, $k_i=k_i(\nu,\mu,\varepsilon)$. 
When the variations in the wavenumber
are not abrupt enough, we may fix a region of interest $\Omega$,
consider (\ref{costH})  as a functional $J(k_i)$ and seek  
coefficient functions $k_i : \Omega \rightarrow {\mathbb R}^+$
minimizing $J(k_i)$. Here, we will mostly work with 
$J(\mathbb R^3 \setminus \overline{\Omega}),$
and occasionally with
$J(\mathbb R^3 \setminus \overline{\Omega}, k_i)$, 
since it is usually more efficient to track
abrupt variations in the wavenumber which would define objects
and then study wavenumber variations inside them if 
required \cite{ip2012}.

The equations governing $E$ act as a constraint in this 
optimization problem.
Nondimensionalizing, we may take $E$ to be a 
dimensionless amplitude obeying
\begin{equation}\label{forward}
\left\{\begin{array}{ll}
\Delta_{\mathbf x} E+ {k}_e^2 E=0,&\quad\mbox{in 
$\mathbb R^3\setminus\overline{\Omega}$},\\
\Delta_{\mathbf x}  E+ {k}_i^2 E=0,
&\quad\mbox{in $\Omega$},\\
E^-= E^+,&\quad \mbox{on $\partial\Omega$},\\
\beta (\partial_{\mathbf{n}}  E)^-= (\partial_{\mathbf{n}} E)^+, 
&\quad \mbox{on $\partial\Omega$},\\
\displaystyle\lim_{r:=|\mathbf{x}|\to \infty} r\big(\partial_r 
(E- E_{inc})- \imath  {k}_e(E- E_{inc})\big)=0.
\end{array}
\right.
\end{equation}
The incident wave is a plane wave in the direction $\hat{\bf d}
= (0,0,1)$, that is, $E_{inc}(\mathbf{x})=E_{inc}(x,y,z)=
e^{\imath {k}_e z}$. The detectors are located at the screen
$z=z_0$, see Figure \ref{fig1}.
The superscripts ``$+$" and ``$-$"  denote limits from the exterior 
and  the interior of $\Omega$ respectively. The symbol
$\partial_{\mathbf n}$ stands for normal derivative.
The condition at infinity is the 
standard Sommerfeld radiation  condition allowing only outgoing 
waves, where $\partial_r$ represents radial derivatives.

Dimensions are restored by setting
$\mathbf{\tilde x}=\mathbf{x} L$,  
$\tilde \Omega=\Omega L$,  
$\tilde E(\mathbf{\tilde x})=E_0 E(\mathbf{x} L)$, 
$\tilde E_{inc}(\mathbf{\tilde x})=E_0 E_{inc}(\mathbf{x} L)
= E_0 e^{\imath \tilde k_e  \hat{\mathbf{d}} \cdot \mathbf{\tilde x}},$
where $L$ is a reference length unit, $E_0$ a reference field, and
\begin{eqnarray} 
\tilde k_e:=k_e/L,\qquad 
\tilde k_i:= k_i/L.
\label{kL}
\end{eqnarray}
The wavenumbers $\tilde k_e$ and $\tilde k_i$ are usually expressed in 
terms of refractive indexes $n_i,n_e$  (ratio of the speed of light in the 
vacuum to  the speed in a particular  medium):
\begin{equation}
\tilde k_e=\frac{2\pi\nu}{c_e} = \frac{2\pi }{\lambda} n_e, \qquad
\tilde k_i= \frac{2\pi\nu}{c_i} = \frac{2\pi }{\lambda} n_i, 
\label{wavenumber}
\end{equation}
where $\lambda$ is the wavelength of the employed light in the
vacuum. Assuming the permittivity $\varepsilon$ and 
permeability $\mu$ to be constant in the
ambient medium and in the scatterers, the local
wave speeds are $c_e={1\over \sqrt{\mu_e \varepsilon_e}}$
for the medium and $c_i={1\over \sqrt{\mu_i \varepsilon_i}}$
for the objects, with $\mu_e= \beta  \mu_i.$
For biological samples $\beta \sim 1$. 
We will set $\beta=1$ in our numerical tests.

For arbitrary shapes $\Omega$ and constant coefficients, 
one can solve an integral  reformulation of any boundary
interior or exterior problem for Helmholtz equations. Here
we make use of an integral representation of the solution
in the exterior domain and deduce a non symmetric 
formulation which combines boundary elements (BEM)
and finite elements (FEM) \cite{siims,fem,selgas}.
We discretize this formulation by means of piecewise
constant boundary elements and continuous $\mathbb P1$
finite elements, so that our solving scheme has order one.
This method allows us to generate the synthetic holograms
employed here for the reconstructions.
Also, notice that for isolated spherical objects $\Omega$ and 
constant wave speeds both in the medium and in the objects,
the scalar system  (\ref{forward}) 
admits explicit solutions, expressed as series expansions
in terms of spherical harmonics 
\cite{sphericalkirsch,sphericalturley}, see Appendix
\ref{sec:explicitforwardadjoint}.
This is a scalar version of the Mie solutions for the vector 
Maxwell equations \cite{bookmaxwell,miecodesschafer}.

To address the general optimization problem for (\ref{costH})
without a priori  information on the number of
scattering objects,  their geometry and  approximate location 
we  need to calculate the derivatives of the 
holography cost functional with respect to domains and coefficients.
This is done in Appendix \ref{sec:derivatives}.

\section{Initial predictions of shapes from recorded holograms}
\label{sec:tholography}

Topological methods provide a way to quantify variations
of shape functionals due to the presence of objects, which
allows us  to construct guesses of objects
from the knowledge of the hologram and the optical properties 
of the ambient medium. The idea is the following.
Consider a functional $J({\cal R})$, defined in a region 
${\cal R}\subset \mathbb R^3$.  Removing from it
a ball of radius $\varepsilon$ centered about $\mathbf x$, 
the expansion
\begin{equation}\label{expansion}
J({\cal R} \setminus \overline{B_\varepsilon (\mathbf{x})})=
J({\cal R})+{4\over 3} \pi \varepsilon^3 D_T(\mathbf{x},{\cal R})
+o(\varepsilon^3),\qquad \varepsilon\to 0,
\end{equation}
holds. $D_T(\mathbf{x},{\cal R})$ is the topological derivative of the 
functional at ${\mathbf x}$.  As explained in Appendix
\ref{sec:derivatives}, it measures the variation of the functional when 
an object is placed at $\mathbf x$ \cite{Sokowloski} and it is used to 
localize abrupt  changes. If $D_T(\mathbf{x},{\cal R})<0$,  then 
$J({\cal R}\setminus \overline{B_\varepsilon (\mathbf{x})})<J({\cal R})$ 
for $\varepsilon>0$ small. 
Therefore, when we locate objects in regions where the topological 
derivative takes large negative values, the functional is expected to 
diminish \cite{feijoo}.
In this way, we are able to predict the number, location  and size of 
the scatterers.

Evaluating the topological derivative at a point using expansion 
(\ref{expansion}) is too expensive for practical purposes.  Instead, 
we use an  expression in terms of adjoint and forward fields, 
see Appendix \ref{sec:derivatives} for a derivation of these 
formulas and technical details. When we have no information on the 
object, we calculate $D_T(\mathbf{x},\mathbb R^3)$ by means of the
explicit formula:
\begin{eqnarray} \hskip 3mm
D_T(\mathbf{x},\mathbb R^3) =
{\rm Re} \left[ 3{1 - \beta \over 2 + \beta}
\nabla E(\mathbf{x})\! \cdot \! \nabla
 \overline{P}(\mathbf{x})
\!+\! (\beta k_i^2  - k_e^2)  E(\mathbf{x})   \overline{P}(\mathbf{x})
\right], \; \mathbf x \in \mathbb R^3,
\label{dtempty}
\end{eqnarray}
where $E$ is the solution of the forward problem: 
\begin{eqnarray} \label{forwardempty}
\left\{
\begin{array}{ll}
\Delta E + {k}_e^2 E=  0 &\quad\mbox{in ${\mathbb R}^3$}, \\
\displaystyle\lim_{r\to \infty} r\big(\partial_r (E-E_{inc}) 
-\imath {k}_e (E-E_{inc}) \big)=0,
\end{array}
\right.
\end{eqnarray}
and $\overline{P}$ is the solution of the conjugate adjoint problem:
\begin{eqnarray} \label{adjointempty}
\left\{
\begin{array}{ll}
\Delta \overline{P} + {k}_e^2 \overline{P}= 2 \sum_{j=1}^N
 ({\mathcal I}-|E|^2) \overline{E} \, \delta_{{\mathbf x}_j}
&\quad\mbox{in ${\mathbb R}^3$},
\\
\displaystyle\lim_{r\to \infty} r\big(\partial_r \overline{P} 
-\imath {k}_e\overline{P} \big)=0,
\end{array}
\right.
\end{eqnarray}
$\delta_{\mathbf x_j}$ being Dirac masses supported at the
receptors. 
For an incident plane wave, $E=E_{inc}$ and $\overline{P}$ is given by:
\begin{eqnarray}\label{adjointemptyH}
\overline{P}({\mathbf x})
=- \sum_{j=1}^N {e^{\imath {k}_e |{\mathbf x}-{\mathbf x_j}|}
\over 4 \pi |{\mathbf x}-{\mathbf x_j}|}
[2 ({\mathcal I}({\mathbf x_j})-|E({\mathbf x_j})|^2 ) 
\overline{E}(\mathbf x_j) ].
\end{eqnarray}
When $\beta\sim 1$, the gradients disappear from (\ref{dtempty})
and $D_T(\mathbf{x},\mathbb R^3) =
(k_i^2  - k_e^2)  E(\mathbf{x})   \overline{P}(\mathbf{x}).$

The topological energy is a companion field of the topological derivative
which has the ability of canceling oscillations as $k_e$ or the number 
of objects grow \cite{topologicalenergy}:
\begin{eqnarray}
E_T(\mathbf{x},{\mathbb R}^3) =  |E(\mathbf{x})|^2  |{P}(\mathbf{x})|^2.
\label{etempty}
\end{eqnarray}
Peaks of the topological energy indicate the location of objects.
Plotting the topological derivatives instead, the regions where large 
negative values are attained represent object shapes.
Our simulations show an effect often observed in microscopy: 
very different resolution in planes orthogonal to the incidence direction of 
the light beam when compared with resolution along the incidence direction.
Whereas $xy$ slices capture correct positions and shapes, the objects
appear to be elongated and shifted in the incidence direction, see Figures
2-4.

\begin{figure}[h!]
\centering
\hskip -2mm (a) \hskip 3cm (b) \hskip 3.1cm (c) \\
\includegraphics[width=4cm]{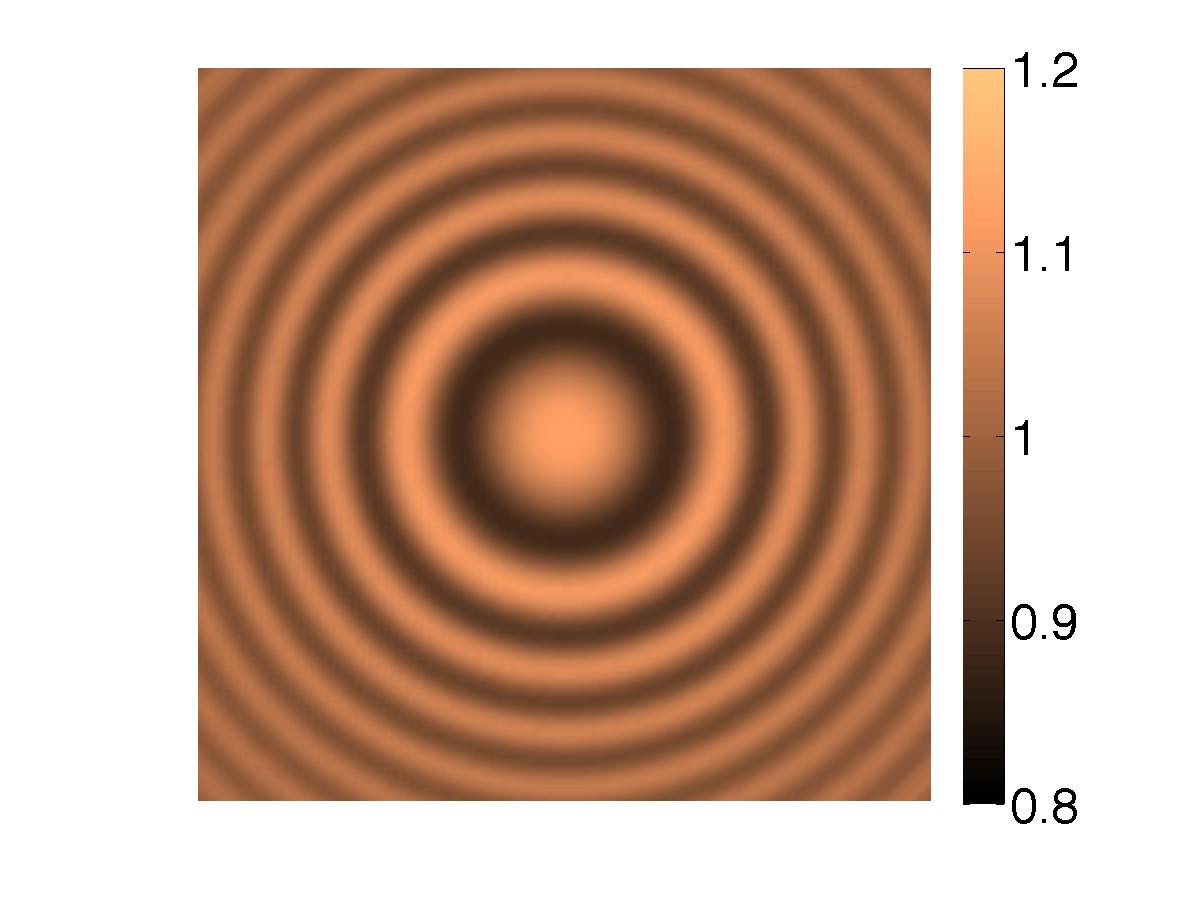} \hskip -5mm
\includegraphics[width=4cm]{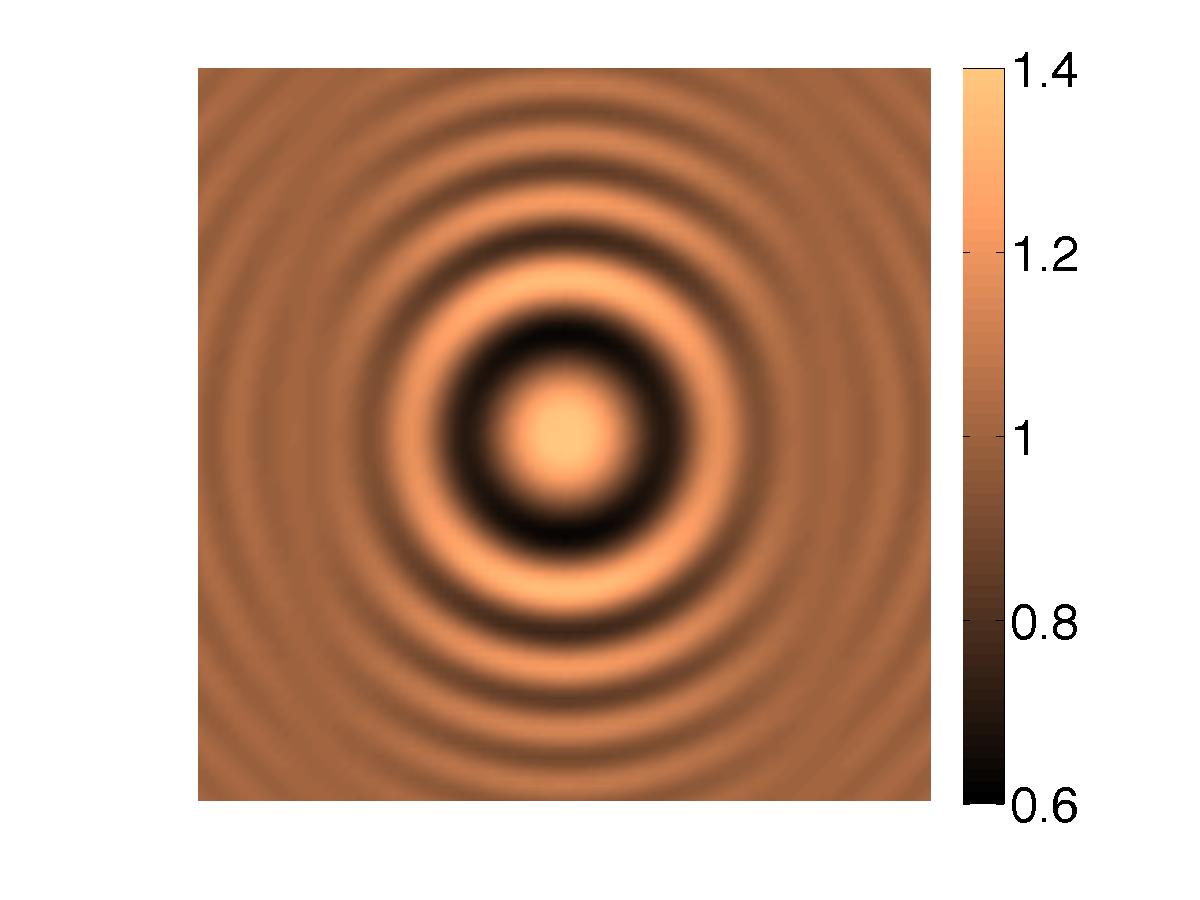} \hskip -5mm
\includegraphics[width=4cm]{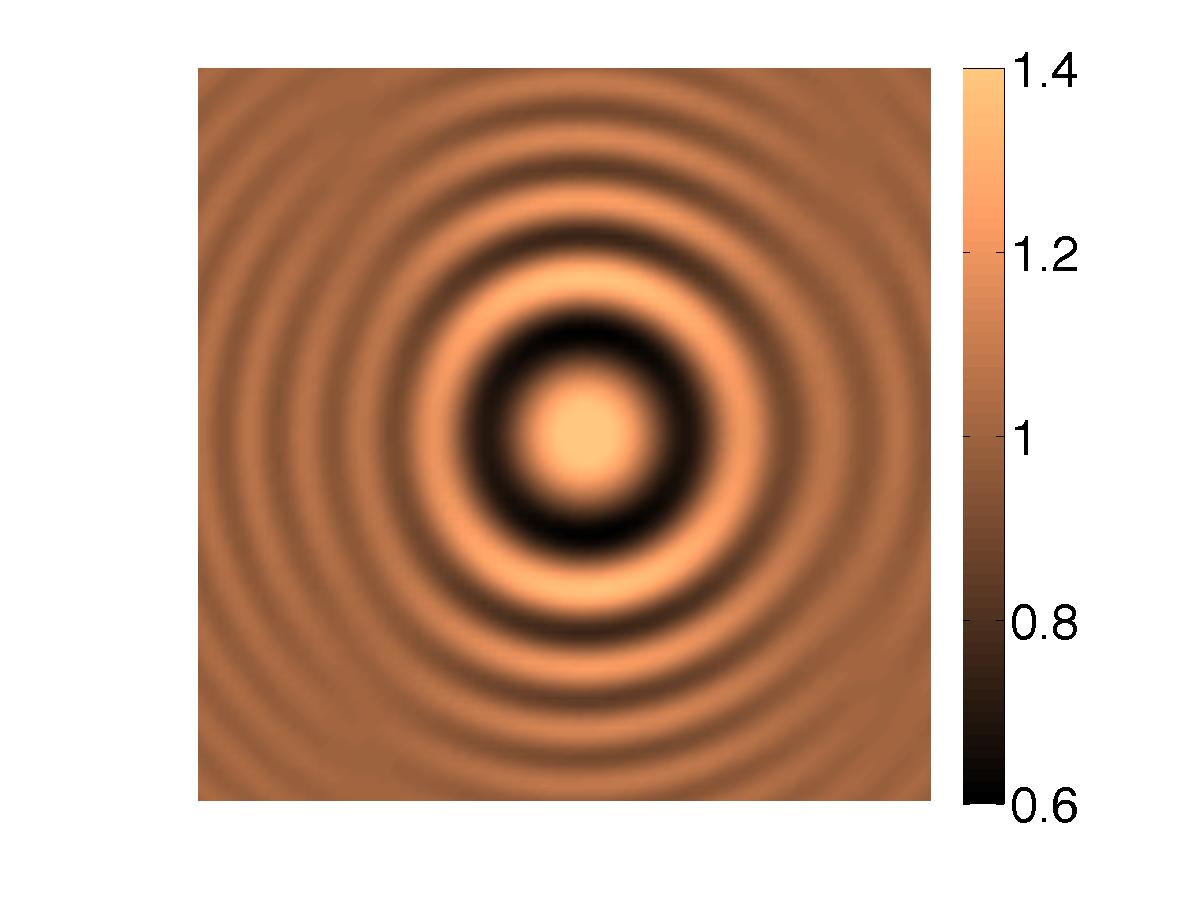} \\
\hskip -2mm (d) \hskip 3cm (e)  \\
\hskip 0.0cm \includegraphics[width=4cm]{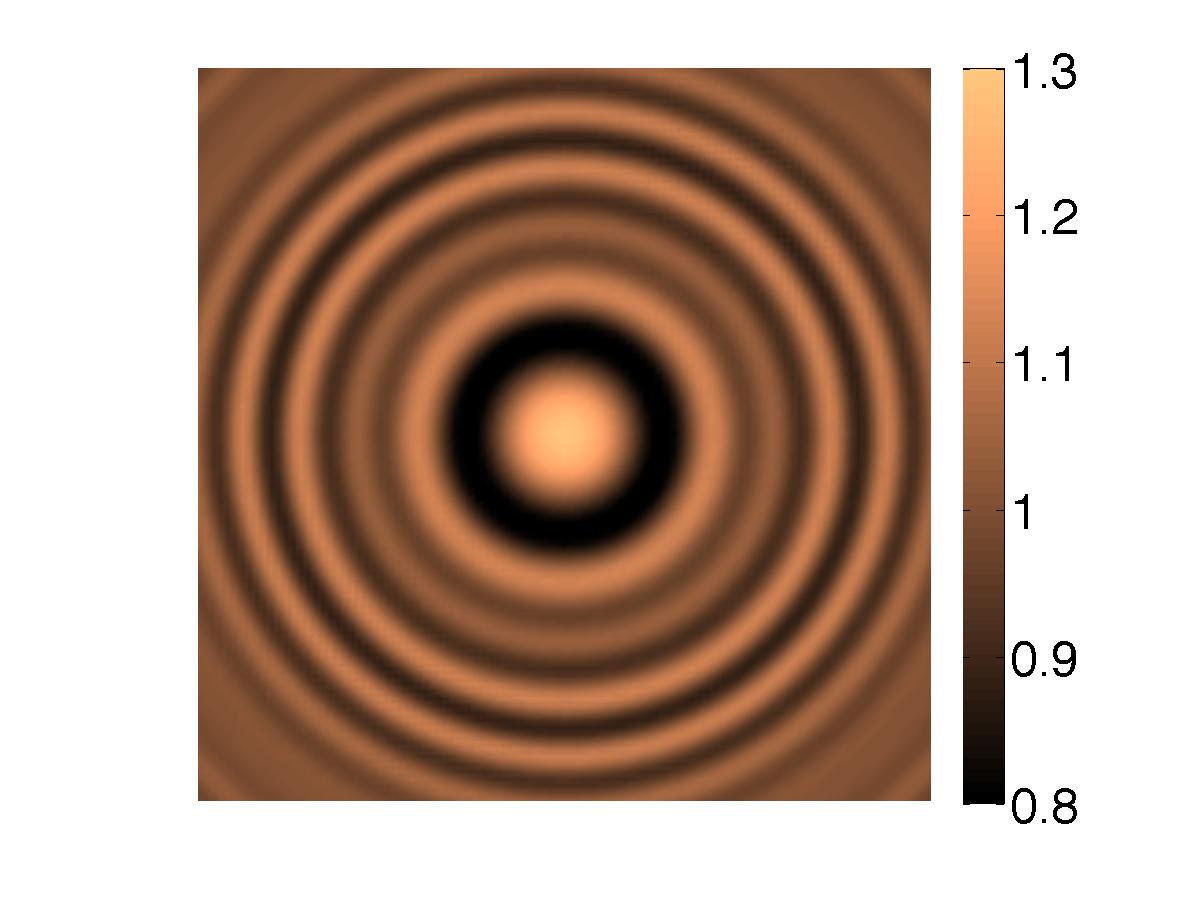}  \hskip -5mm
\includegraphics[width=4cm]{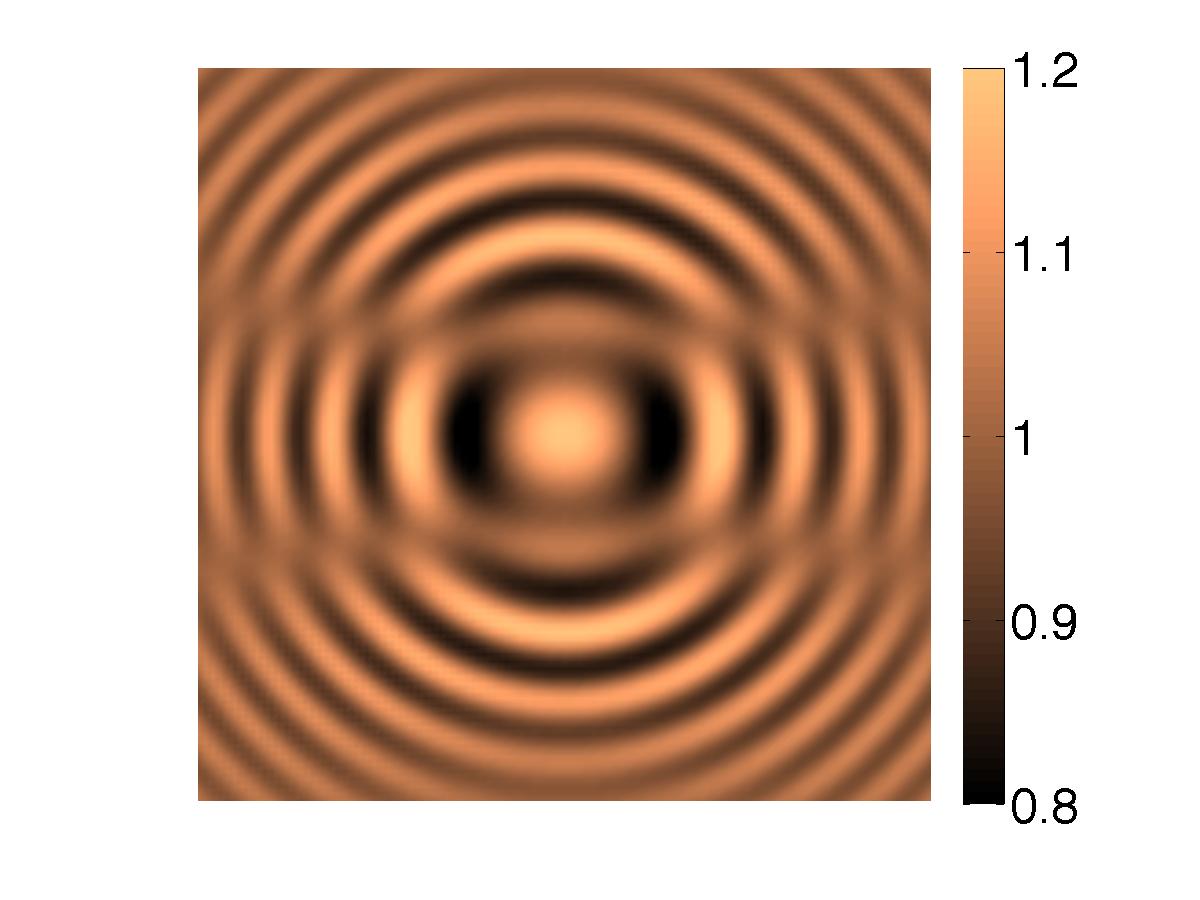} \\
\caption{Synthetic holograms ${\cal I}=| E_{sc}+E_{inc}|^2$
for the shapes considered in Figures
\ref{fig3}-\ref{fig4}: (a) Sphere, 
(b) Spherocylinder,
(c) Pear-shaped object, 
(d) Two spheres of the same material, aligned along the $z$ axis,
(e) Two spheres of the same material, aligned along the $y$ axis.
}
\label{fig5}
\end{figure}

The technique is tested in the imaging setting represented in Figure \ref{fig1} with light of wavelength $660 \, {\rm nm}$ (red) and $405 \, {\rm nm}$ (violet). 
The light emitter is placed at $z=0$. A hologram recording screen of size 
$10 \mu{\rm m} \times 10 \mu{\rm m}$ is  located at $z=10 \mu{\rm m}$.  
The problem is nondimensionalized choosing $ L=1 \, \mu$m, representative of the average object size considered. The original permittivities are typical of biological media and $\beta=1$.  
Many bacteria display spherical or rod-like
shapes and their size lies in the range of a few microns. They adopt the forms depicted in Fig \ref{fig2}(c) and Fig \ref{fig2}(f) when they are moving or dividing. Viruses also adopt simple geometrical shapes, with typical sizes ranges from $500$ nm down to $30$ nm.
Here, the data are synthetic holograms generated solving the forward problem (\ref{forward}) for the true objects by BEM-FEM, as described in \cite{siims,fem,selgas}.
Figure \ref{fig5} displays some of the synthetic holograms ${\cal I}=|  E_{sc}+E_{inc}|^2$  used in the holography cost functional (\ref{costH}) and for the calculation of the adjoint (\ref{adjointemptyH}).
For specific objects, we have checked that holograms generated by BEM-FEM, DDA \cite{ddayurkin} and/or the series solutions  in Appendix \ref{sec:explicitforwardadjoint} agree. The relative error in the difference is of order $10^{-1}$ for our typical choices of the wavelength and the mesh size. 
The  numerical error can be considered as noise;  indeed, we have checked that we
do not commit any numerical crime by seeing that the introduction of random noise
in the data does not lead to significant change in our reconstructions.
Topological methods have been shown to be very robust to the presence of noise.

\tikzstyle{block1} = [rectangle, draw, fill=blue!20, 
    text width=22em, text centered, rounded corners, minimum height=2em]
\tikzstyle{block2} = [rectangle, draw, fill=blue!20, 
    text width=22em, text centered, rounded corners, node distance=1.1cm, 
    minimum height=2em]
\tikzstyle{line} = [draw, -latex']
\tikzstyle{cloud} = [draw, ellipse,fill=red!20, node distance=7cm,
    minimum height=2em, text width=5em]
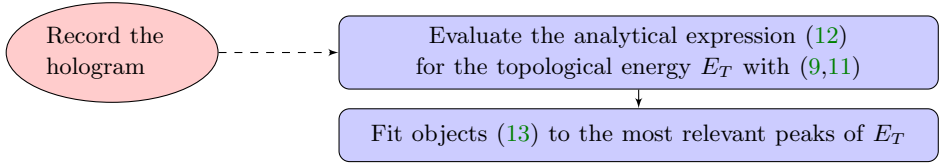
\begin{figure}[h!]
\centering
\begin{tikzpicture}[node distance = 1.5cm, auto]
    \node [block1] (ETapprox) {\small Evaluate the analytical expression (\ref{etempty}) for the topological energy $E_T$ with   (\ref{forwardempty},\ref{adjointemptyH})};
    \node [cloud, left of=ETapprox] (measures) {\small Record the hologram};
    \node [block2, below of=ETapprox] (Omega0) 
    {\small Fit objects (\ref{initialguess}) to the   most  relevant peaks of $E_T$};
    \path [line,dashed] (measures) -- (ETapprox);
    \path [line] (ETapprox) -- (Omega0);
\end{tikzpicture}
\caption{Procedure to recover objects from the hologram
in our imaging setting, for the cost functional (\ref{costH}) and
sizes similar to or smaller than the wavelength. }
\label{fig6}
\end{figure}

\begin{figure}[b!]
\centering
\hskip -4mm (a) \hskip 3.3cm (c) \hskip 3.3cm (e) \\
\includegraphics[width=4cm]{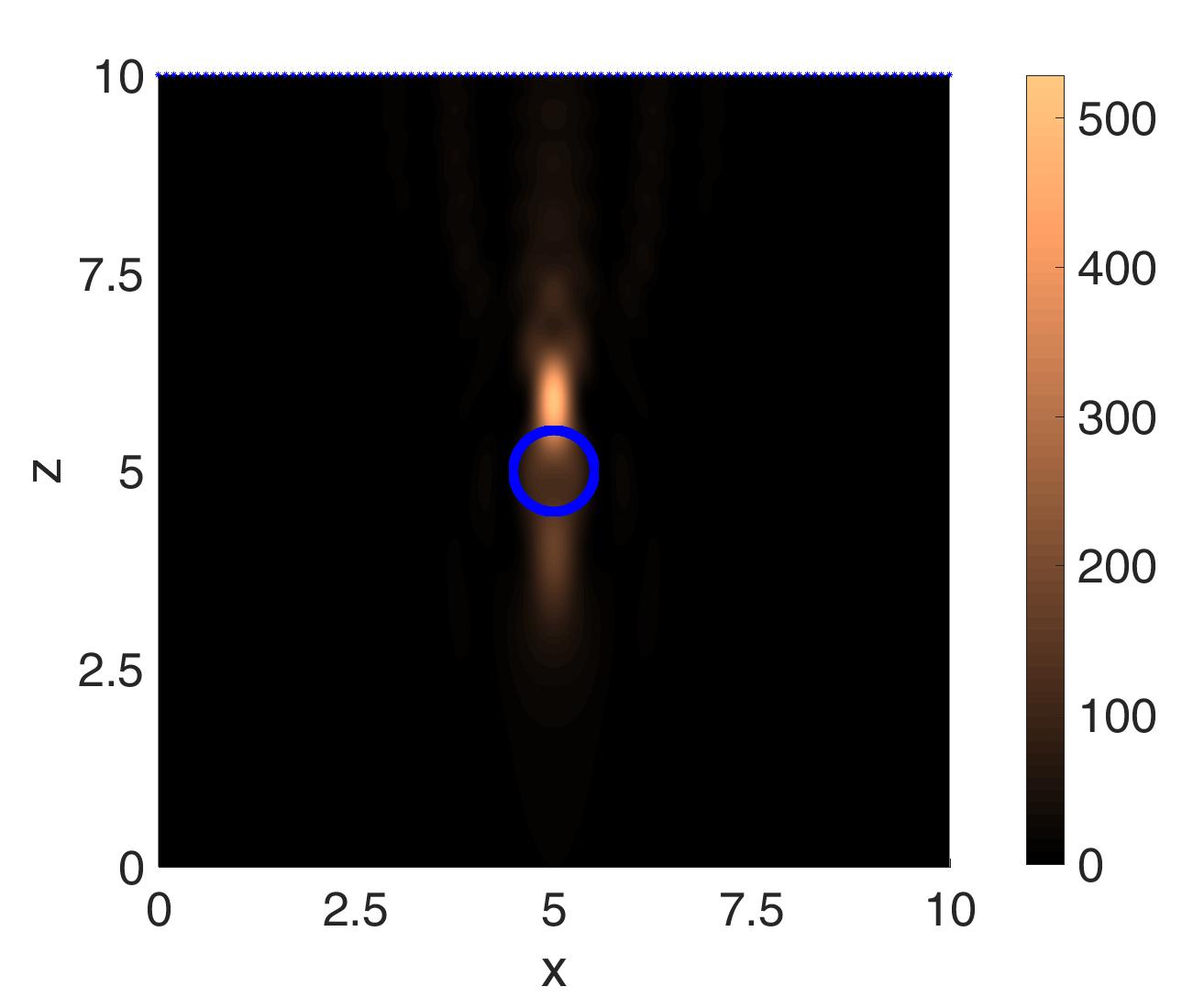} 
\hskip -2mm
\includegraphics[width=4cm]{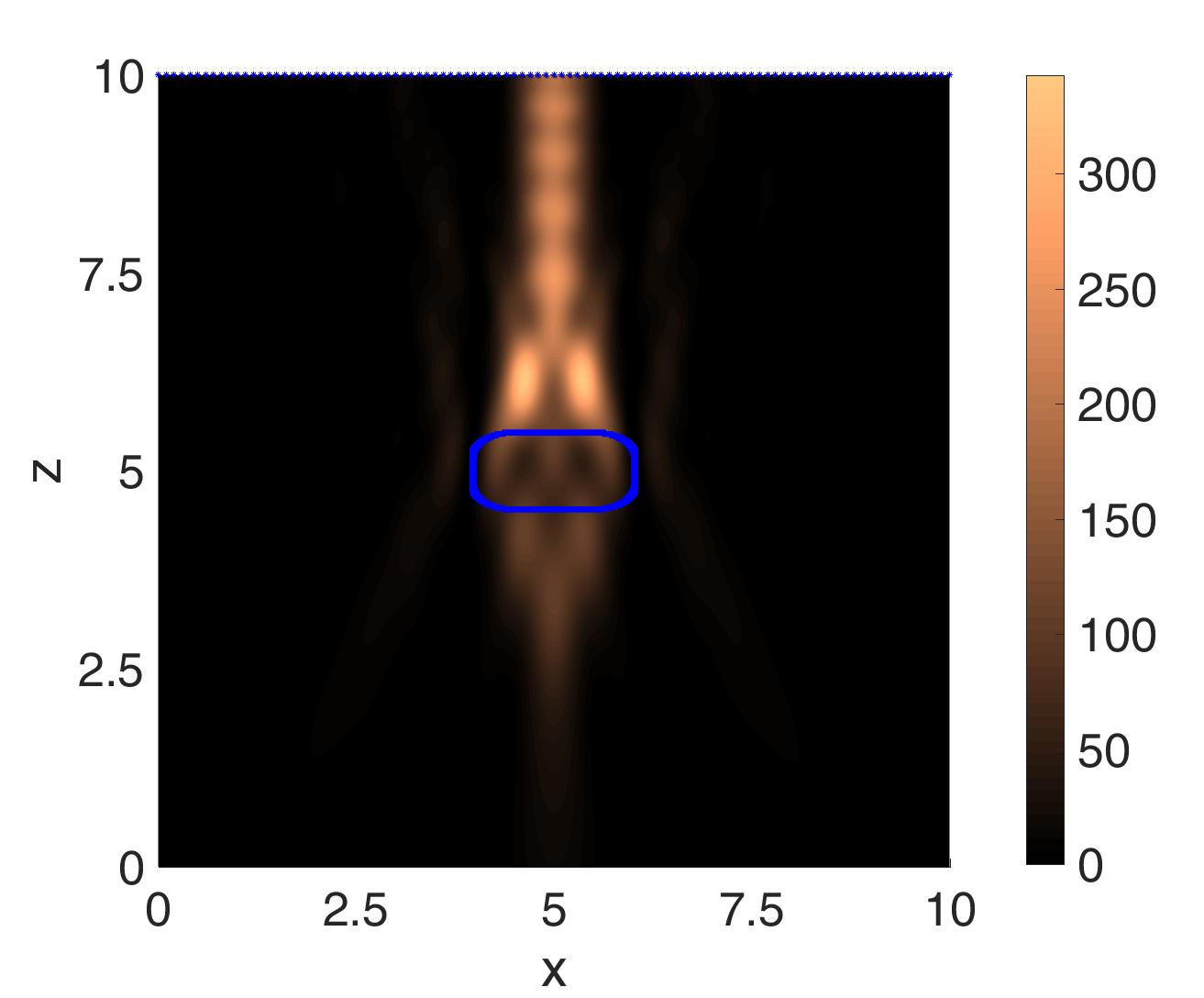} 
\hskip -2mm
\includegraphics[width=4cm]{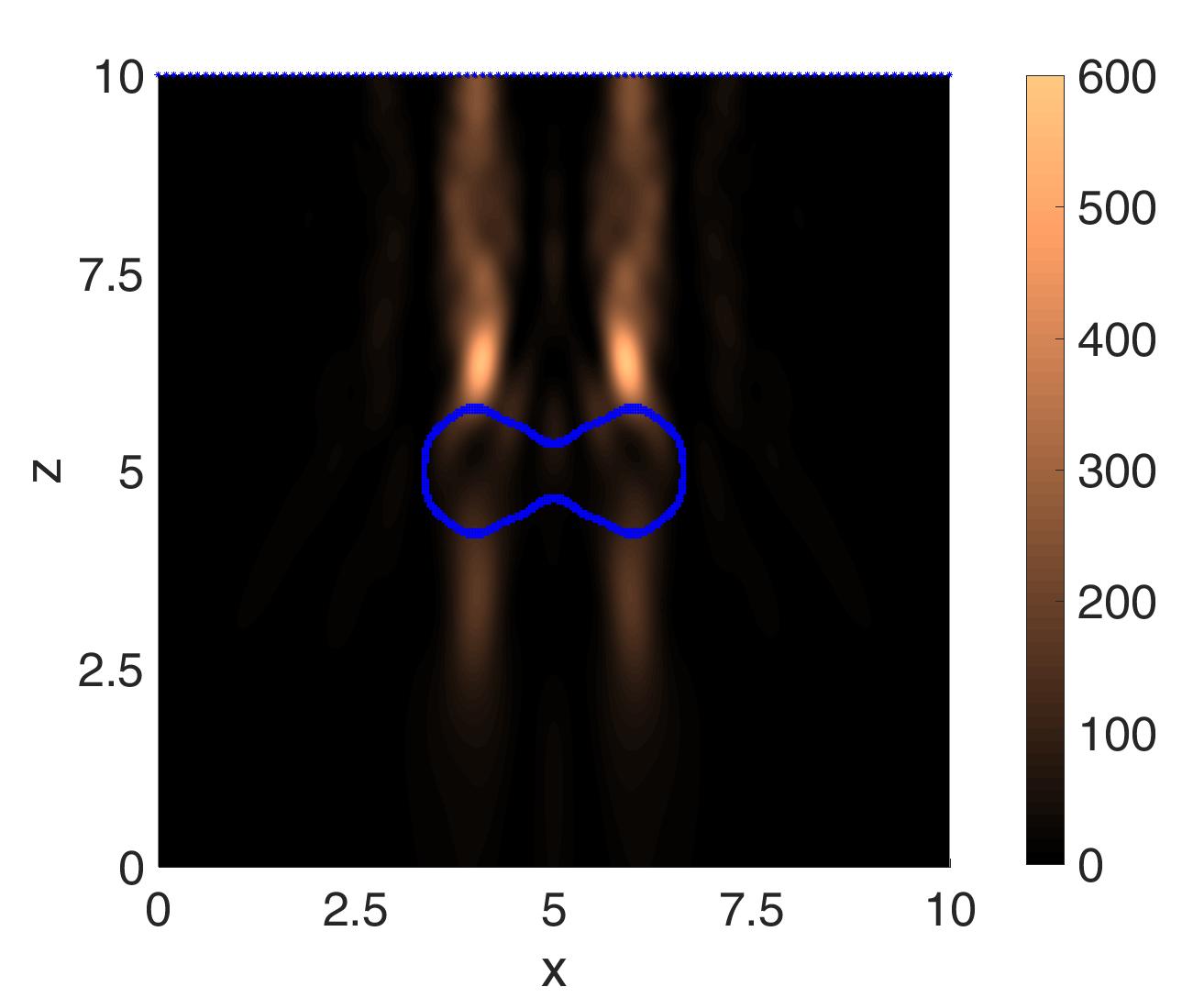} \\
\hskip -4mm (b) \hskip 3.3cm (d)  \hskip 3.3cm (f) \\
\includegraphics[width=4cm]{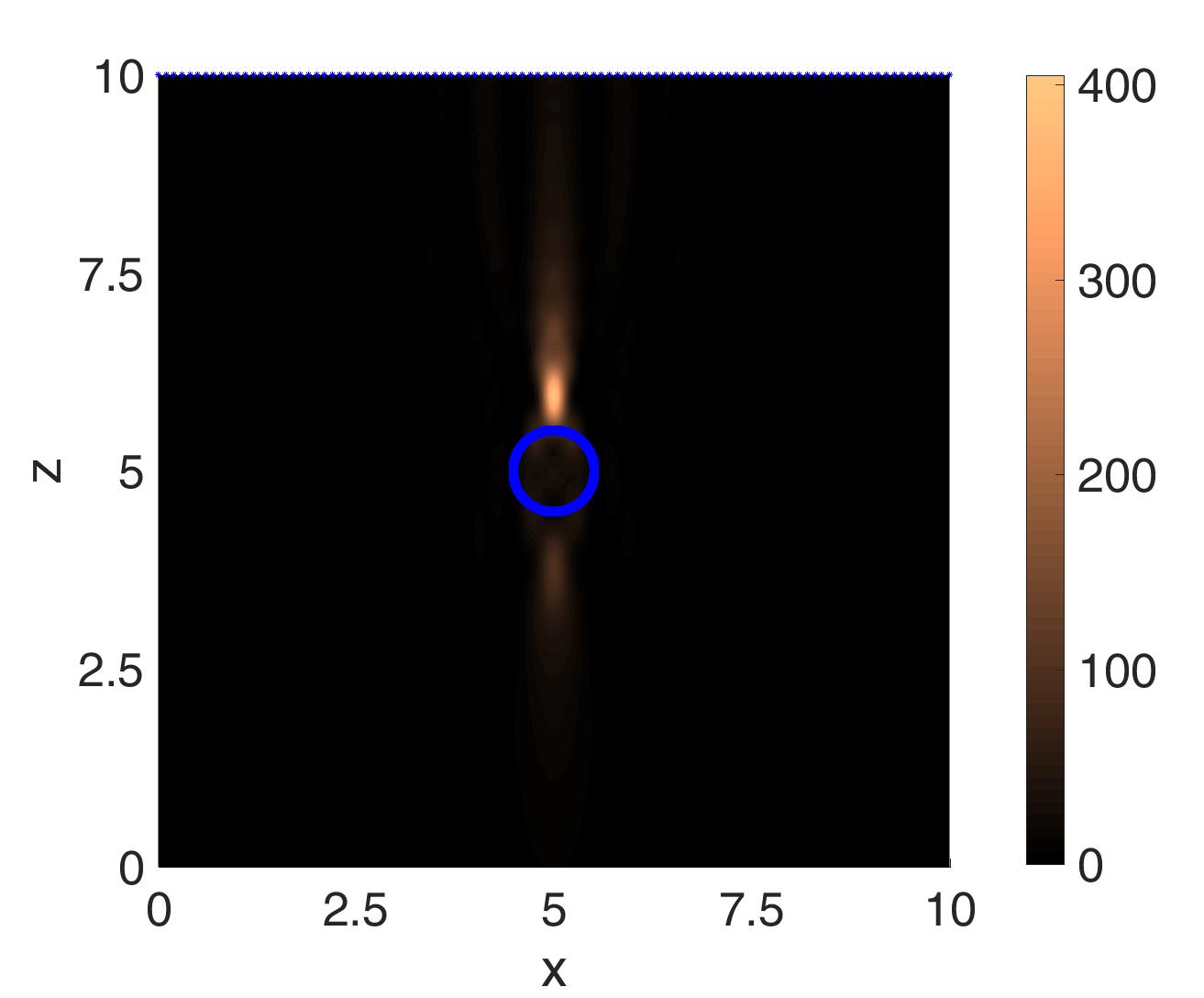} 
\hskip -2mm
\includegraphics[width=4cm]{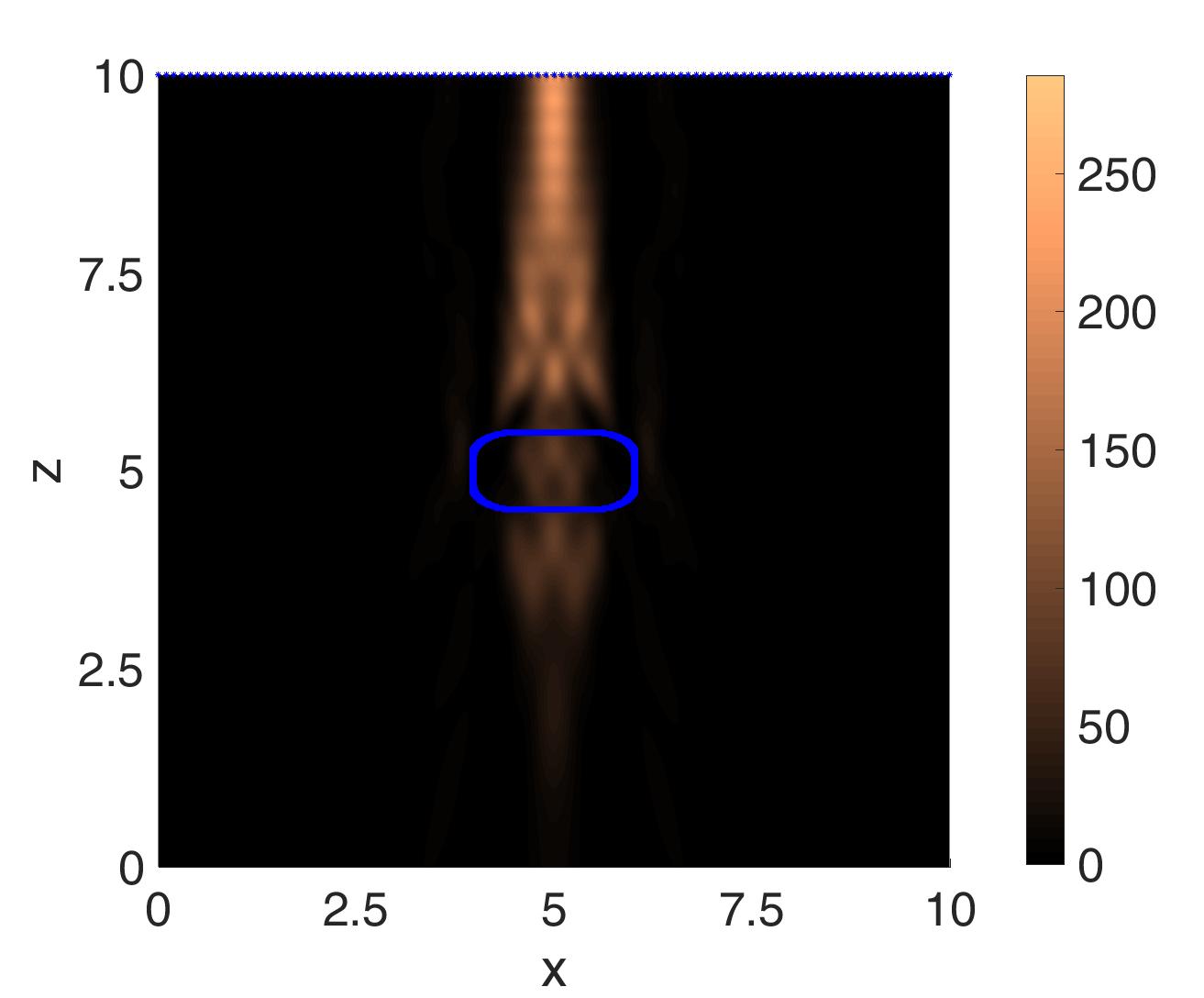} 
\hskip -2mm
\includegraphics[width=4cm]{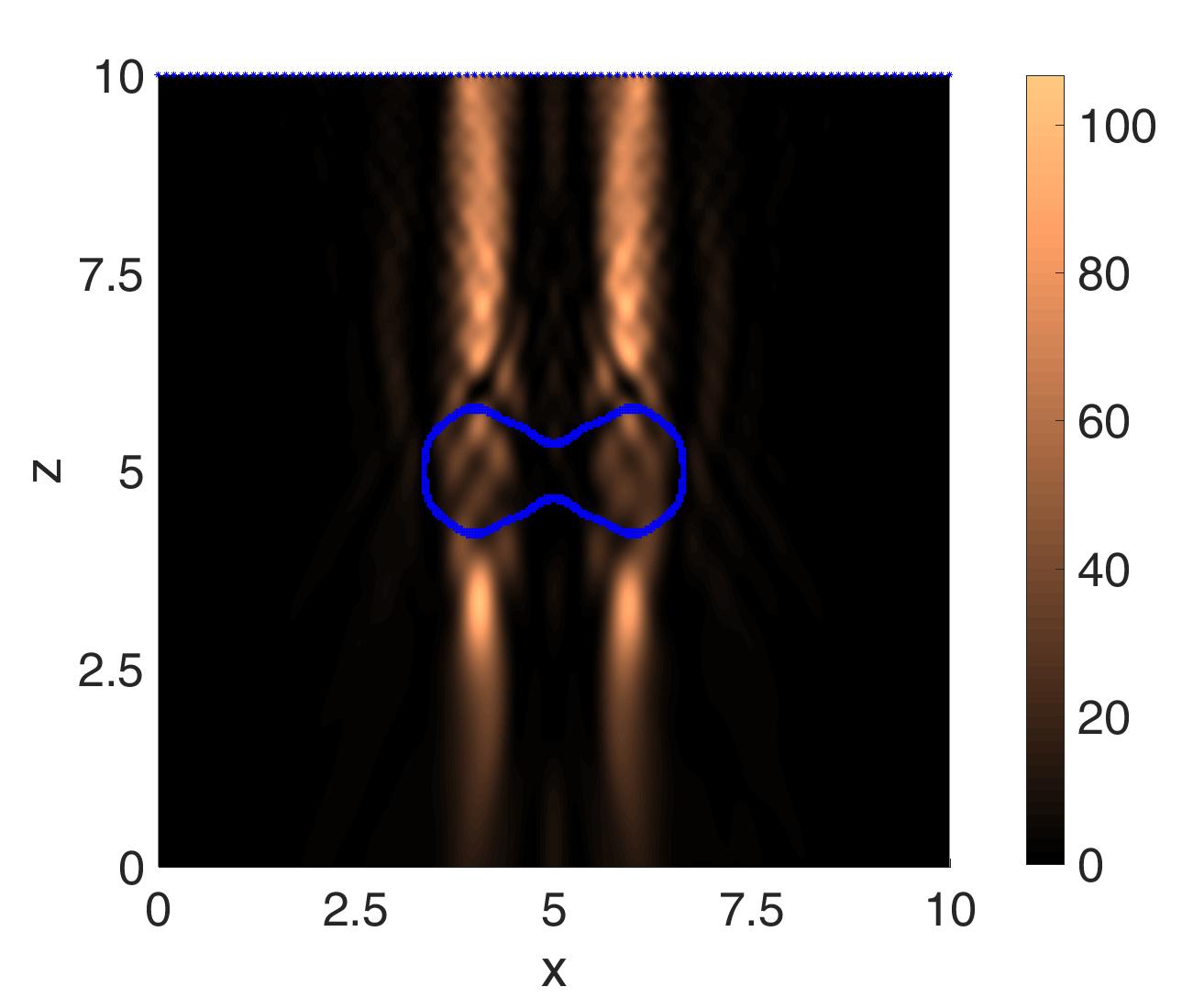} 
\caption{
Slices $y=5$  of the topological energies
for the holography cost functional (\ref{costH})
computed evaluating 
(\ref{etempty}) with (\ref{forwardempty}) and 
(\ref{adjointemptyH})
for the objects considered in Fig. \ref{fig2} (d)-(f).
(a),(b)  Sphere.
(c),(d)   Spherocylinder.
(e),(f)  Sand clock shaped object.
The  top row corresponds to
${k}_e= 12.6$ and ${k}_i=15.12$.
The bottom row corresponds to ${k}_e= 20.60$ 
and ${k}_i=24.79$.
}
\label{fig7}
\end{figure}

Then, a simple strategy to define an initial guess $\Omega^{(0)}$ for the
scatterer is
\begin{equation}
    \Omega^{(0)}:=\{{\bf x}\in {\mathbb R}^3 \,|\, E_T({\bf x},{\mathbb R}^3)\geq C_0\}
    \label{initialguess}
\end{equation}
where $C_0$ is a positive constant, chosen so that 
$J({\mathbb R}^3 \setminus\overline{\Omega}^{(0)})<J({\mathbb R}^3)$.  
Otherwise, we increase the constant $C_0$. The process is
sketched in Figure \ref{fig6}. Alternatively, we might set
$\Omega^{(0)}:=\{{\bf x}\in {\mathbb R}^3 \,|\, D_T({\bf x},{\mathbb R}^3)\leq -C_0\}.
$
In our tests, initialization (\ref{initialguess}) is usually easier to
implement and reduces the shift towards the screen in the object 
location.

Notice that neither the definition of the topological energy
(\ref{etempty}) nor the analytical expressions for the forward and
adjoint fields (\ref{forwardempty}), (\ref{adjointempty}), 
(\ref{adjointemptyH}) depend on ${k}_i$ and $\beta$.
Therefore, formula (\ref{initialguess}) provides a first approximation 
to the objects without knowing their optical properties beforehand.
For biological samples $\beta \sim 1$ and the gradient term in 
expression (\ref{dtempty}) disappears. In consequence, the  
function ${\rm Re} [E(\mathbf{x})   \overline{P}(\mathbf{x})]$ may 
also yield a guess of the objects ignoring the variations of ${k}_i$.

We have tried sizes in the range $10$ nm - $1 \, \mu$m 
with similar results.
However, the behavior of the topological fields change
as the wavelength diminishes and the size of the object grows, a
phenomenon that is still not well understood in spite of recent
theoretical work \cite{guzinalarge}.
Whereas the $xy$ slices still provide information on the shape, as 
shown in Fig. \ref{fig2} (d)-(f), the peaks of the topological fields  
begin to concentrate in the illuminated and  the dark side of the 
boundary in the incidence direction. Figure \ref{fig7} illustrates the 
transition to this new behavior. Once it takes place, as in 
Fig. \ref{fig7}(f), this marks the location $[z_1,z_2]$ of the 
object  in the direction $z$ reducing the offset and loss of axial
resolution.
The shape is reconstructed plotting slices of the topological
fields for values $z=z'$ in that range $[z_1,z_2]$. Bright areas in such 
slices reproduce the true shape and the object is approximated by 
gluing them  together, see reference \cite{siims} for details. 

In the next sections, we will restrict our study to situations in which the 
dimensionless wavenumbers $k_i$, $k_e,$ are not large enough
to reach this transition regime (the wavelength is large enough compared
 to the object size), so that iterative  refinements 
to correct the offset, the number of objects and their
parameters  may be  developed.

\section{Geometry and parameter correction}
\label{sec:shape}

The approximations to the scatterers obtained in Section \ref{sec:tholography} 
can be sharpened by an iterative procedure for small enough sizes, depending on the wavelength.  We consider here the test problem of approximating two objects whose permittivities may be known or unknown, equal or different.  We focus  on two  
cases to clarify axial and horizontal resolution: two similar 
spheres aligned along the incidence direction and two spheres in an 
orthogonal plane with different permittivities in each sphere.

Consider the configurations with multiple balls along the $z$ axis in Figure \ref{fig4}(d). Using expression (\ref{initialguess}) we define an approximation $\Omega^{(0)}$ to the scatterers. We may take
$C_0= \alpha_0 \max_{{\bf x}\in\mathcal{R}_{obs}}E_T(\mathbf x,\mathbb R^3),$
where $0<\alpha_0<1$  and $\mathcal{R}_{obs}$ is the region where we
are looking for objects. 
However, depending on the value of $\alpha_0$ we are left with one or two, or more, elongated objects.  Let us fix a value $\alpha_0$ for which only the upper object, the one closer to the screen is detected. The corresponding shape $\Omega^{(0)}$ is elongated along the $z$ axis. This guess $\Omega^{(0)}$ may improve updating the topological derivative field to consider its presence.
A new approximation $\Omega^{(1)}$ is constructed from $\Omega^{(0)}$ following the iterative scheme:
\begin{eqnarray}
    \Omega^{(n+1)}:= \{{\bf x}\in \Omega^{(n)}  \,|\, 
    D_T({\bf x},{\mathbb R}^3\setminus
    \overline{\Omega^{(n)}},k_i)< c_{n+1}\}
    \label{updatedguess2td} \\
\hspace*{1cm} \cup \, \{{\bf x}\in  \mathbb R^3 \setminus\overline{\Omega^{(n)}}   \,|\, D_T({\bf x},{\mathbb R}^3\setminus \overline{\Omega^{(n)}},k_i)<- C_{n+1}\},
    \nonumber
\end{eqnarray}
see Figure \ref{fig8}(a). For $\beta \sim 1$, the topological derivative is given by
\begin{eqnarray} \hskip 3mm
D_T(\mathbf{x},\mathbb R^3 \setminus \overline {\Omega^{(n)}},k_i) =
{\rm Re} \left[( k_i^2  - k_e^2)  E_n(\mathbf{x})   \overline{P}_n(\mathbf{x})
\right], \qquad \mathbf x \in \mathbb R^3,
\label{dtfull}
\end{eqnarray}
where the forward field $E_n$ is a solution of (\ref{forward}) with object
$\Omega^{(n)}$ and the conjugate adjoint field $\overline{P}_n$  obeys
\begin{eqnarray} \label{adjointobject}
\left\{
\begin{array}{ll}
\Delta \overline{P}_n + {k}_e^2 \overline P_n= 2 \sum_{j=1}^N
({\mathcal I}-|E_n|^2) \overline{E}_n\delta_{{\mathbf x}_j}
&\quad\mbox{in ${\mathbb R}^3\setminus\overline{\Omega^{(n)}}$}, \\
\Delta \overline P_n + {k}_i^2 \overline P_n =0,&\quad\mbox{in $\Omega^{(n)}$},\\
\overline P_n^-= \overline{P}_n^+,&\quad \mbox{on $\partial \Omega^{(n)}$},\\
\beta \partial_{\mathbf{n}}  \overline{P}_n^-= 
\partial_{\mathbf{n}} \overline{P}_n^+, &\quad 
\mbox{on $\partial  \Omega^{(n)}$},\\
\displaystyle\lim_{r\to \infty} r\big(\partial_r \overline{P}_n
-\imath {k}_e \overline{P}_n \big)=0.
\end{array}
\right.
\end{eqnarray} 
The positive constants $C_{n+1}$, $c_{n+1}$ in (\ref{updatedguess2td}) are selected to ensure a decrease in the shape functional: $J({\mathbb R}^3 \setminus\overline{\Omega^{(n+1)}})<J({\mathbb R}^3 \setminus\overline{\Omega^{(n)}})$. 
Notice that for $\mathbf x \notin \Omega$ expansion (\ref{expansion}) holds and  the cost functional decreases adding to $\Omega$ points of negative topological derivative. On the contrary,  for $\mathbf x \in \Omega$ we have
$J((\mathbb R^3 \setminus \overline{\Omega}) \cup B_\varepsilon)=
J(\mathbb R^3 \setminus \overline{\Omega})-{4\over 3} \pi \varepsilon^3 D_T(\mathbf{x},\mathbb R^3 \setminus \overline{\Omega}) +o(\varepsilon^3)$ and we must remove from $\Omega$ points with positive topological derivatives to decrease.\footnote{This is a matter of choice
in the sign of the definition.}

The approximations $\Omega^{(n)}$ provided by (\ref{initialguess}) or (\ref{updatedguess2td}) have irregular shapes. 
To define their boundaries, we may fit starshaped parametrizations
to each component, or adjust surfaces employing blobby molecules. 
This later approach, combined with BEM-FEM to evaluate forward and adjoint 
fields for $\Omega^{(n)}$ was taken in Ref. \cite{siims} 
for a different functional. Alternatively, one can use discrete dipole 
approximations that just need a  discretized representation of objects 
\cite{holotom}.  We resort here to blobbly molecule fitting
and the previously described non symmetric BEM-FEM scheme \cite{siims,fem,selgas}.

    \tikzstyle{block} = [rectangle, draw, fill=blue!20, 
    text width=15em, text centered, rounded corners, node distance=1.2cm,
    minimum height=2em]
    \tikzstyle{block1} = [rectangle, draw, fill=blue!20, 
    text width=15em, text centered, rounded corners, node distance=6cm,
    minimum height=2em]
    \tikzstyle{block2} = [rectangle, draw, fill=blue!20, 
    text width=15em, text centered, rounded corners, node distance=1.4cm, 
    minimum height=2em]
   \tikzstyle{block3} = [rectangle, draw, fill=blue!20, 
    text width=15em, text centered, rounded corners, node distance=6cm, 
    minimum height=2em]
\tikzstyle{line} = [draw, -latex']

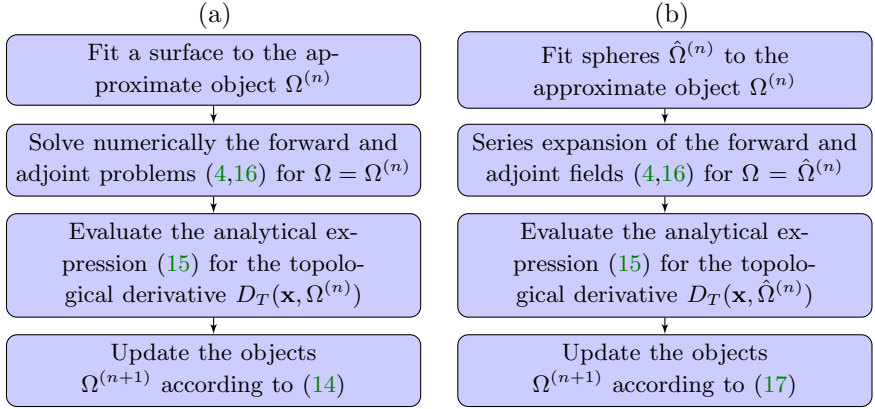
\begin{figure}[!h]
\centering
(a) \hskip 5.5cm (b) \\
\begin{tikzpicture}
    \node [block] (OEscapprox) { \small Fit a surface to the approximate object $ \Omega ^{(n)}$};
    \node [block1, right of=OEscapprox] (OEscapprox2) { \small Fit spheres
    $\hat \Omega^{(n)}$ to the approximate object $\Omega ^{(n)}$};
    \node [block, below of=OEscapprox] (auxprobs) {\small Solve numerically the forward and adjoint problems (\ref{forward},\ref{adjointobject}) for $\Omega=\Omega ^{(n)}$};
    \node [block1, right of=auxprobs] (auxprobs2) {\small Series expansion of  the forward and adjoint fields (\ref{forward},\ref{adjointobject}) for $\Omega=
    \hat \Omega ^{(n)}$};
    \node [block2, below of=auxprobs] (DTimprove) {\small Evaluate the analytical expression (\ref{dtfull}) for the topolo- gical derivative $D_T(\mathbf x,\Omega^{(n)})$};
    \node [block3, right of=DTimprove] (DTimprove2) {\small Evaluate the analytical expression (\ref{dtfull}) for the topolo- gical derivative $D_T(\mathbf x,\hat \Omega^{(n)})$};
    \node [block2, below of=DTimprove] (Omegan) {\small Update the objects $\Omega ^{(n+1)}$ according to (\ref{updatedguess2td})};
    \node [block3, right of=Omegan] (Omegan2) {\small Update the objects $\Omega ^{(n+1)}$ according to (\ref{updatedguess2tdhat})};
    \path [line] (OEscapprox) -- (auxprobs);
    \path [line] (auxprobs) -- (DTimprove);
    \path [line] (DTimprove) -- (Omegan);
    \path [line] (OEscapprox2) -- (auxprobs2);
    \path [line] (auxprobs2) -- (DTimprove2);
    \path [line] (DTimprove2) -- (Omegan2);
\end{tikzpicture}
\caption{Procedure to sharpen the approximate scatterers for cost functional (\ref{costH}):
(a) Keeping arbitrary shapes and computing the adjoint and forward fields by BEM-FEM. (b) When we are only interested in the number of objects and their position, we may reduce the computational complexity by an intermediate computation with spheres.}
\label{fig8}
\end{figure} 

When we are only interested in determining the number of objects and their location, we may reduce the computational complexity by fitting balls $\hat \Omega^{(n)}$ to the current approximation $\Omega^{(n)}$ and setting
\begin{eqnarray}
    \Omega^{(n+1)}:= \{{\bf x}\in \hat \Omega^{(n)}  \,|\, 
    D_T({\bf x},{\mathbb R}^3\setminus
    \overline{\hat \Omega^{(n)}},k_i)< c_{n+1}\}
    \label{updatedguess2tdhat} \\
\hspace*{1cm} \cup \, \{{\bf x}\in  \mathbb R^3 \setminus\overline{\hat \Omega^{(n)}}   \,|\, D_T({\bf x},{\mathbb R}^3\setminus \overline{\hat \Omega^{(n)}},k_i)<- C_{n+1}\},
    \nonumber
\end{eqnarray}
Here, $D_T({\bf x},{\mathbb R}^3\setminus \overline{\hat \Omega^{(n)}},k_i)$ may be evaluated solving the forward and adjoint problems (\ref{forward}), (\ref{adjointobject}) for $\hat \Omega^{(n)}$ using the expansions in Appendix \ref{sec:explicitforwardadjoint}. 
We can iterate the procedure fitting spheres $\hat \Omega^{(n+1)}$ to the components of $\Omega^{(n+1)}$  defined by (\ref{updatedguess2tdhat}) again.
This is done by centering them at points where the minimum value of the topological derivative is attained and chosing as diameter of each component the smallest diameter  in the three space directions.  The procedure is summarized in Fig. \ref{fig8}(b).

\begin{figure}[!h]
\centering
\hskip -5mm (a) \hskip 6cm (b) \\
\includegraphics[height=5cm]{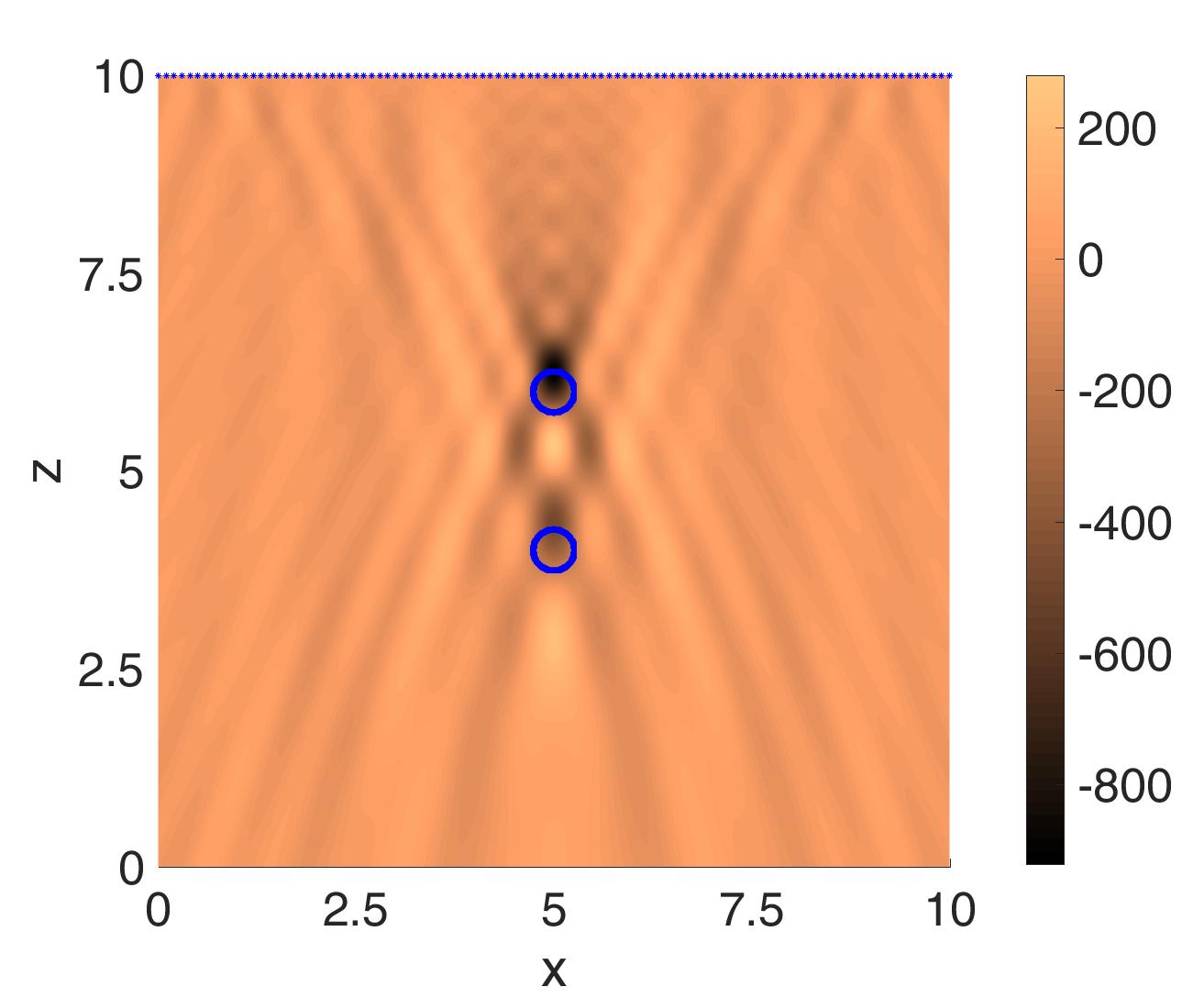}  
\includegraphics[height=5cm]{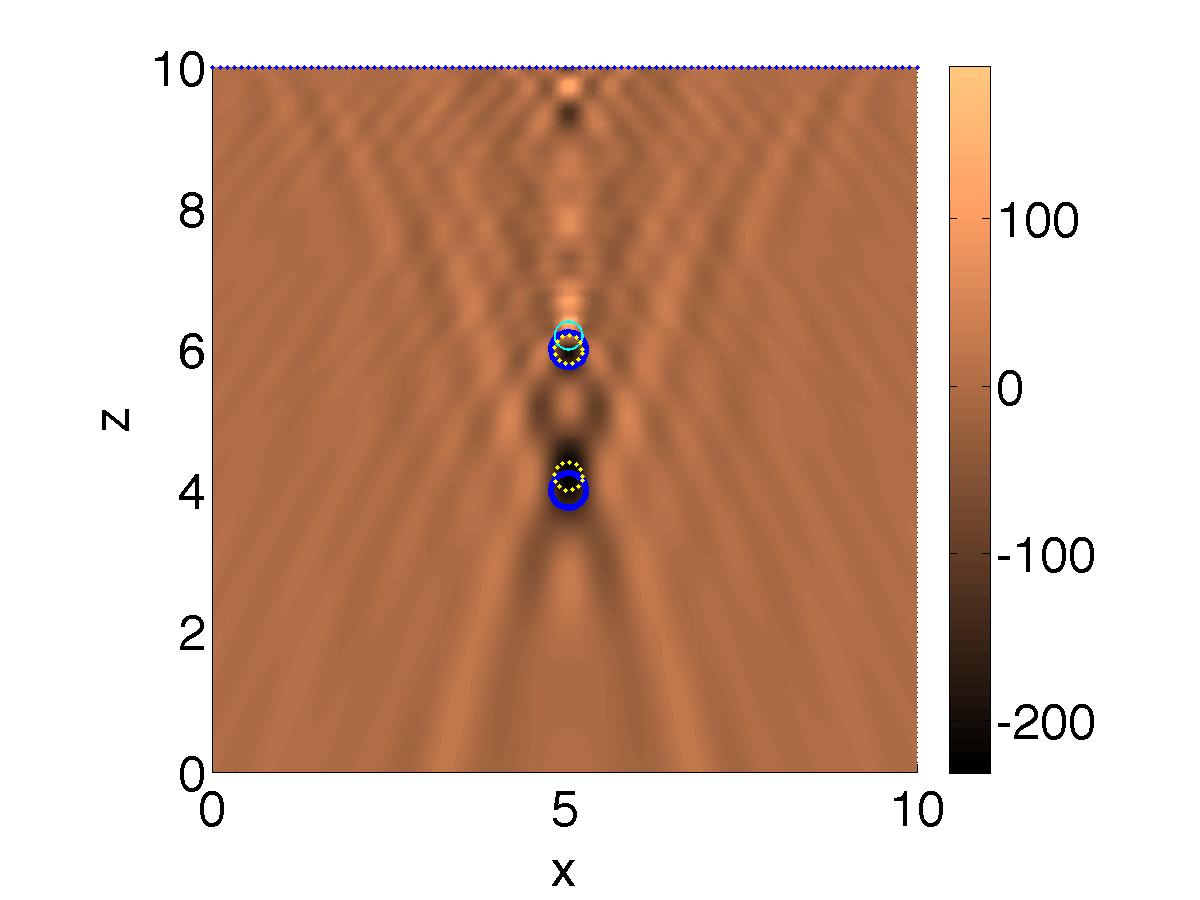} \\
\hskip -5mm (c) \hskip 6cm (d) \\
\includegraphics[height=5cm]{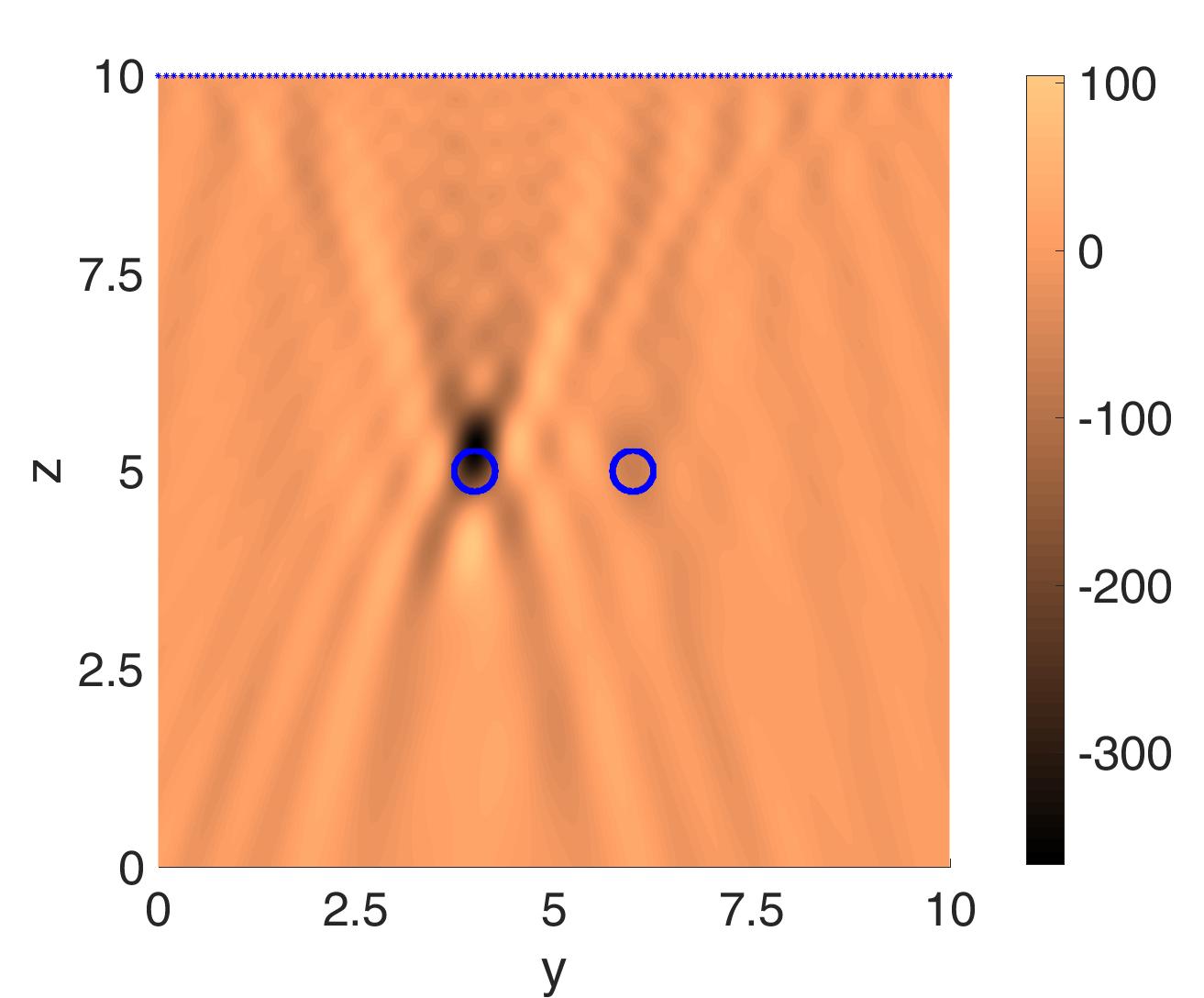}  
\includegraphics[height=5cm]{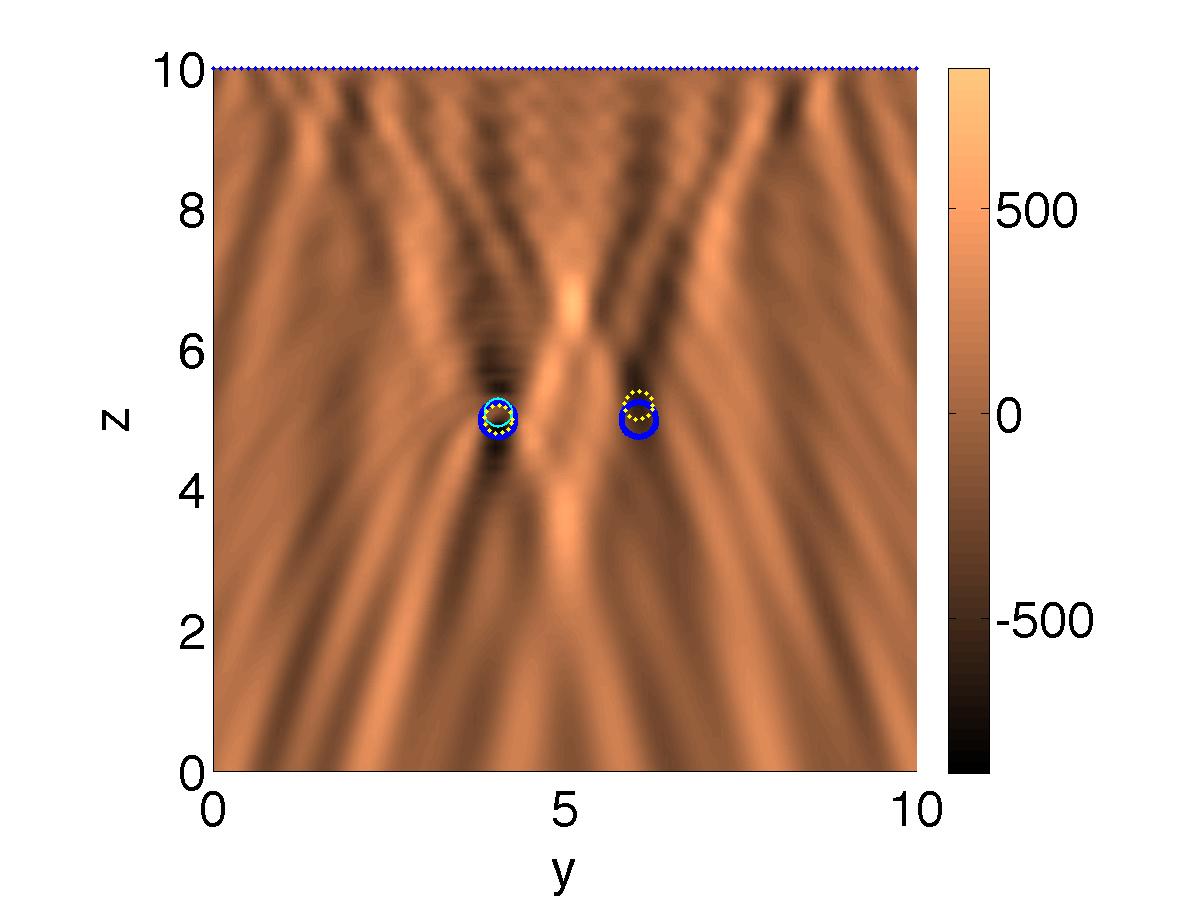} \\
\caption{Iteration using topological derivatives of the holography cost functional (\ref{costH}) in geometries with multiple spheres.
Two spheres along the $z$ axis with ${k}_e= 12.6$ and ${k}_i=15.12$ in both spheres: 
(a) Slice $y=5$ of the topological derivative computed evaluating (\ref{dtempty}) with (\ref{forwardempty}) and (\ref{adjointemptyH}). Only the upper object  is clearly distinguished.
(b) Same as (a) but evaluating (\ref{dtfull}) with (\ref{forward}) and (\ref{adjointobject}) when the approximate object is the sphere defined by the thin cyan contour and $k_i$ is taken as known. The lower object is now visible.
Two spheres along the $y$ axis with ${k}_e= 12.6$, ${k}_i=15.12$ in the leftmost sphere and 
${k}_i=13.75$ in the rightmost sphere: (c) Slice $x=5$ of the topological derivative computed evaluating (\ref{dtempty}) with (\ref{forwardempty}) and (\ref{adjointemptyH}). Only the leftmost object  is clearly distinguished.
(d) Same as (c) but evaluating (\ref{dtfull}) with (\ref{forward}) and (\ref{adjointobject}) when the approximate object is the sphere defined by the thin cyan contour and the approximate value of $k_i$ in it is $k_i^{(0)}=15.76$. The rightmost object is now visible.
}
\label{fig9}
\end{figure}

We have treated the configuration in Figure \ref{fig4}(d)  in both ways. Figure \ref{fig9}(a) plots the slice $y=5$ of the topological derivative $D_T({\bf x},\mathbb R^3)$. Notice that only the object closer to the screen is clearly distinguished.  Let us first apply the strategy in Fig. \ref{fig8}(b). Figure \ref{fig9}(b) displays the slice $y=5$ of the topological derivative
$D_T({\bf x},{\mathbb R}^3\setminus \overline{\hat \Omega^{(0)}},k_i)$ 
when $\hat \Omega^{(0)}$ is the sphere represented by a thin cyan contour in Figure \ref{fig9}(b),  which has been fitted to $\Omega^{(0)}$ defined by (\ref{initialguess}). 
The two thicker solid blue contours correspond to the true objects. Following the rule (\ref{updatedguess2tdhat}), the updated object  $\Omega^{(1)}$ should loose points in the upper half, 
where the topological derivative becomes positive and large, and gain neighboring points below the lower half, where the topological derivative becomes negative and large. Moreover, an additional prominent region where large  negative values are attained is identified. Fitting two balls to $\Omega^{(1)}$ we obtain $\hat \Omega^{(1)}$ represented by the two dotted yellow lines,  which 
we use to construct a new object $\Omega^{(2)}$ defined by (\ref{updatedguess2tdhat}),  improving the position of the component  placed farther from the screen.  Instead, if we iterate with the procedure described in Fig. \ref{fig8}(a), keeping the approximations $\Omega^{(n)}$, $n=0,1,$ defined by (\ref{initialguess})-(\ref{updatedguess2td}) and calculating $D_T(\mathbf x, \Omega^{(n)})$ by BEM-FEM, we also see the second object appear and the position improves, but the shapes remain elongated in the $z$ direction. Shapes in $xy$ slices are correct though.

Consider now two spheres along the $y$ axis. Figures \ref{fig4}(a)-(b) show that both are identified when $k_i$ takes similar values in them. Increasing  the  gap between the values of $k_i$, only the sphere corresponding to the largest value is detected initially, see the topological derivative in Figure \ref{fig9}(c).  
The topological energy has a similar aspect, and, as said before, does not use the value of $k_i$.  We exploit it to initialize $\Omega^{(0)}$ by
formula (\ref{initialguess}), recovering only the object with higher contrast.

Assuming the values of $k_i$ unknown, we next
attempt to recover the number of objects and their parameters combining 
topological derivatives and descent techniques for the coefficients, as 
indicated in Fig. \ref{fig10}(a).
We initialize ${k}_i$ as a perturbation ${k}_{i}^{(0)}$ of the ambient
dimensionless wavenumber ${k}_e$, and iterate as follows.
Given an approximation of the scatterers $\Omega^{(n)}$, we
update the value of ${k}_{i}^{(n)}$ to obtain ${k}_{i}^{(n+1)}$
by a descent technique.
The parameter to be optimized appears in the forward problem 
governing the electric field. A derivative of the cost functional
with respect to it is given by (\ref{derk}). Choosing 
$k_{i}^{(n+1)} = {k}_{i}^{(n)} + \delta  \psi_n$, with $\delta >0$ small
and
\begin{equation}\label{psi1}
 \psi_n(\mathbf x) =    - {\rm Re} \left[E_n(\mathbf x) \overline{P}_n(\mathbf x)\right],  \ \
 \qquad  {\mathbf{x}} \in \Omega^{(n)},
\end{equation}
the functional decreases. $E_n$ and $P_n$ are forward and adjoint 
fields with object $\Omega^{(n)}$ and coefficient $k_i^{(n)}$
computed by BEM-FEM.
If $\Omega^{(n)}= \bigcup_{\ell=1}^L \Omega_\ell^{(n)}$ and
we are looking for piecewise constant parameters, we may take
\begin{equation}\label{psi2}
 \psi_n(\mathbf x)  =  - {1\over {\rm meas} (\Omega^{(n)}_\ell)} 
 {\rm Re} \int_{\Omega^{(n)}_\ell} E_n \overline{P}_n,  \ \
 \qquad  {\mathbf{x}} \in \Omega^{(n)}_\ell, \; \ell=1,\ldots, L.
\end{equation}
Setting $k_{i}^{(n,0)}=k_i^{(n)}$ we calculate:
\begin{equation}\label{parametrosk}
{k}_{i}^{(n,m+1)}={k}_{i}^{(n,m)}+\delta \psi_{n,m}
\end{equation}
with $\delta>0$ small enough to ensure $J({k}_{i}^{(n,m+1)})<
J({k}_{i}^{(n,m)}).$
We have defined  $\psi_{n,m}$  only in $\Omega^{(n)}.$
However, formulas (\ref{psi1}) and (\ref{psi2}) make sense 
everywhere, and (\ref{parametrosk})
defines $ {k}_{i}^{(n,m+1)}$ for ${\mathbf{x}} \in {\mathbb{R}}^{3}.$
After a number of iterations $M$, we stop, fix $k_{i}^{(n+1)}=
k_{i}^{(n,M)}$ and update the approximation of the objects to obtain
$\Omega^{(n+1)}$ by the scheme described in Fig. Ê\ref{fig8}(a).
From a computational point of view, updating $k_i^{(n)}$ is less expensive
than updating $\Omega^{(n)}$ because the computational domain
is fixed and already meshed. Therefore, it seems advisable
at first sight to take $M>1$.

In view of  (\ref{parametrosk}), once the
approximation $\Omega^{(0)}$ to the objects is proposed by (\ref{initialguess}), we 
 initialize ${k}_{i}^{(0)}={k}_{e} + \delta \psi_0$ where $\psi_0$ is defined
by (\ref{psi2}), $E_0$ and $P_0$ being the forward and adjoint
fields computed in the whole space by (\ref{forwardempty}) and 
(\ref{adjointempty}).  
Then formula (\ref{parametrosk}) provides
a new approximation $k_i^{(1)}$. We stopped the iteration at $M=10$
and computed $D_T(\mathbf x,\Omega^{(0)},k_i^{(1)})$  by solving
(\ref{forward}),(\ref{adjointobject}) by BEM-FEM 
to produce $\Omega^{(1)}$ with (\ref{updatedguess2td}). 
The location of the first object is corrected, and a new object is clearly seen,  
so that $\Omega^{(1)}$ is formed by two elongated components. The  new
component, however, is  slightly displaced forward in the incidence
direction. When we seek  $k_i^{(2)}$ defined
in both components we see that small variations in the previous 
process lead to very different results in the new one (corresponding to
the object with lowest contrast).

    \tikzstyle{block} = [rectangle, draw, fill=blue!20, 
    text width=15em, text centered, rounded corners, node distance=1.55cm,
    minimum height=2em]
    \tikzstyle{block1} = [rectangle, draw, fill=blue!20, 
    text width=15em, text centered, rounded corners, node distance=6cm,
    minimum height=2em]
    \tikzstyle{block2} = [rectangle, draw, fill=blue!20, 
    text width=15em, text centered, rounded corners, node distance=1.6cm, 
    minimum height=2em]
   \tikzstyle{block3} = [rectangle, draw, fill=blue!20, 
    text width=15em, text centered, rounded corners, node distance=6cm, 
    minimum height=2em]
\tikzstyle{line} = [draw, -latex']

\begin{figure}
\centering
(a) \hskip 5.5cm (b) \\
\begin{tikzpicture}
    \node [block] (OkEscapprox) {\small Start from an approximation of the objects 
$\Omega ^{(n)}$ and the permittivity ${k}_i^{(n)}$};
    \node [block1, right of=OkEscapprox] (OkEscapprox2) {\small Start from an approximation of the objects $\Omega ^{(n)}$ and the permittivity ${k}_i^{(n)}$};
    \node [block, below of=OkEscapprox] (auxprobs) {\small Solve numerically the forward and adjoint problems (\ref{forward},\ref{adjointobject}) with $\Omega=\Omega ^{(n)}$ and $k_i={k}_i^{(n)}$};
    \node [block1, right of=auxprobs] (auxprobs2) {\small Fit spheres 
    $\hat \Omega^{(n)}$ to $\Omega^{(n)}$ and piecewise constants $\hat k_i^{(n)}$ to 
   ${k}_i^{(n)}$};
    \node [block, below of=auxprobs] (ki_improve) {\small Evaluate the expressions (\ref{psi1}) or (\ref{psi2}), and (\ref{parametrosk}) to update the approximation of 
    ${k}_i^{(n)}$ to ${k}_i^{(n+1)}$};
    \node [block1, right of=ki_improve] (ki_improve2) {\small Plot landscapes for
    $J(k_i)$  in a parameter region by solving (\ref{forward}) explicitly with 
     $\Omega=\hat \Omega^{(n)}$};
    \node [block2, below of=ki_improve] (Omega_improve) {\small Update $\Omega ^{(n)}$    to generate $\Omega ^{(n+1)}$
    following the scheme in figure \ref{fig8}(a) keeping $k_i^{(n+1)}$ fixed};
    \node [block3, right of=Omega_improve] (Omega_improve2) {\small If a minimum $\hat{k}_i^{(n+1)}$ is located update $\Omega ^{(n)}$    to generate $\Omega ^{(n+1)}$
    as in figure \ref{fig8}(b) keeping $\hat{k}_i^{(n+1)}$ fixed};
    \path [line] (OkEscapprox) -- (auxprobs);
    \path [line] (auxprobs) -- (ki_improve);
    \path [line] (ki_improve)--(Omega_improve);
    \path [line] (OkEscapprox2) -- (auxprobs2);
    \path [line] (auxprobs2) -- (ki_improve2);
    \path [line] (ki_improve2)--(Omega_improve2);
\end{tikzpicture}
\caption{Strategy to predict the permittivities of the objects
for cost functional (\ref{costk}):  (a) General procedure allowing for irregular domains and variable coefficients. (b) When we are interested
in piecewise constant guesses or in studying the structure of the
minima we may simplify the computational complexity by intermediate
computations with spheres.}
\label{fig10}
\end{figure}
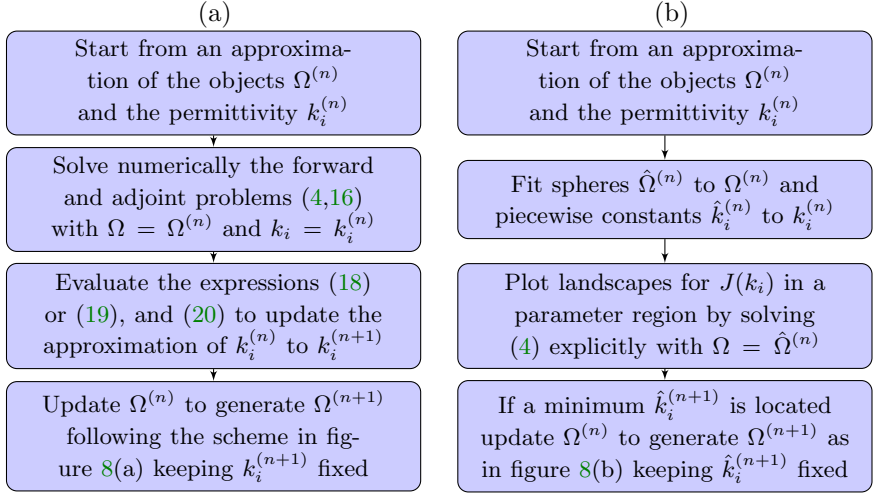

To understand this behavior, we revisit the approximation using the strategy in Fig. \ref{fig10} (b). We fit a ball $\hat \Omega^{(0)}$ to $\Omega^{(0)}$ defined by (\ref{initialguess}), represented by the thin cyan contour in Figure \ref{fig9}(d).  To produce a guess of $k_i$, we fix $\hat \Omega^{(0)}$ and evaluate the functional (\ref{costH}) for a range of $k_i$ using Appendix \ref{sec:explicitforwardadjoint}.
It reaches a minimum value at $k_i^{(0)}=15.76.$ Figure \ref{fig9}(d) depicts the topological derivative $D_T(\mathbf{x},{\mathbb R}^3 \setminus \overline{\hat \Omega^{(0)}},k_i^{(0)}).$ The rightmost object is now clearly detected. The position of the leftmost one is corrected removing points from its top and adding them at the bottom, using (\ref{updatedguess2tdhat}) to generate a new guess $\Omega^{(1)}$, and then fitting spheres $\hat \Omega^{(1)}$ to it. We are left with two balls represented by the dotted yellow lines.
To find a value for $k_i$ at the second sphere, we may fix the two spheres, use the known approximation of $k_i$ at the preexisting one, and minimize (\ref{costH}) with respect to the value of $k_i$ in the second one. However, we fail to find isolated local minima. A continuum exists. An isolated global minimum is only encountered when the offset in the location of the second object is reduced.
Even if updating $k_i^{(n)}$ is less expensive than updating $\Omega^{(n)}$, as indicated before, for convergence reasons it may be advisable to choose $M=1$. However, this problem may disappear when the contrast with the ambient medium is more alike for the different objects.

\section{Approximation of the electric field at the recording screen}
\label{sec:electricholography}

The previous sections propose strategies to recover
holographied objects from recorded in-line holograms ${\cal I}$
without a priori information.
We have obtained initial guesses of objects,
which can be improved by iteration to correct the number
of components and their position. However, while the shapes
in $xy$ planes are correct, they remain elongated in the
$z$ direction. Ref. \cite{siims} showed in a similar setting and
using the full electric field $E_{meas}$ as data that the elongation
of the shape in the incidence direction could be removed
by iteration. As said before, the electric field $E_{meas}$ 
cannot be measured in practice, instead we may develop
mathematical methods to approximate it numerically
from the hologram ${\cal I}$.
We study here the possibility of approximating the full
electric field $E_{meas}$ at the recording screen from the holograms 
${\cal I}$ and then using this approximated field $E_{approx}$ 
to reconstruct the objects by topological methods. 

\subsection{Initial approximation of the electric field and the scatterers}
\label{sec:crude}
Consider the imaging setting depicted in Fig. \ref{fig1}, with light 
of wavelength  $660$ nm emitted at a distance of $10 \, \mu$m from 
the recording screen.
The incident wave $E_{inc} \sim 1$ at the detector screen. 
Setting $E_{inc} =1$ in the expansion
\begin{eqnarray}
{\cal I}= |E_{meas}|= |E_{sc} + E_{inc}|^2 =  |E_{sc}|^2 + |E_{inc}|^2 
+ E_{sc} \overline{E_{inc}} + \overline{E_{sc}} E_{inc},
\end{eqnarray}
we may infer a rough approximation of the  scattered field at the 
detectors. If we neglect the quadratic term $ |E_{sc}|^2$, we find:
\begin{eqnarray}
E_{scapprox,j}^{(0)} = ({\cal I}(\mathbf x_j)-1)/2 \sim 
{\rm Re} (E_{sc}), \quad j=1,\ldots,N.
\label{Ein1-1} 
\end{eqnarray}
Alternatively, we may set ${\rm Im} (E_{sc}) = 0$ and look for 
a solution of 
$ {\rm Re} (E_{sc})^2 + 2 {\rm Re} (E_{sc}) + 1 - {\cal I} =0$
decaying to zero at the borders, which yields
$E_{scapprox}^{(0)} = {-1 +\sqrt{\cal I}}.$ 
The true total electric field  $E_{meas}$ at the detectors being unknown,
we approximate it by
\begin{eqnarray}
E_{approx,j}^{(0)}=E_{scapprox,j}^{(0)}+E_{inc}(\mathbf x_j),
\quad j=1,\ldots, N,  \label{Emeasapprox}
\end{eqnarray}
and consider the cost functional:
\begin{eqnarray}
J(\Omega) = {1\over 2}  \sum_{j=1}^N
|E(\mathbf{x}_j)-E_{approx,j}^{(0)}|^2. \label{costEapprox}
\end{eqnarray}
The topological derivative and energy in $\mathbb R^3$ are given by 
(\ref{dtempty}) and (\ref{etempty}), respectively, 
with forward field $E=E_{inc}$ and conjugate adjoint field \cite{siims}
\begin{eqnarray}\label{adjointemptyE}
\overline{P}({\mathbf x})
=- \sum_{j=1}^N {e^{\imath {k}_e |{\mathbf x}-{\mathbf x_j}|}
\over 4 \pi |{\mathbf x}-{\mathbf x_j}|}
(\overline{E_{approx,j}^{(0)}-E(\mathbf x_j)}).
\end{eqnarray}
We may use them to obtain initial guesses of the objects,
as indicated in Fig. \ref{fig11}. The results are 
similar to the initial predictions of scatterers obtained 
in Section \ref{sec:tholography} using the holography cost functional
(\ref{costH}), and in Ref. \cite{siims} using the cost functional
(\ref{costEapprox}) with the true synthetic
data $E_{meas}$ instead of $E_{approx}$.
This illustrates the fact that topological methods
are very robust to noise. The $xy$ sections still suggest the true shape
whereas the $xz$ sections locate the region occupied by the
object, showing a certain elongation and  shift towards the screen.

\tikzstyle{block1} = [rectangle, draw, fill=blue!20, 
    text width=22em, text centered, rounded corners, minimum height=2em]
\tikzstyle{block2} = [rectangle, draw, fill=blue!20, 
    text width=22em, text centered, rounded corners, node distance=1.25cm, 
    minimum height=2em]
\tikzstyle{line} = [draw, -latex']
\tikzstyle{cloud} = [draw, ellipse,fill=red!20, node distance=7cm,
    minimum height=2em, text width=5em]
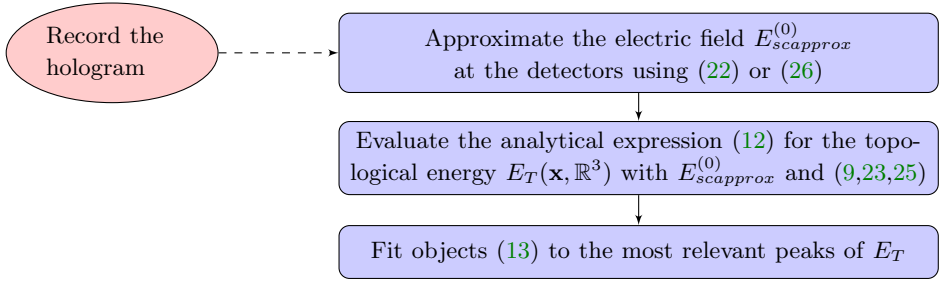
\begin{figure}
\centering
\begin{tikzpicture}[node distance = 1.4cm, auto]
    \node [block1] (Escapprox) {\small Approximate the electric field $E^{(0)}_{scapprox}$ at the detectors using (\ref{Ein1-1}) or (\ref{Ein2-1})};
    \node [cloud, left of=Escapprox] (measures) {\small Record the hologram};
    \node [block1, below of=Escapprox] (ETapprox) {\small Evaluate the analytical expression (\ref{etempty}) for the topological energy $E_T(\mathbf x,
    \mathbb R^3)$ with $E^{(0)}_{scapprox}$ and  (\ref{forwardempty},\ref{Emeasapprox},\ref{adjointemptyE})};
    \node [block2, below of=ETapprox] (Omega0) {\small Fit objects (\ref{initialguess}) to the most  relevant peaks of $E_T$};
    \path [line,dashed] (measures) -- (Escapprox);
    \path [line] (Escapprox) -- (ETapprox);
    \path [line] (ETapprox) -- (Omega0);
\end{tikzpicture}
\caption{Procedure to recover objects using a crude approximation of the electric field at the detectors, obtained from the hologram using cost functional (\ref{costEapprox}).}
\label{fig11}
\end{figure}

For light of wavelength $660$ nm we have approximated the electric field by
 at the detectors exploiting that 
$E_{inc} \sim 1$ at a screen located a distance $z_0=z_i+z_p=10$ 
from the light emitter.   Similarly, for light of wavelength $405$ nm
$E_{inc} \sim -\imath $ at the detectors.
Then $\overline{E_{inc}} \sim  \imath $ and
$ E_{sc} \overline{E_{inc}} + \overline{E_{sc}} E_{inc}
\sim - 2 {\rm Im}(E_{sc})$. If we set $ |E_{sc}|^2 
\sim 0$, we obtain:
\begin{eqnarray}
E_{scapprox}^{(0)} =  {\rm Im} (E_{sc}) =  (1-{\cal I})/2,  
\label{Ein2-1} 
\end{eqnarray}
Alternatively, we may set ${\rm Re} (E_{sc}) \sim 0$ and look for 
a solution of 
$ {\rm Im} (E_{sc})^2 - 2 {\rm Im} (E_{sc}) + 1 - {\cal I} =0$
decaying to zero at the borders, which yields
$E_{scapprox}^{(0)} = {1 - \sqrt{\cal I} }. $

%
%
%
%

A similar strategy may be used if $E_{inc} \sim -1$ or $E_{inc} \sim \imath $ at the screen, or for other values.

\subsection{Improved approximation of the electric field}
\label{sec:improved}
To be able to sharpen object guesses we need better approximations 
of the electric field. To obtain them, we use the hologram definition:
\begin{eqnarray}
{\cal I} =  |E_{sc} + E_{inc}|^2 =  |E_{inc}|^2 +  |E_{sc}|^2  
+ E_{sc} \overline{E_{inc}} + \overline{E_{sc}} E_{inc},
\label{hdefinition}
\end{eqnarray} 
which can be rewritten as 
\begin{eqnarray}
{\cal I} \!-\!  |E_{inc}|^2
  = {\rm Re}(E_{sc})^2 \!+\! {\rm Im}(E_{sc})^2 
\!+\! 2 [{\rm Re}(E_{sc}) \cos(k_e z_0) 
\!+\! {\rm Im}(E_{sc}) \sin(k_e z_0) ],
\label{hcartesiano}
\end{eqnarray}
where $z_0$ is the distance from the emitter to the
recording screen, or
\begin{eqnarray}
{\cal I} - |E_{inc}|^2
 = r^2 + r [e^{\imath \phi} e^{-\imath k_e z_0}
+  e^{-\imath \phi} e^{\imath k_e z_0}], \quad 
E_{sc}=r e^{\imath \phi}.
\label{hpolar}
\end{eqnarray}
At each fixed receptor, equations (\ref{hcartesiano}) and
(\ref{hpolar}) are undetermined: one equation 
for two unknowns. Setting the real or imaginary parts
equal to zero we obtain the crude approximations employed 
before. However, the true fields are solutions with non zero 
real and imaginary parts.

Our procedure to approximate the complex electric field
at the detectors knowing the hologram is sketched in 
Figure \ref{fig12}. We define an error functional to quantify
the difference between the true hologram and the hologram
that would be observed using a prediction of the electric
field at the receptors. Then, we combine a gradient method 
to reduce the error in our prediction of the electric field and 
a Gaussian  filter to smooth out strong variations. To initialize
the optimization procedure avoiding convergence to the already 
known solutions with zero real or imaginary part, we use the 
electric field scattered by spheres placed at the peaks of the 
topological energy fields computed using the explicit expressions  
in Appendix \ref{sec:explicitforwardadjoint}.
This procedure yields the approximations to the electric
fields at the recording screen depicted in 
Figure \ref{fig13}. We detail the steps next.

\tikzstyle{block1} = [rectangle, draw, fill=blue!20, 
    text width=24em, text centered, rounded corners, minimum height=2em]
\tikzstyle{block2} = [rectangle, draw, fill=blue!20, 
    text width=24em, text centered, rounded corners, node distance=1.0cm,
    minimum height=2em]
\tikzstyle{block3} = [rectangle, draw, fill=blue!20, 
    text width=24em, text centered, rounded corners, node distance=1.4cm,
    minimum height=2em]
\tikzstyle{line} = [draw, -latex']
\tikzstyle{cloud} = [draw, ellipse,fill=red!20, node distance=7cm,
    minimum height=2em, text width=5em]

\begin{figure}[!h]
\centering
\begin{tikzpicture}[node distance = 1.2cm, auto]
    \node [block1] (Escapprox) {\small
    Approximate the electric field $E^{(0)}_{scapprox}$ 
    at the detectors using (\ref{Ein1-1}) or (\ref{Ein2-1})};
    \node [cloud, left of=Escapprox] (measures) {\small Record the hologram};
    \node [block1, below of=Escapprox] (ETapprox) {\small Evaluate the analytical expression (\ref{etempty}) for the topological energy $E_T$ with $E^{(0)}_{scapprox}$ and  (\ref{forwardempty},\ref{Emeasapprox},\ref{adjointemptyE})};
    \node [block2, below of=ETapprox] (Omega0) {\small Fit spheres to the most relevant peaks (\ref{initialguess})  of $E_T$};
    \node [block2, below of=Omega0] (Escanal) {\small Compute analytically the electric field $E^{(0)}_{scsynth}$ scattered by the spheres at the detectors using (\ref{series})};
    \node [block3, below of=Escanal] (Escimprove) {\small Alternate gradient optimization    (\ref{decrease}) starting from $E^{(0)}_{scsynth}$ with Gaussian filtering (\ref{filtertrue}) to generate an electric field $ E^{(1)}_{scapprox}$ that fits the recorded hologram};
    \node [block3, below of=Escimprove] (ETimprove) {\small Evaluate the analytical expression (\ref{etempty}) for the topological energy $E_T$ with $E^{(1)}_{scapprox}$ and  (\ref{forwardempty},\ref{Emeasapprox1},\ref{adjointemptyE})};
    \node [block2, below of=ETimprove] (Omegaimprove) {\small Fit objects (\ref{initialguess}) to the most relevant peaks of $E_T$};
    \path [line,dashed] (measures) -- (Escapprox);
    \path [line] (Escapprox) -- (ETapprox);
    \path [line] (ETapprox) -- (Omega0);
    \path [line] (Omega0) -- (Escanal);
    \path [line] (Escanal) -- (Escimprove);
    \path [line] (Escimprove) -- (ETimprove);
    \path [line] (ETimprove) -- (Omegaimprove);
\end{tikzpicture}
\caption{Strategy to obtain a sharper approximation of the electric field at 
the detectors from the hologram, as well as a  possibly better description 
of the geometry  of the scatterers.}
\label{fig12}
\end{figure}
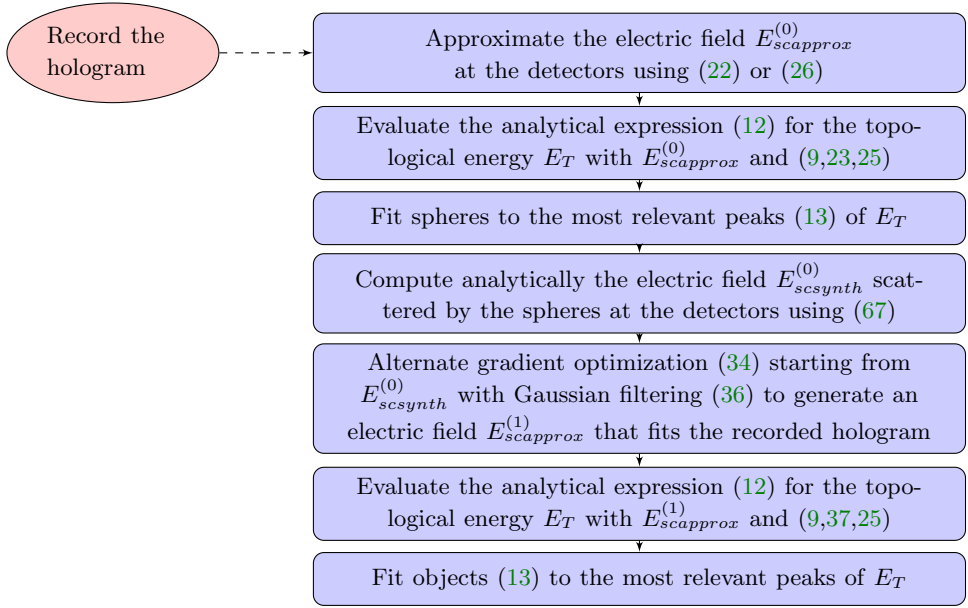

\subsubsection{Gradient optimization}
\label{sec:gradient}

We define the following optimization problem: 
Find $E_{scapprox} \in \mathbb C^N$ minimizing
\begin{eqnarray}
F(E_{scapprox}) = \sum_{j=1}^N |  {\mathcal I}({\bf x}_j) - |
 E_{scapprox,j} + E_{inc}({\bf x}_j)|^2 |^2,
\label{costEG}
\end{eqnarray}
where ${\mathcal I}$ is the hologram recorded for the true
scatterers.

Since the real and imaginary parts of our fields are strongly 
oscillatory, we use the polar representation of complex numbers
and resort to a gradient technique.
Let us set $E_{scapprox,j}= r_j e^{\imath \phi_j}$,
${\bf r}=(r_1,\ldots,r_N)$, ${\boldsymbol \phi}=
(\phi_1,\ldots,\phi_N)$, and
\begin{eqnarray}
F({\bf r},{\boldsymbol \phi})
= \sum_{j=1}^N |  {\mathcal I}({\bf x}_j)  - |
  r_j e^{\imath \phi_j} + E_{inc}({\bf x}_j)|^2 |^2.
\label{costF}
\end{eqnarray}  
Then, the partial derivatives of the cost function are given by:
 \begin{eqnarray}
 {\partial F \over \partial r_j} = - 4 [{\mathcal I}({\bf x}_j) - |
  r_j e^{\imath \phi_j} + E_{inc}({\bf x}_j)|^2 ]
  [(r_j e^{\imath \phi_j} + E_{inc}({\bf x}_j)) e^{-\imath \phi_j}
  \nonumber \\ [-1ex]
  +  (r_j e^{-\imath \phi_j} + \overline{E_{inc}({\bf x}_j)}) 
  e^{\imath \phi_j}],
  \label{partialFr}\\ 
  {\partial F \over \partial \phi_j} = - 4 [{\mathcal I}({\bf x}_j) - |
  r_j e^{\imath \phi_j} + E_{inc}({\bf x}_j)|^2 ]
  [-\imath (r_j e^{\imath \phi_j} + E_{inc}({\bf x}_j)) r_j e^{-\imath \phi_j}
  \nonumber \\   [-1ex]
  +  \imath (r_j e^{-\imath \phi_j} + \overline{E_{inc}({\bf x}_j)})
   r_j e^{\imath \phi_j}].
  \label{partialFphi}
 \end{eqnarray}
We generate sequences along which the error functional
decreases setting
\begin{eqnarray}
(r_1,\ldots,r_N,\phi_1,\ldots,\phi_N)^{new}=
(r_1,\ldots,r_N,\phi_1,\ldots,\phi_N)^{old} \nonumber \\  
- \delta ({\partial F \over \partial r_1},\ldots,{\partial F \over \partial r_N},
 {\partial F \over \partial \phi_1}, \ldots, {\partial F \over \partial \phi_N})
 (r_1,\ldots,r_N,\phi_1,\ldots,\phi_N)^{old}.
\label{decrease}
\end{eqnarray}
The parameter $\delta >0$ might be chosen to take different values 
along different directions, though we will keep it uniform in our tests.
We have set $\delta=0.1$.

To implement the gradient procedure we need to select an adequate
initial value for the electric field. To do so, we exploit the information
about the scatterers provided by the strategy sketched in 
Figure \ref{fig11}.
We have two options:
\begin{itemize}
\item Use (\ref{initialguess}) 
to obtain a guess $\Omega^{(0)}$ of the scatterer geometry.
The resulting shape for $\Omega^{(0)}$ is arbitrary. 
We then solve the forward problem (\ref{forward}) 
numerically to evaluate the corresponding scattered electric 
field $E_{scsynth}^{(0)}$ at the detectors.  This can be done
employing BEM-FEM  as in \cite{siims}.
\item Fit balls $\Omega^{(0)}$ to the shapes predicted 
by   (\ref{initialguess}).
We then solve the forward problem (\ref{forward}) 
using series expansions in terms of spherical harmonics
to evaluate the corresponding electric field $E_{scsynth}^{(0)}$
at the detectors, see Appendix \ref{sec:explicitforwardadjoint}.
This choice reduces the computational complexity.
\end{itemize}
The resulting scattered fields $E_{scsynth}^{(0)}$, however, do 
not satisfy (\ref{hdefinition}). We use $E_{scsynth}^0$ as starting 
point to  optimize (\ref{costEG}). 
The scheme starts from $(r_1,\ldots,r_N,\phi_1,\ldots,\phi_N)^{0},$
obtained expressing
$E_{scsynth,j}^{(0)}= r_j^{0} e^{\imath \phi_j^{0}}$,
$j=1,\ldots,N$. Then, we update these values using
identity (\ref{decrease}), where the derivatives are
given by formulas (\ref{partialFr})-(\ref{partialFphi}).
The procedure stops when  the cost 
functional (\ref{costEG})
falls below a fixed threshold value.

Computing the electric field scattered by a sphere 
$E_{scsynth}^{(0)}$ requires the knowledge of the 
parameter ${k}_i$, that is, its permittivity.
However, we only use this field as a starting value for the
optimization procedure which avoids converging to purely real or
imaginary solutions.  
We have checked that the output of the gradient scheme is not 
really sensitive to variations of ${k}_i$ in a wide range of values.
In case ${k}_i$ was unknown, setting it equal to a small
perturbation of ${k}_e$ would produce a reasonable approximation
of the electric field at the detectors.

\subsubsection{Gaussian filter}
\label{sec:filter}

The corrections provided by the gradient procedure described in 
Section \ref{sec:gradient} are local, in the sense that the evolution
at one detector is not influenced by the evolution at the neighboring
ones. Since the fields are oscillatory, this may result in spurious
local spikes. Large errors at some points will compensate very 
small errors at other points. To overcome this problem 
we successively distribute the error using a Gaussian filter
\cite{gaussian1,gaussian2}.

Given a Gaussian $G(x,y)=A e^{\frac{-((x-x_0)^2+(y-y_0)^2)}{2\sigma^2}}$ 
centered at $(x_0,y_0)$ and a function 
$E_{scapprox}(x,y)$, we define the discrete convolution as
\begin{eqnarray} \hskip 6mm
E_{scfilter}(x,y)=E_{scapprox}*G(x,y)= 
\sum_{i=-\infty}^\infty \sum_{j=-\infty}^\infty G(i,j) E_{scapprox}(x-i,y-j). 
\label{filter}
\end{eqnarray}
For numerical purposes we select a finite approximation, in the
form of a $3\times 3$ matrix:
\begin{eqnarray}
G(i,j)=Ae^{-\frac{(i-2)^2+(j-2)^2}{2\sigma^2}}, \quad
A^{-1}=\sum_{i=1}^3\sum_{j=1}^3 e^{-\frac{(i-2)^2+(j-2)^2}{2\sigma^2}}.
 \nonumber
\end{eqnarray}
The filtered field is:
\begin{eqnarray}
E_{scfilter}(i,j)=\sum_{k=1}^3\sum_{l=1}^3 G(k,l)E_{scapprox}(i-2+k,j-2+l),
\label{filtertrue} \\
G=\left( 
\begin{matrix}
    0.0113  &  0.0838 &   0.0113\\
    0.0838  &  0.6193 &   0.0838\\
    0.0113  &  0.0838 &   0.0113\\
\end{matrix}
\right).  \nonumber
\end{eqnarray}
This function is smoother than the original one and
can be used as starting point for a new gradient procedure.
Better results might be obtained using adaptive filters, which 
vary in space.

\begin{figure}[!h]
\centering
\hskip -1mm (a) \hskip 2.1cm (b)  \hskip 2.1cm (c) \hskip 2.1cm (d) \\
\includegraphics[width=3cm]{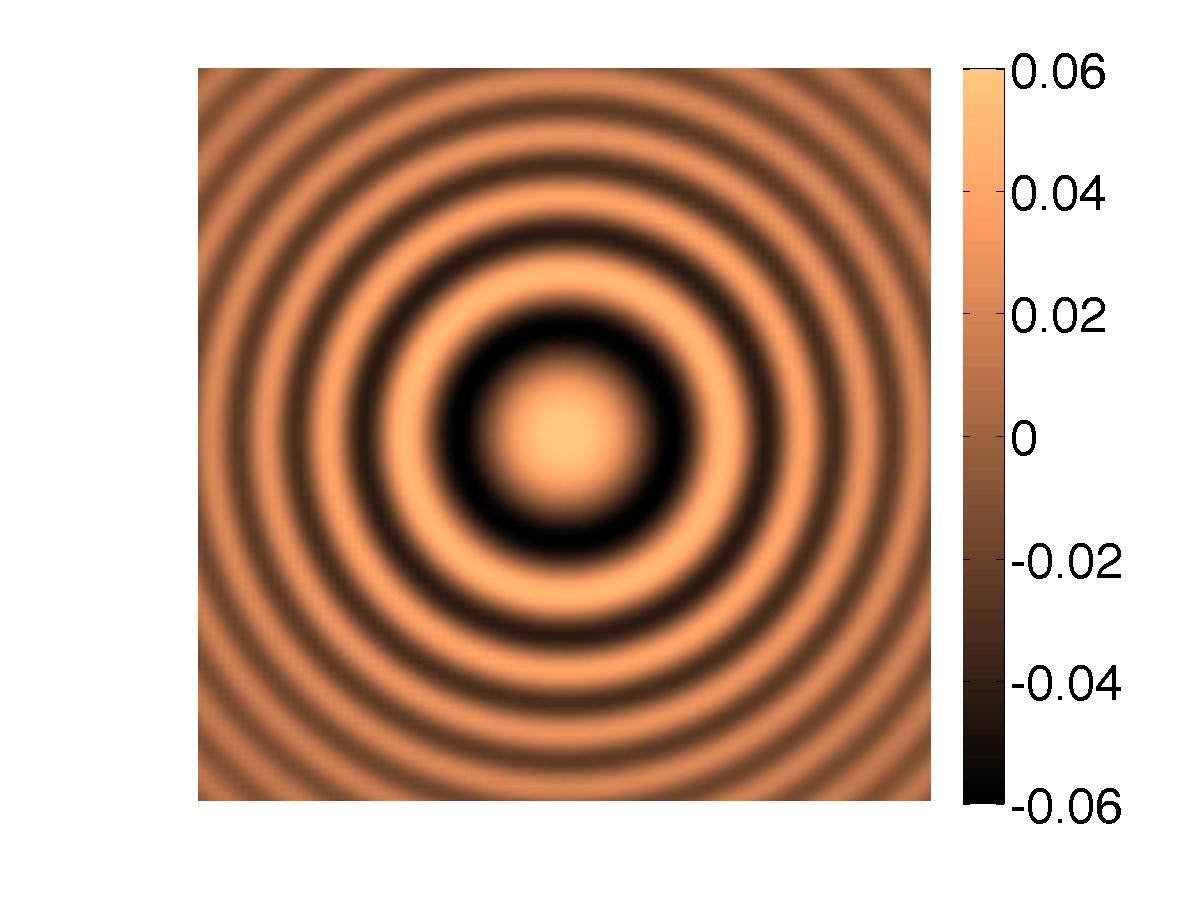} \hskip -4mm
\includegraphics[width=3cm]{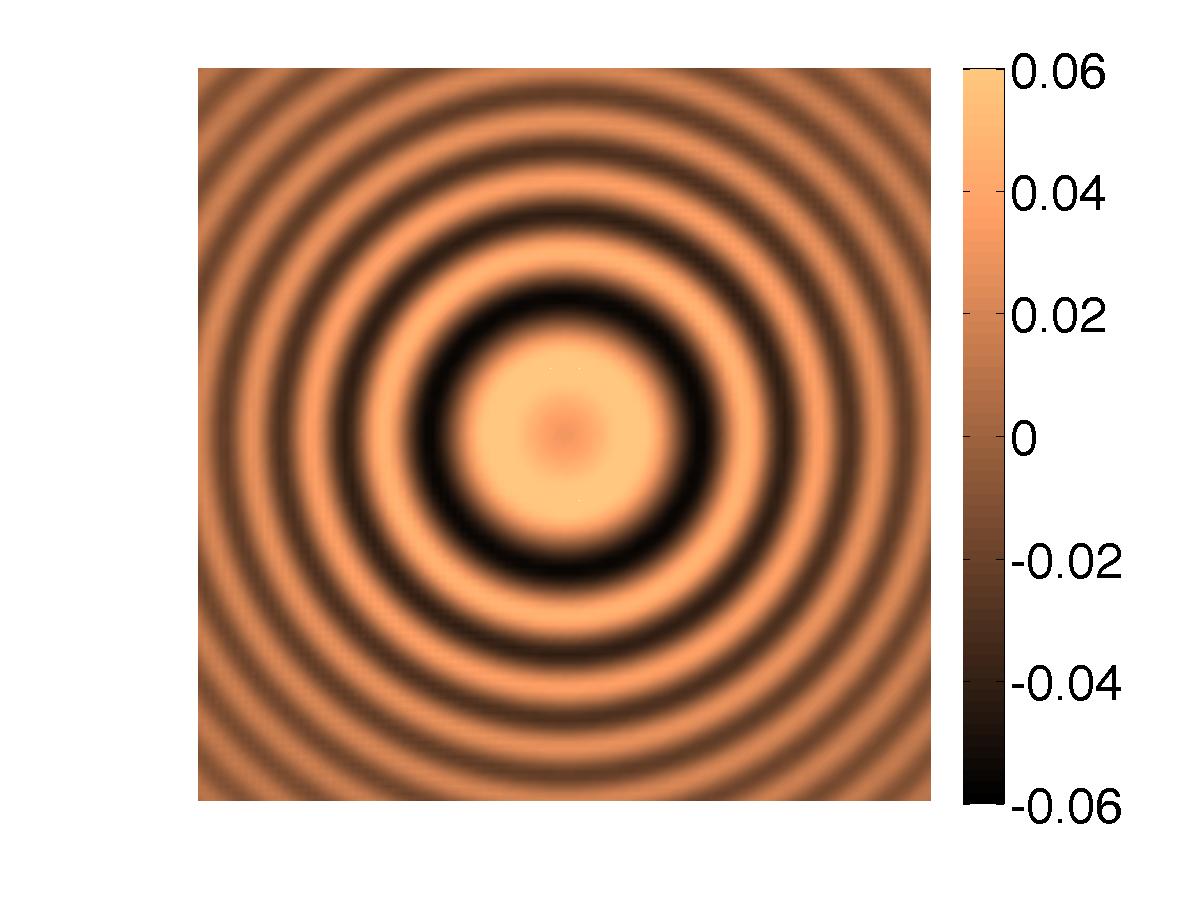} \hskip -4mm
\includegraphics[width=3cm]{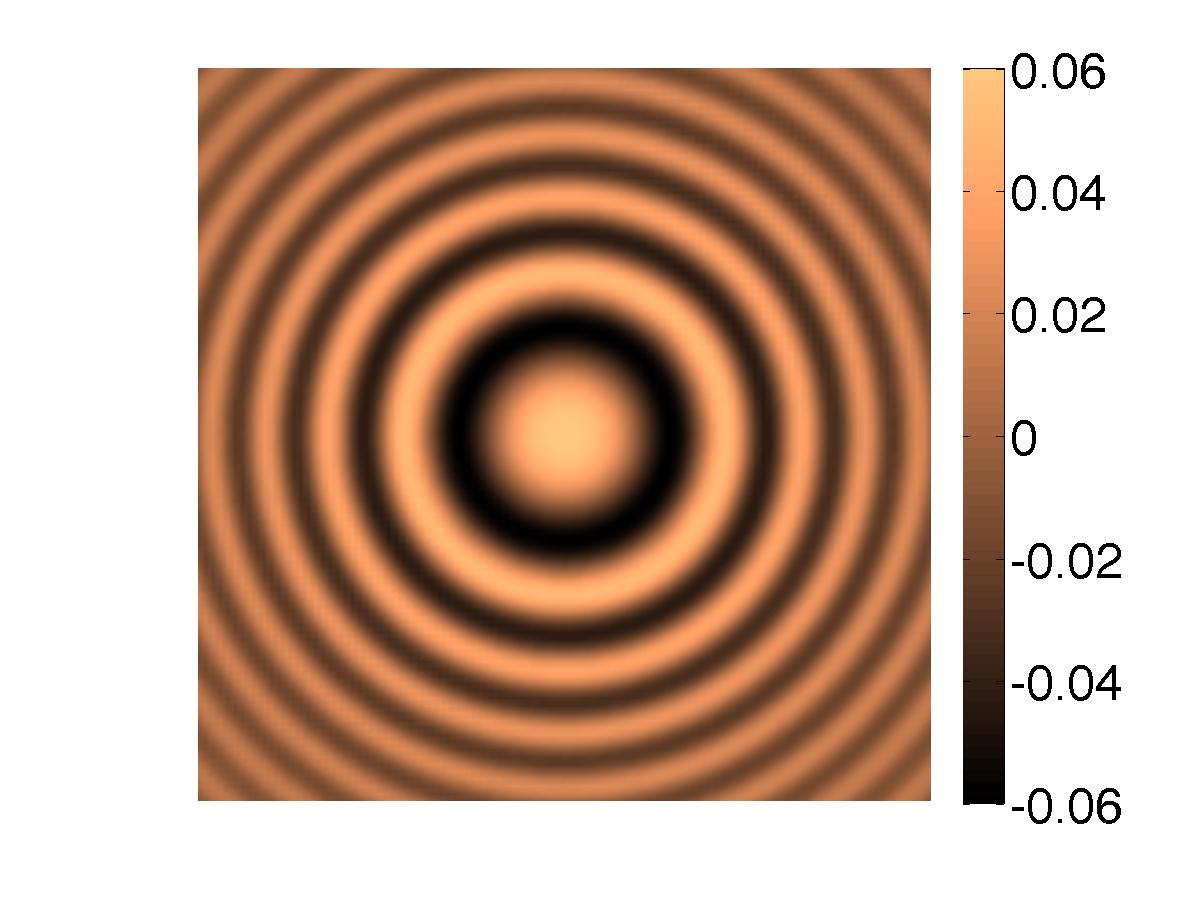} \hskip -4mm
\includegraphics[width=3cm]{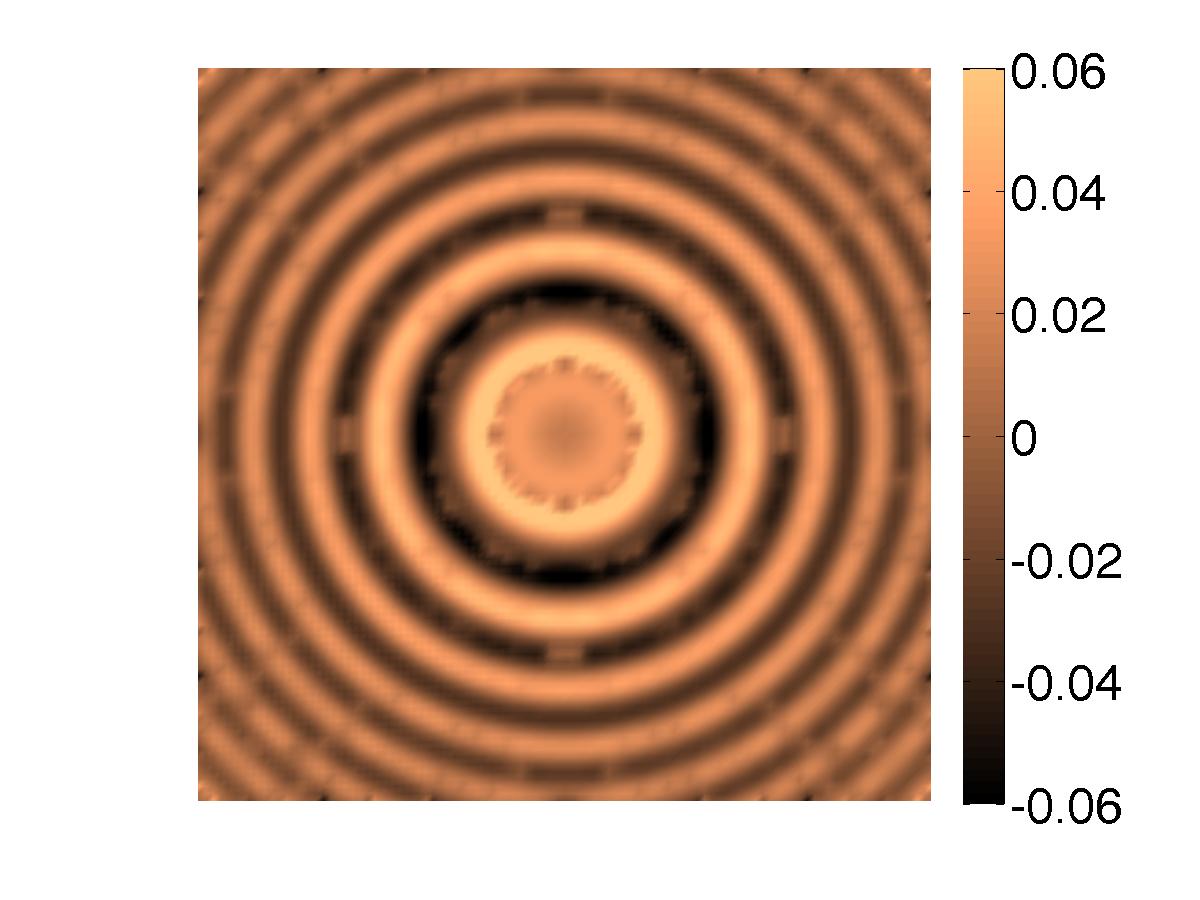} \\
\hskip -1mm (e) \hskip 2.1cm (f)  \hskip 2.1cm (g) \hskip 2.1cm (h) \\
\includegraphics[width=3cm]{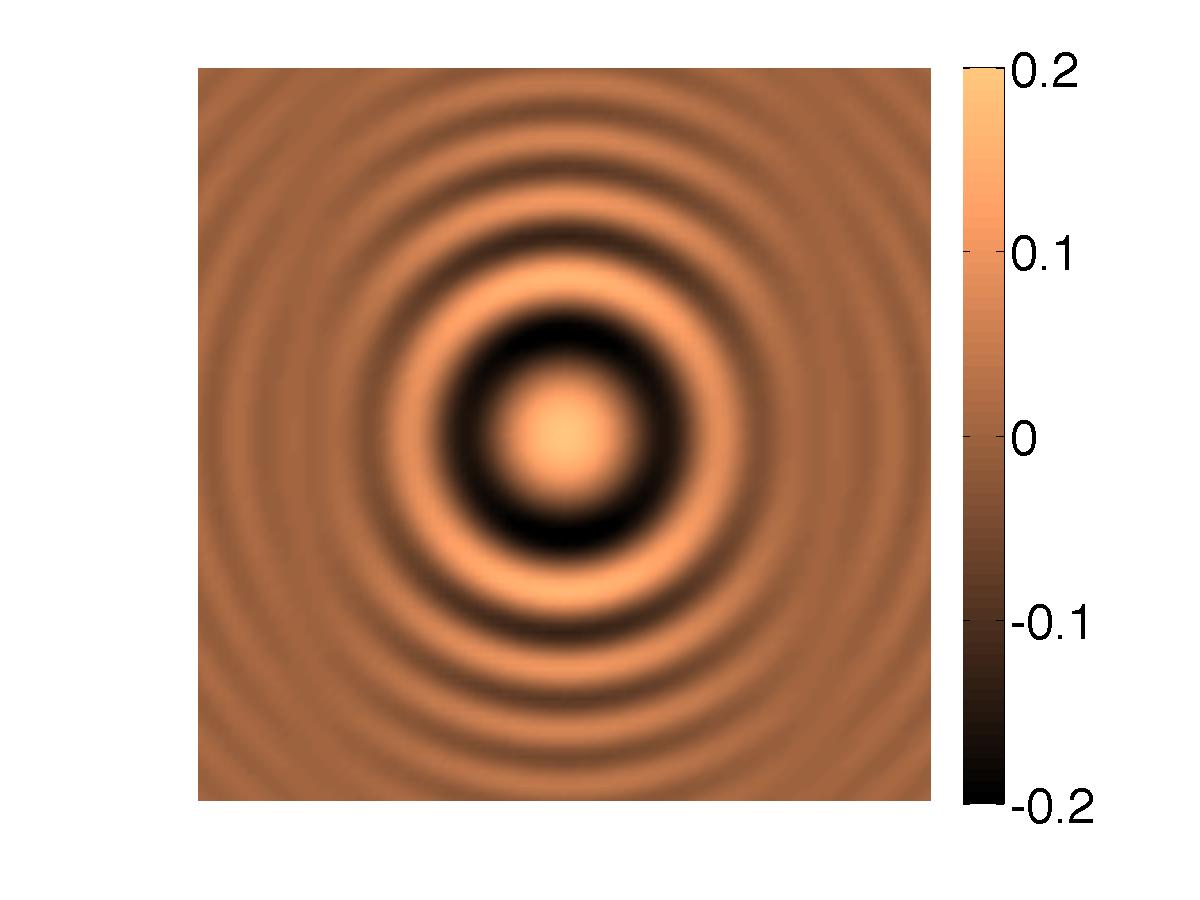}  \hskip -4mm
\includegraphics[width=3cm]{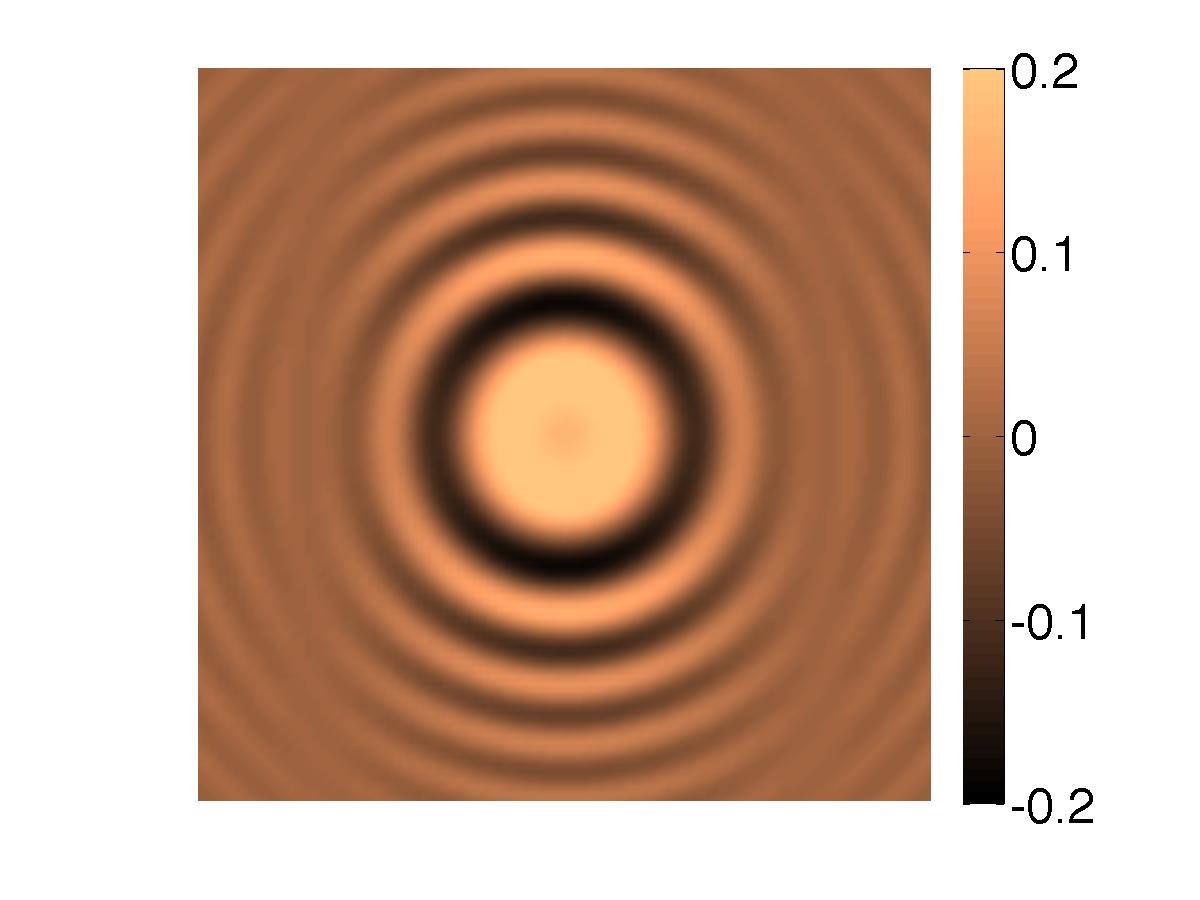}  \hskip -4mm
\includegraphics[width=3cm]{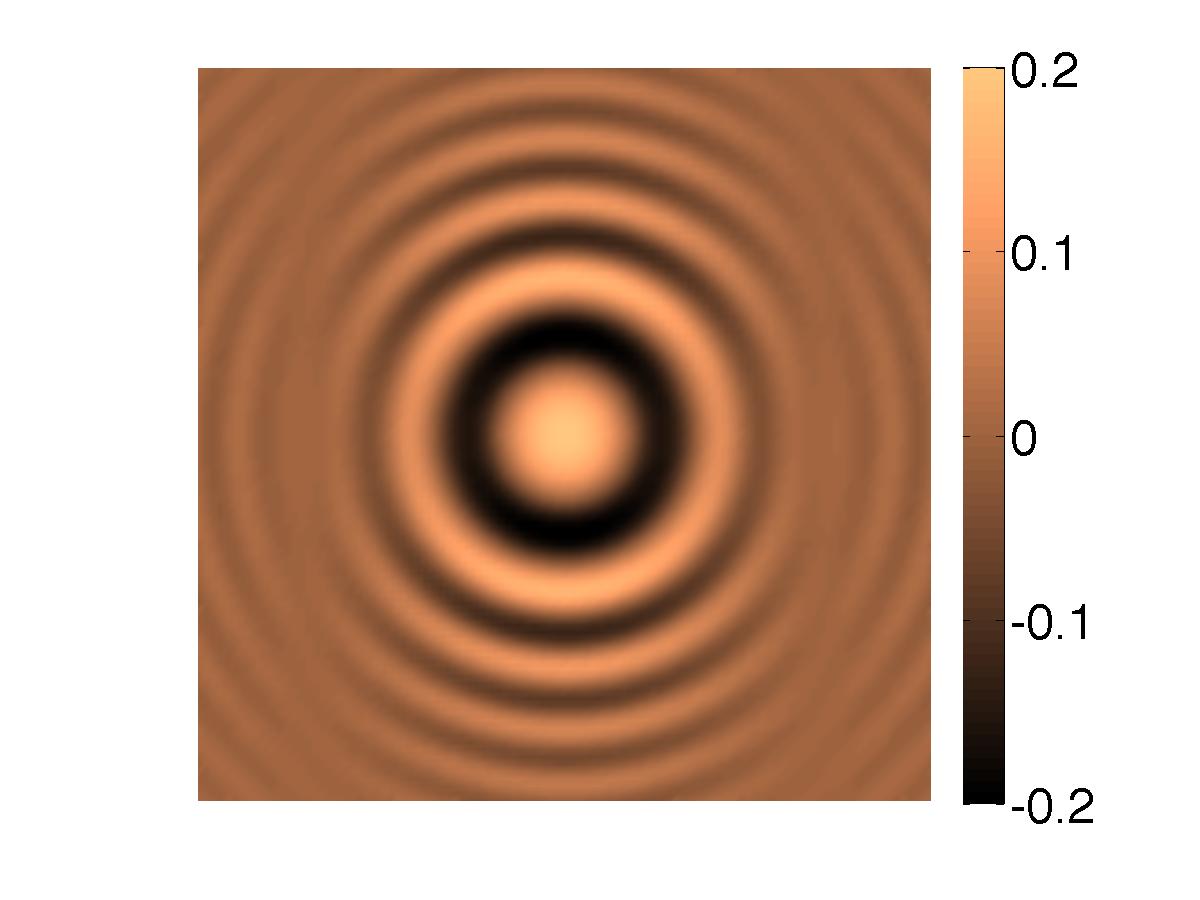}  \hskip -4mm
\includegraphics[width=3cm]{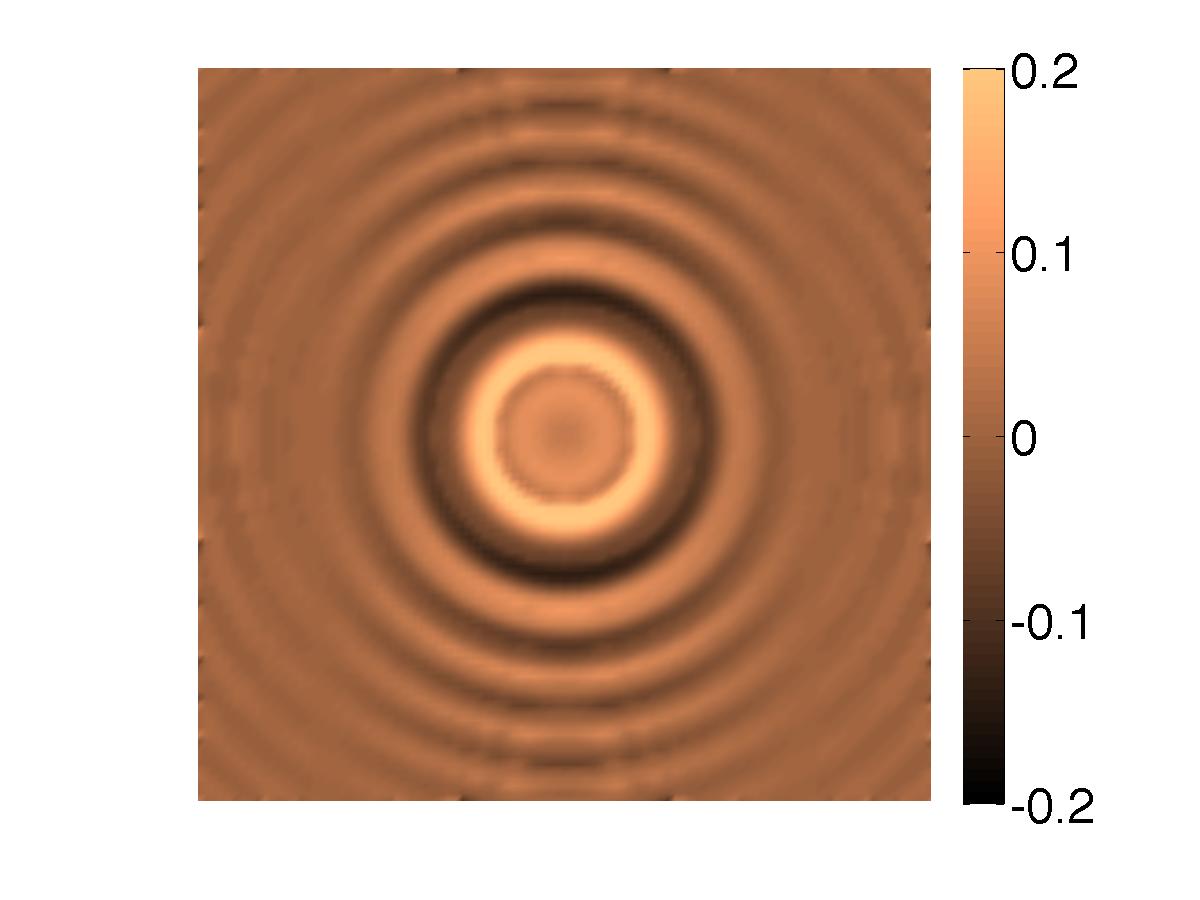} \\
\hskip -1mm (i) \hskip 2.1cm (j)  \hskip 2.1cm (k) \hskip 2.1cm (l) \\
\includegraphics[width=3cm]{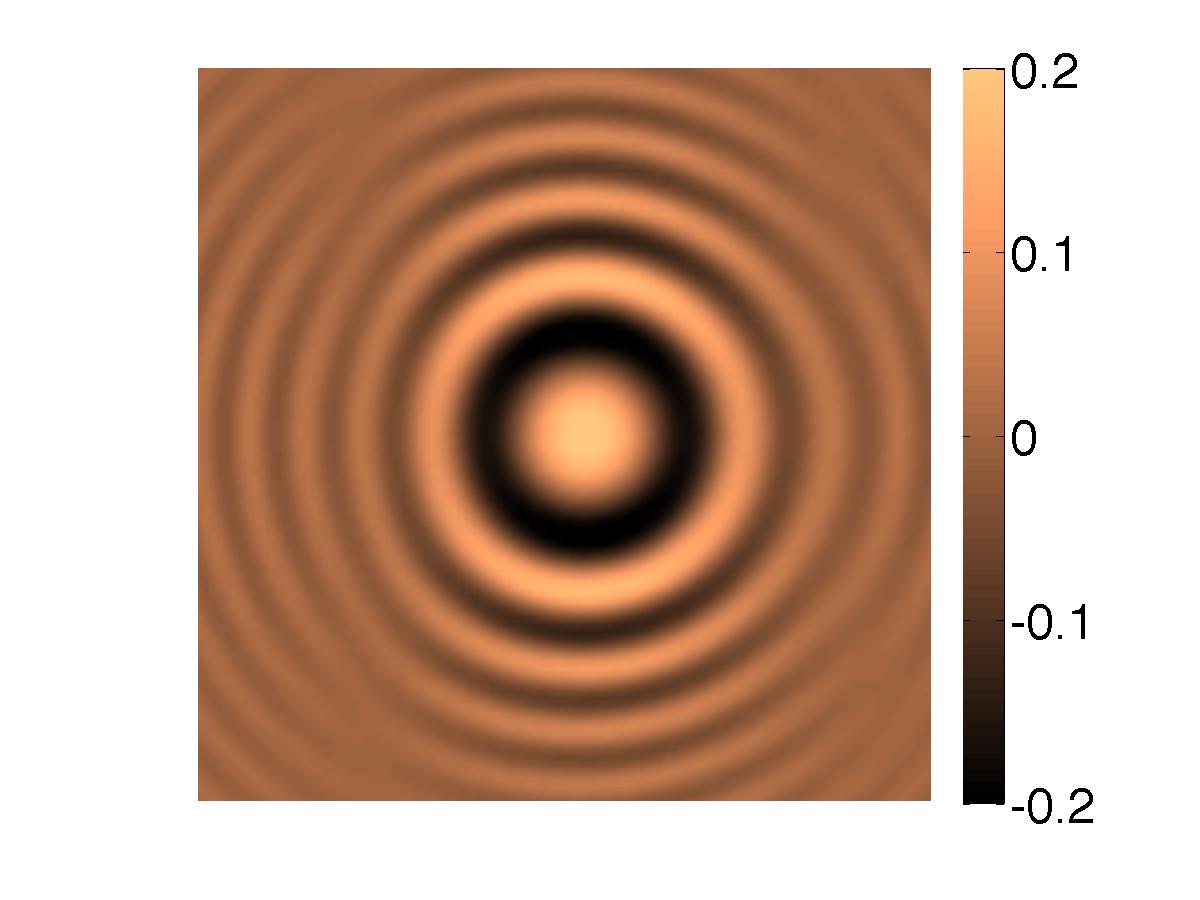}  \hskip -4mm
\includegraphics[width=3cm]{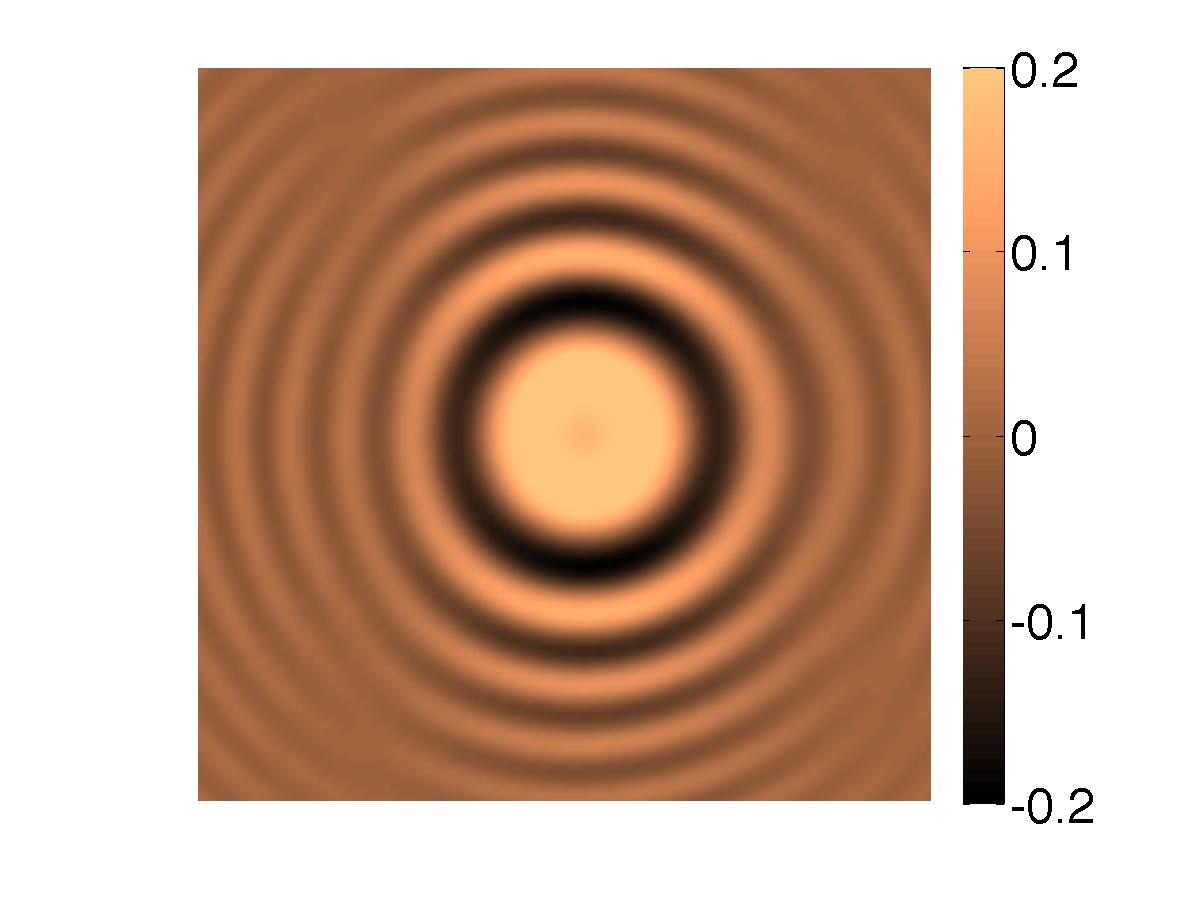}  \hskip -4mm
\includegraphics[width=3cm]{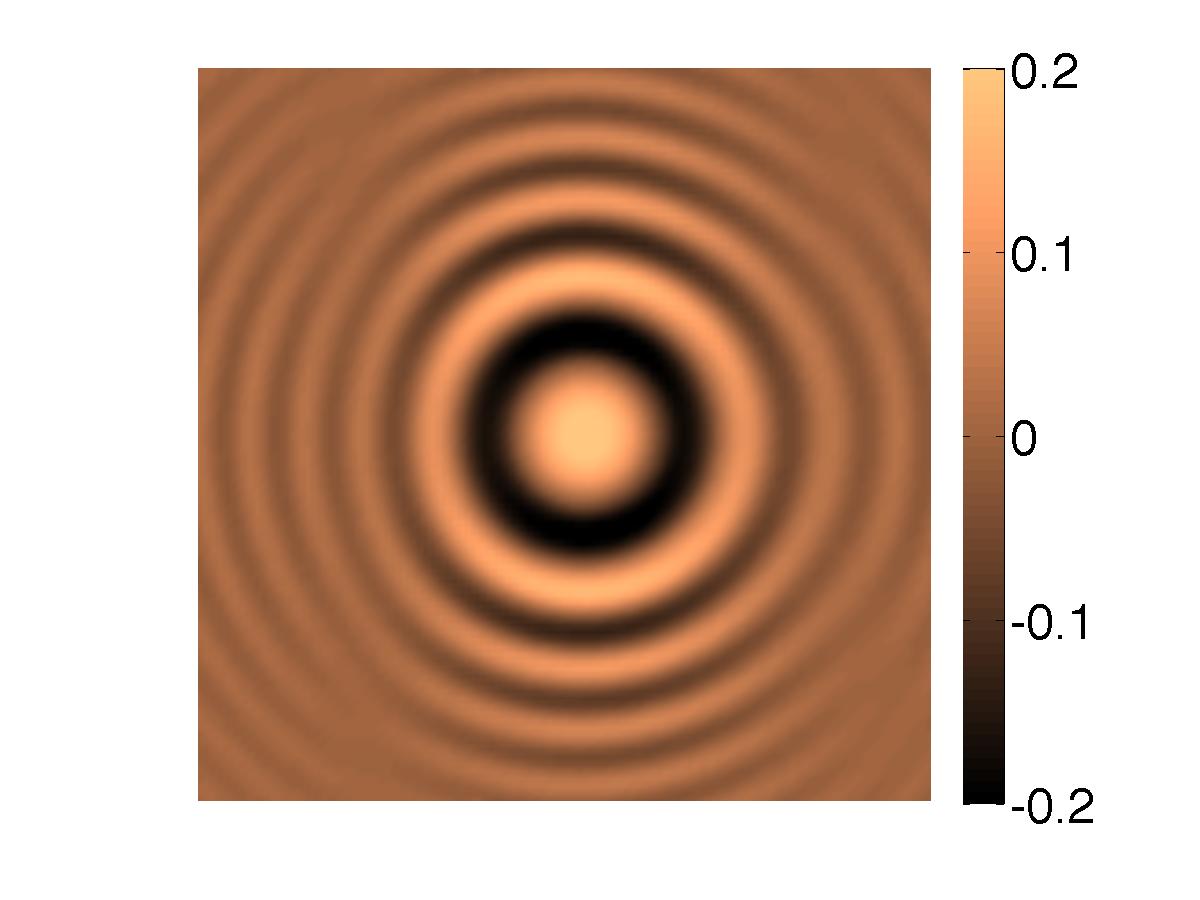}  \hskip -4mm
\includegraphics[width=3cm]{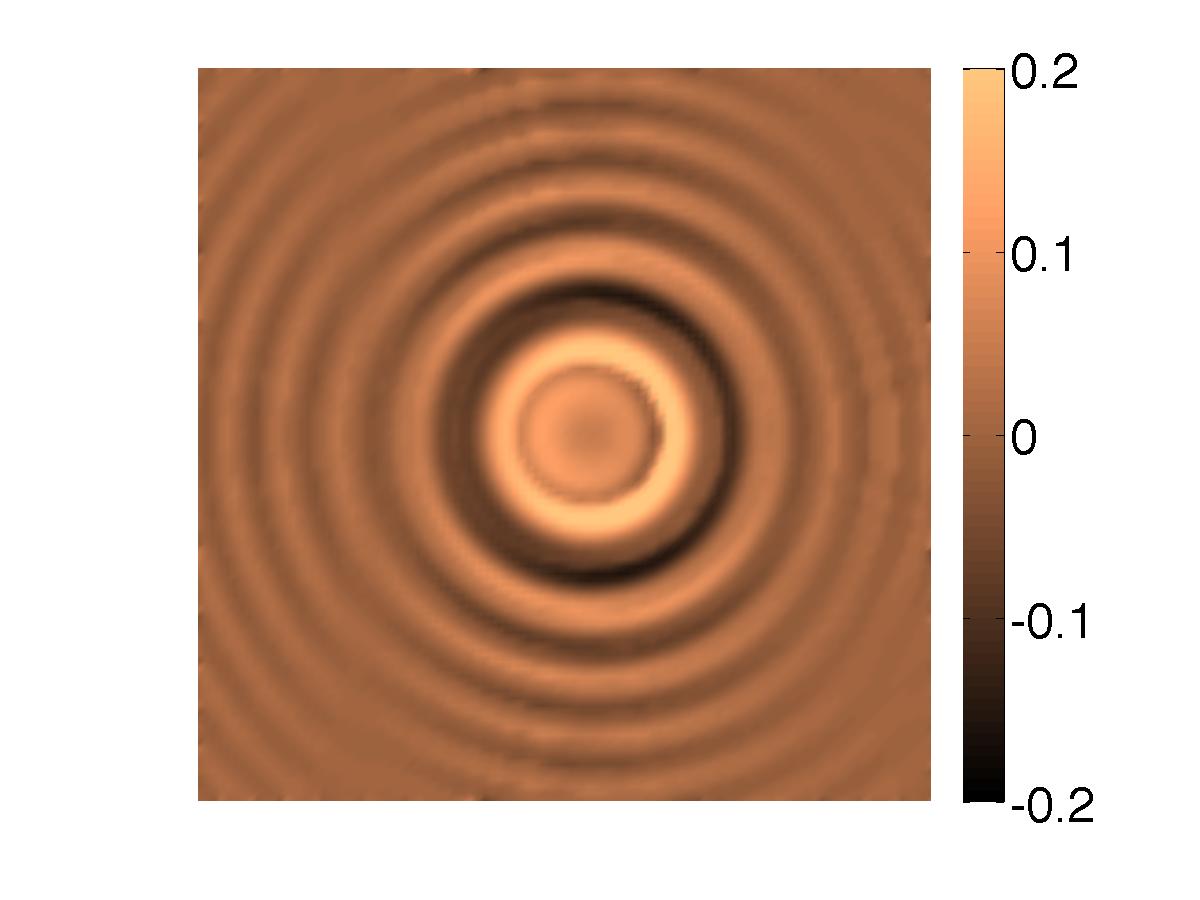} \\
\hskip -1mm (m) \hskip 2.1cm (n)  \hskip 2.1cm (o) \hskip 2.1cm (p) \\
\includegraphics[width=3cm]{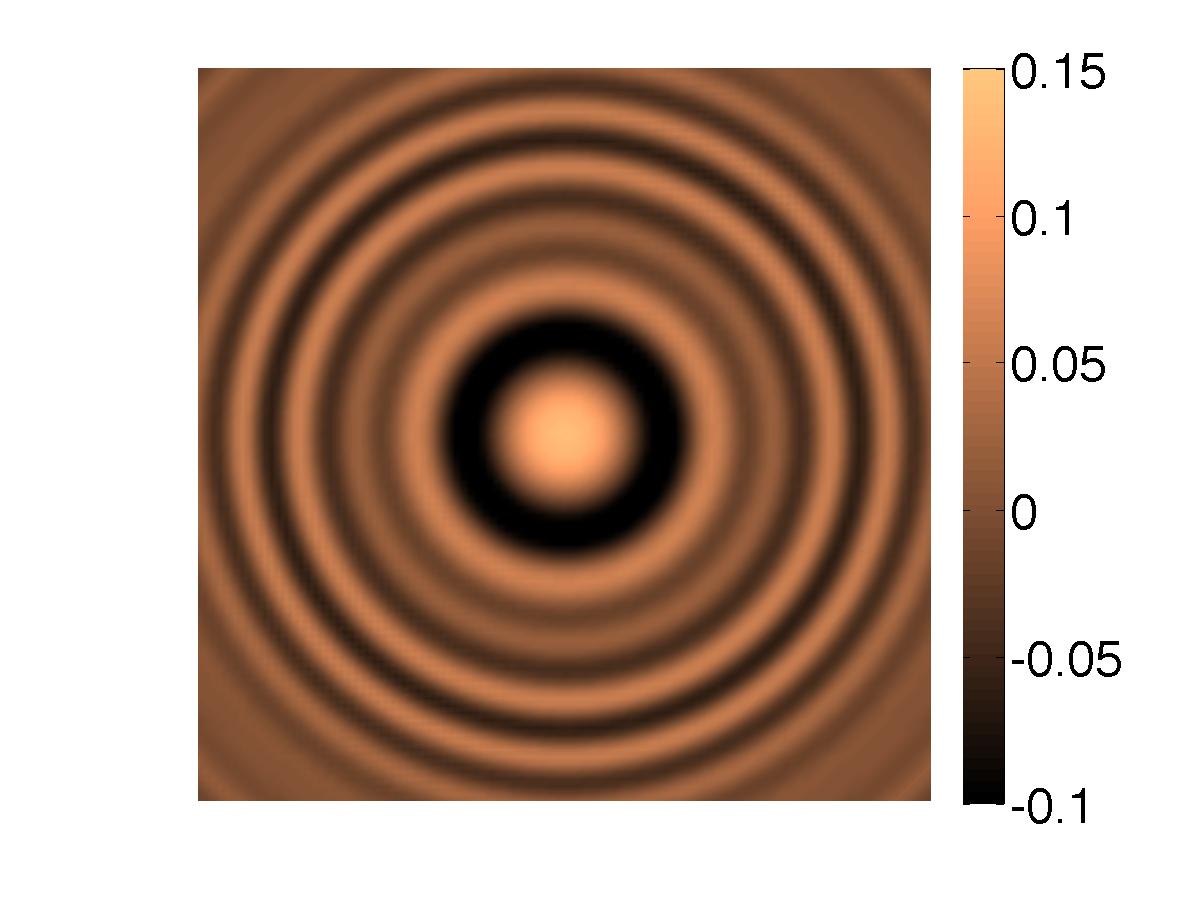}  \hskip -4mm
\includegraphics[width=3cm]{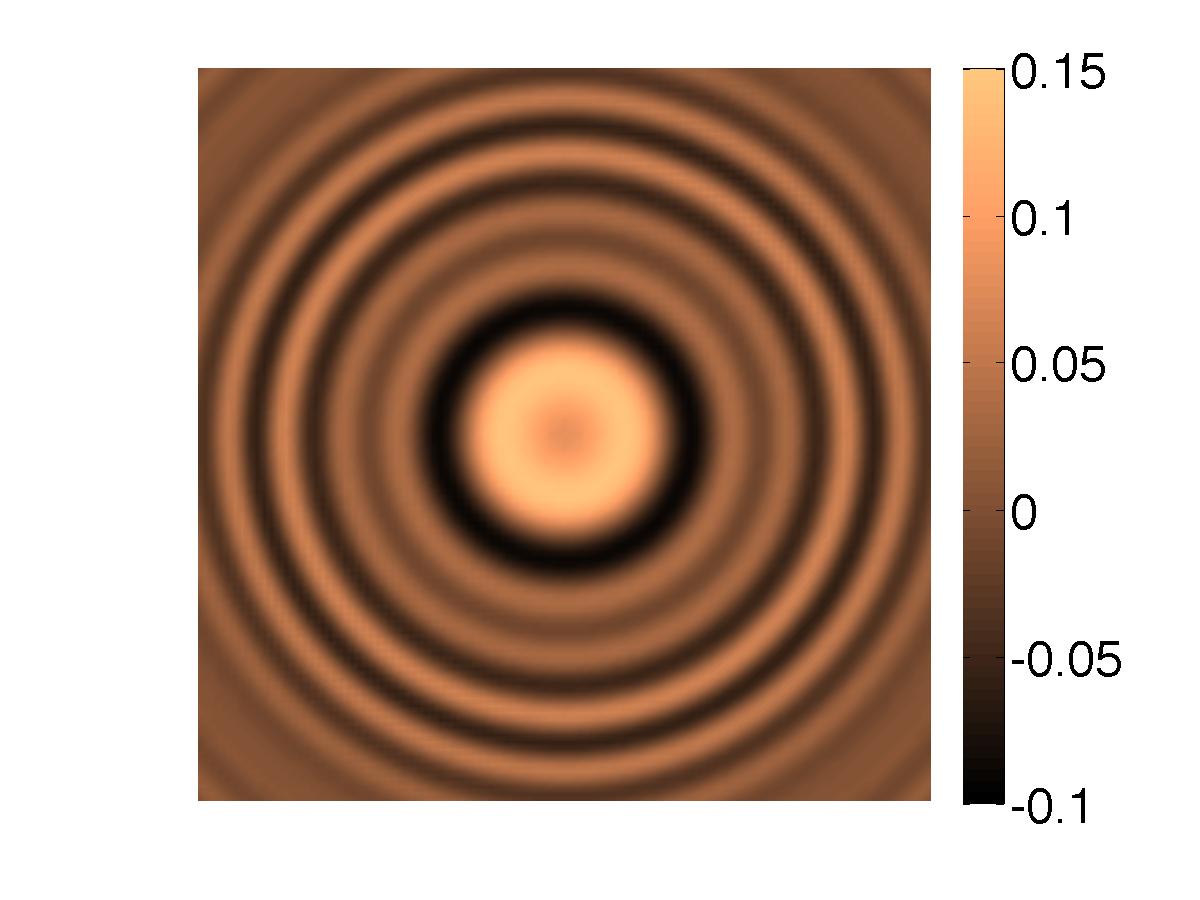}  \hskip -4mm
\includegraphics[width=3cm]{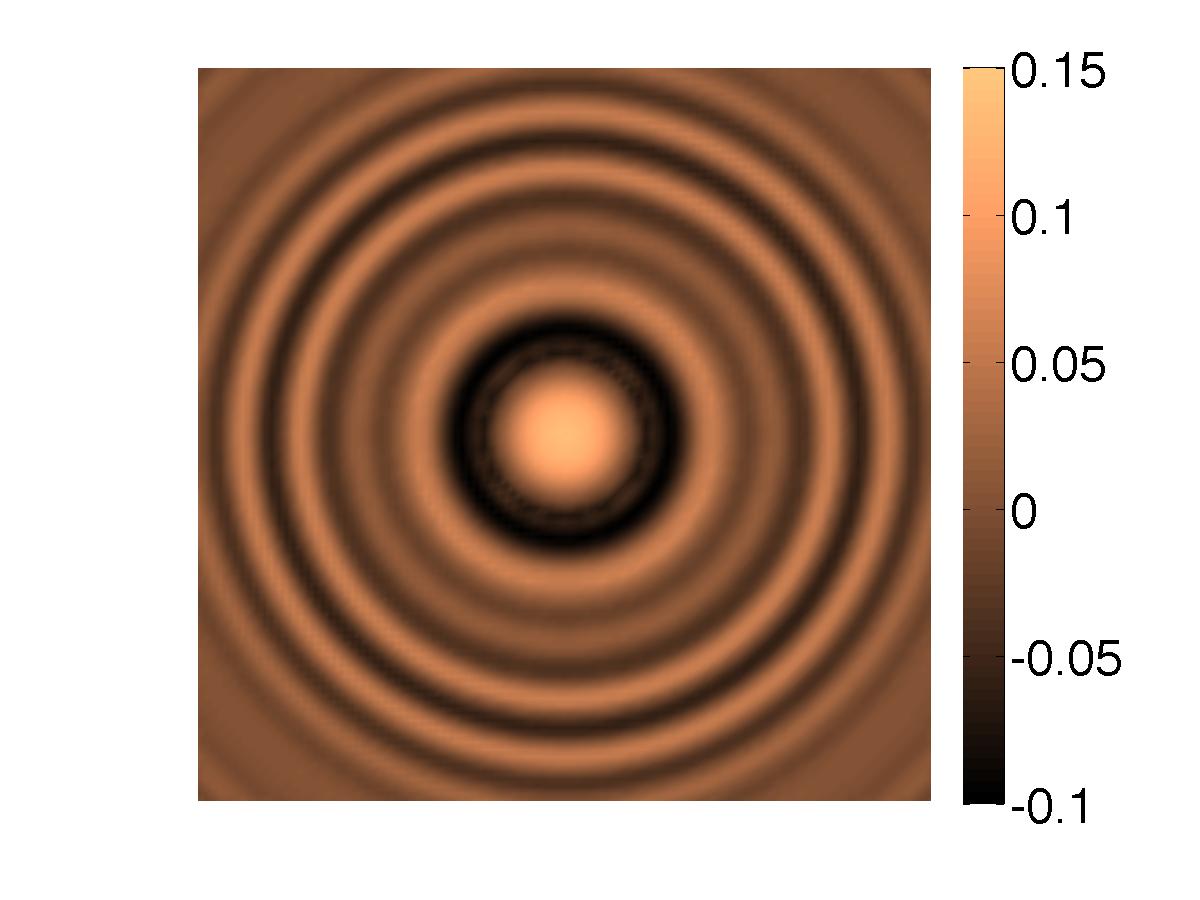}  \hskip -4mm
\includegraphics[width=3cm]{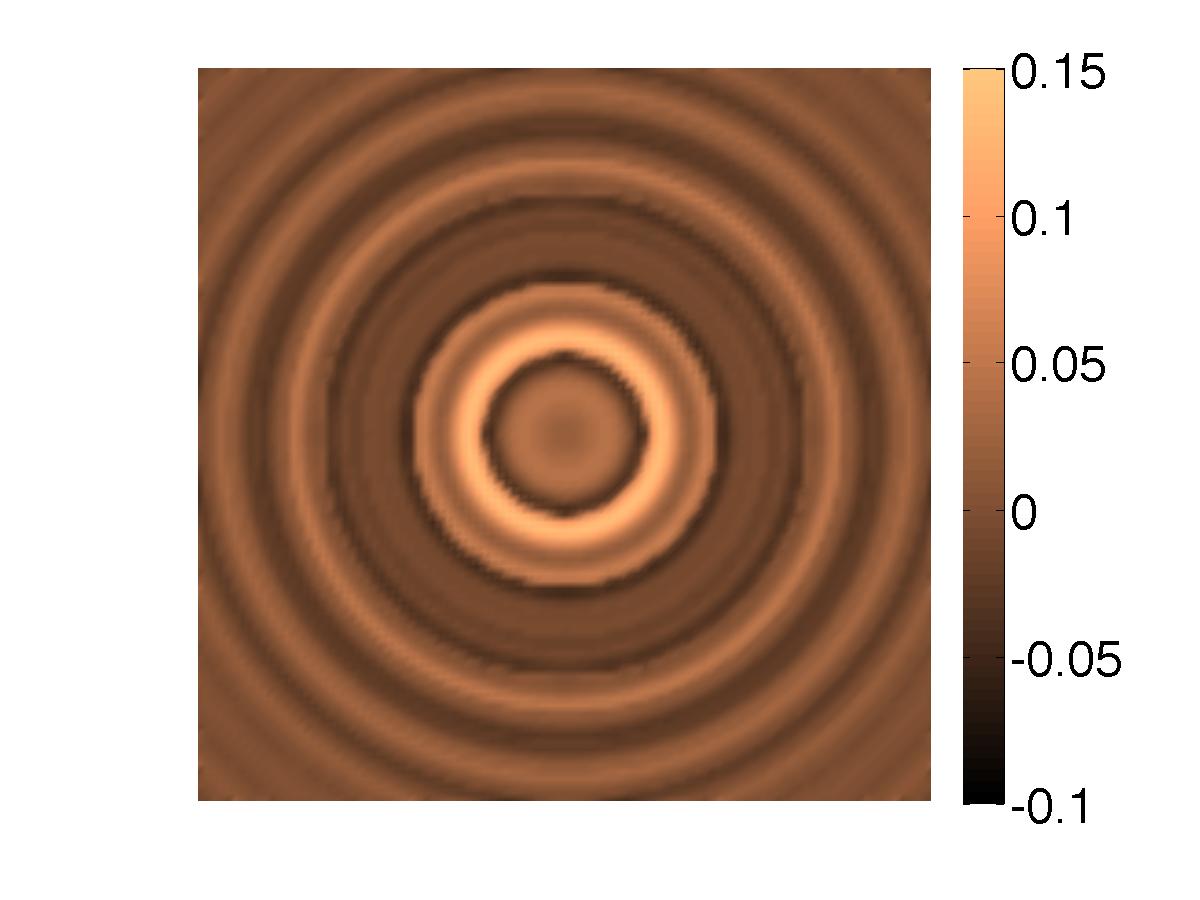} \\
\hskip -1mm (q) \hskip 2.1cm (r)  \hskip 2.1cm (s) \hskip 2.1cm (t) \\
\includegraphics[width=3cm]{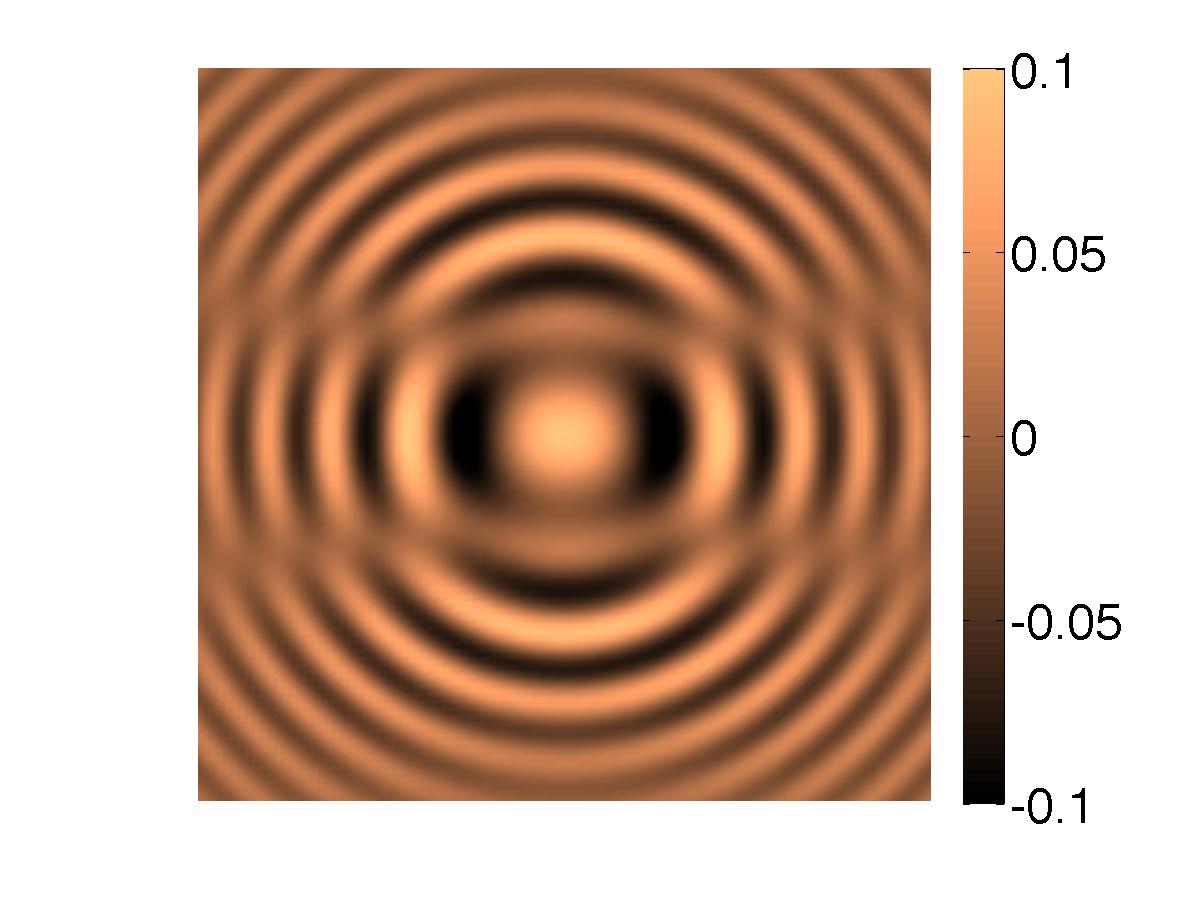}  \hskip -4mm
\includegraphics[width=3cm]{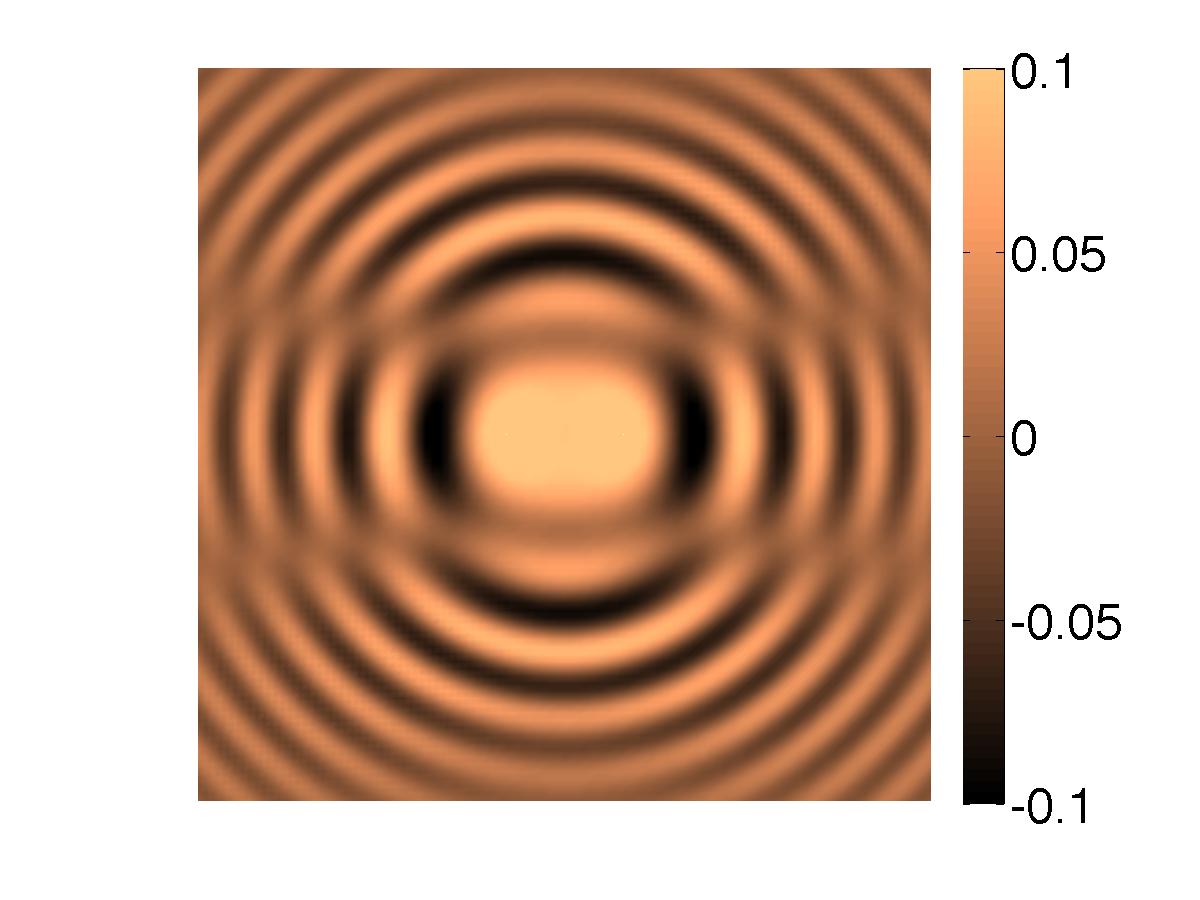}  \hskip -4mm
\includegraphics[width=3cm]{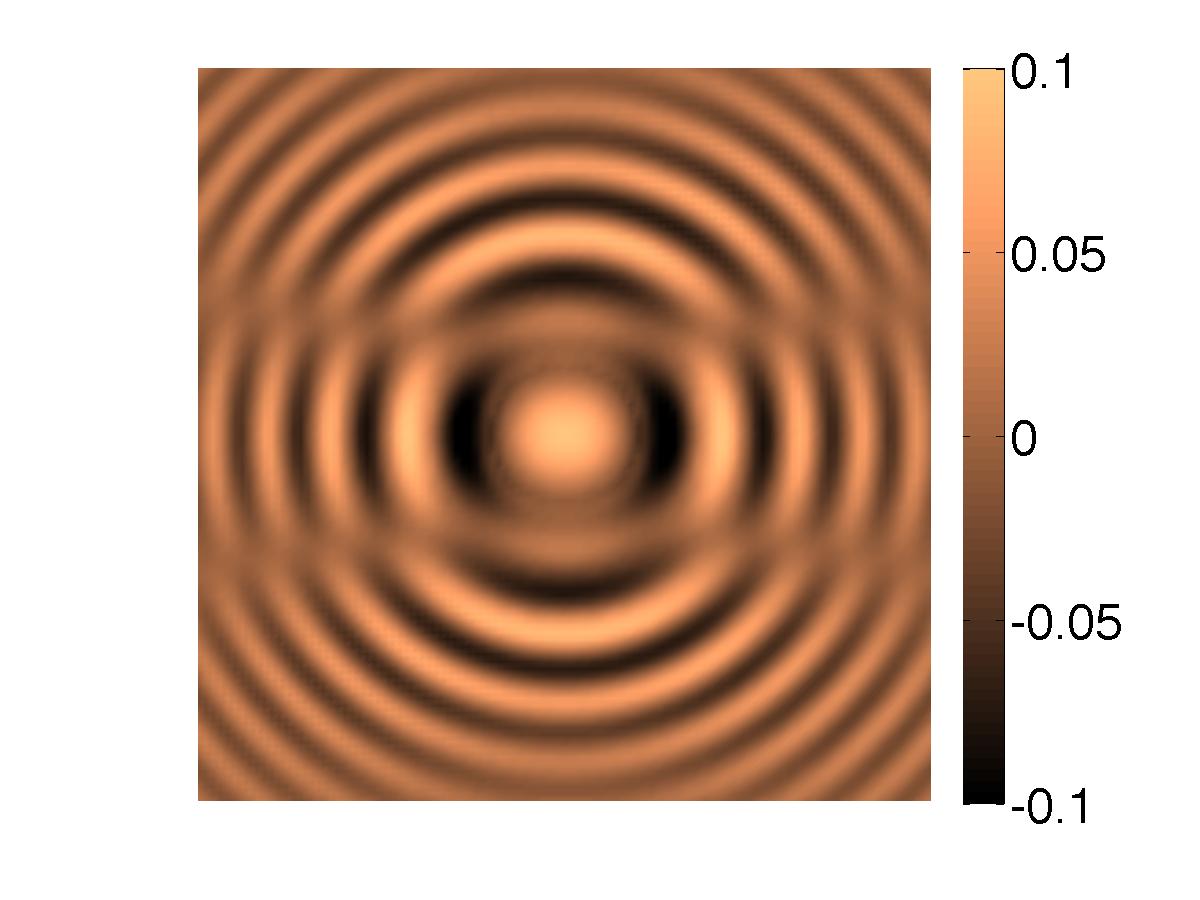}  \hskip -4mm
\includegraphics[width=3cm]{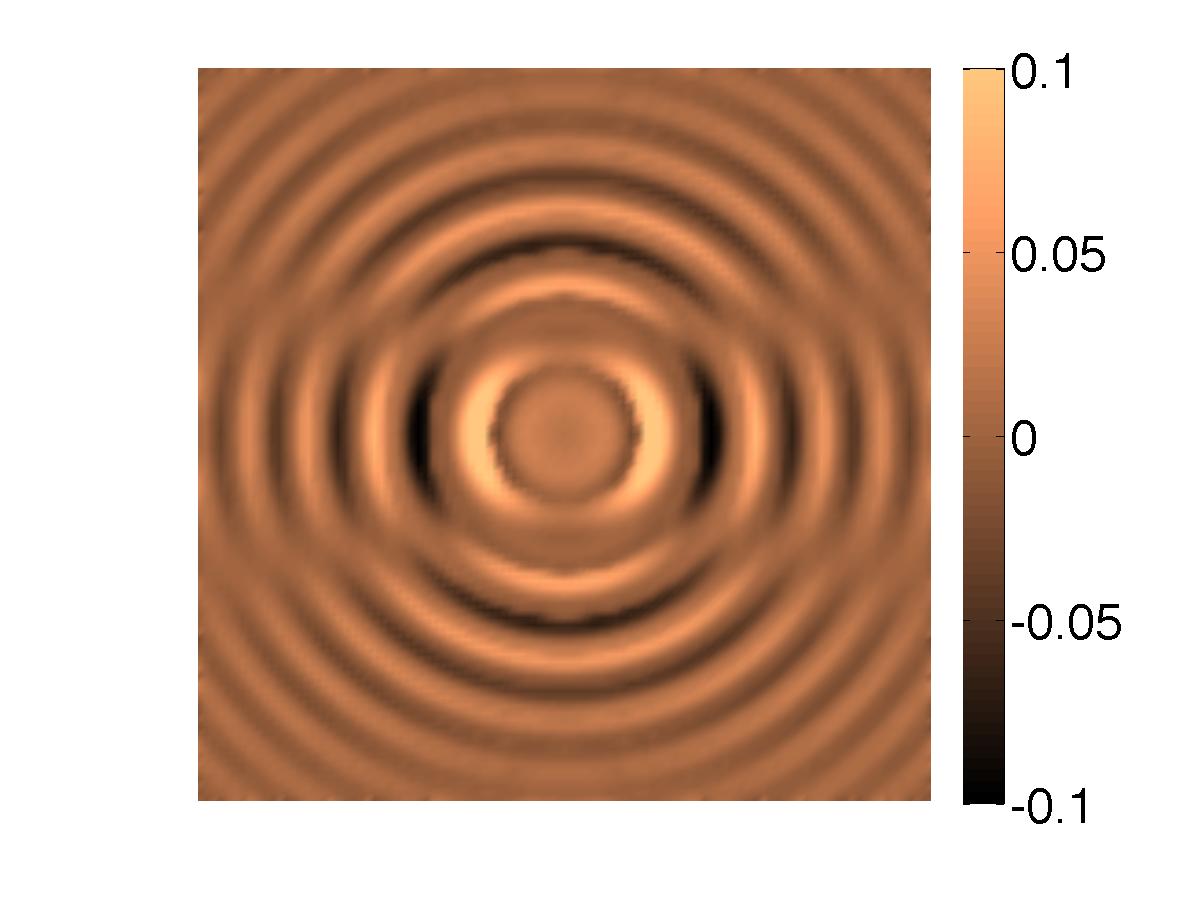}  \\
\hskip 5mm $Re(E_{sc})$ \hskip 1.2cm 
$Im(E_{sc})$ \hskip 1.1cm 
$Re(E_{scapprox}^{(1)})$  \hskip 0.5cm 
$Im(E_{scapprox}^{(1)})$ \\
\caption{True versus approximated real and imaginary
parts of the scattered electric fields for objects
with ${k}_i=15.12$ in a medium with ${k}_e=12.6$.
(a)-(d) Sphere of diameter $0.5$ ($500$ nm).
(e)-(h) Spherocylinder of diameter $0.5$ 
and length $1$ along the $x$ axis.
(i)-(l) Pear-like object of maximum diameter
$0.5$  along the $x$ axis.
(m)-(p) Two spheres of diameter $0.5$ 
along the $z$ axis.
(q)-(t) Two spheres of diameter $0.5$ 
along the $y$ axis.
}
\label{fig13}
\end{figure}

\subsubsection{Combined gradient and filter corrections}
\label{sec:combined} 

We will resort to an optimization strategy combining gradient 
variations  and Gaussian filtering,
{ see Figure \ref{fig12}:} 
\begin{itemize}
\item $E_{scsynth}^{(0)}$ is the initial approximation 
obtained from $E_{scapprox}^{(0)}$.
\item $E_{scgradient}^{(1)}$ is the outcome of applying
the gradient procedure (\ref{decrease}) starting
from $E_{scsynth}^{(0)}$.
\item At each stage $n$
\begin{itemize} 
\item $E_{scfilter}^{(n)}$ is the outcome of convolving 
$E_{scgradient}^{(n)}$ with the Gaussian filter (\ref{filtertrue}).
\item $E_{scgradient}^{(n+1)}$ is the outcome of applying
the gradient procedure (\ref{decrease}) starting
from $E_{scfilter}^{(n)}$.
\end{itemize}
\end{itemize}
The final result  is the new approximation of the scattered
electric field $E_{scapprox}^{(1)}$. Then, the approximation of the
total measured field $E_{meas}$ would be:
\begin{eqnarray}
E_{approx}^{(1)}=E_{scapprox}^{(1)}+E_{inc},  \label{Emeasapprox1}
\end{eqnarray}
which can be used to obtain better approximations of the
objects by topological methods replacing $E_{approx}^{(0)}$
in functional (\ref{costEapprox}).

\subsubsection{Numerical results}
\label{sec:numerics}

Figure \ref{fig13} compares the scattered electric fields of 
different objects recorded at the detectors with the approximation
$E_{scapprox}^{(1)}$ produced by the above algorithm, 
summarized in Figure \ref{fig12}.
The true electric fields are computed synthetically, solving the forward 
problem (\ref{forward}) for the chosen object by BEM-FEM.
For the isolated sphere, we have combined $16$ steps, stopping
in each of them the gradient algorithm when the cost functional 
became smaller than $10^{-9}$. 
We define the  $\ell^2$ error when comparing two functions $f$ and $g$
as $\left(\sum_{j=1}^N |f({\mathbf x}_j) - g({\mathbf x}_j) |^2 \right)^{1/2}.$
The $\ell^2$ error between the true and approximated electric field
for a single ball falls below $10^{-1}$.
For the remaining shapes, this is the case for the real parts. The 
imaginary parts show a reasonable qualitative agreement. However,
the quantitative relative error in the central part is of order one. Alternating
one or two gradient iterations with one Gaussian convolution during 
$16$ steps provides better and faster results than waiting until the 
cost functional is very small at each step unless we deal with single 
spheres. In our tests, the worst approximation of the imaginary part 
corresponds to a geometry of two spheres aligned along the $z$ axis, see
Figure \ref{fig13}(p). Notice that only one sphere has 
been used to produce this approximation of the electric field, fitted to 
the most prominent bright area in Figure \ref{fig4}(d). It improves once
we detect the second component of the object by the procedure
in Fig. \ref{fig12} and use it to generate a new prediction 
$E_{scapprox}^{(2)}.$

The fields depicted in Figure \ref{fig13} have been computed using
the known value of ${k}_i$ to evaluate $E_{scsynth}^{(0)}$ and
then $E_{scapprox}^{(1)}.$
In biological and soft matter applications ${k}_i$ is usually 
slightly larger than ${k}_e$. We have checked that 
$E_{scapprox}^{(1)}$ does not  change significantly varying
${k}_i$ in an interval $[{k}_e+0.5, {k}_e+6]$. This
fact suggests that we would be able to approximate the
electric field at the detectors even if ${k}_i$ was unknown.

We have exploited the predicted electric fields $E_{approx}^{(1)}$ 
to produce first guesses of the scatterers following the procedure
sketched  in Figure \ref{fig12} with  cost functionals
of the form (\ref{costEapprox}). The results are similar to those
in Figures \ref{fig2}-\ref{fig4}, \ref{fig7} and \ref{fig9}. 
When we seek to sharpen the approximation of 
general shapes in the
$z$ direction by iteration following the scheme in Fig. \ref{fig8}(a), 
the subsequent domains still keep some elongation in the
$z$ direction. The performance of the method in this respect seems
to be limited by the quality of the approximation of the electric field.
Since the use of the true synthetic electric fields 
as data may remove the elongation \cite{siims}, improved predictions of the
objects would be expected by succesfully improving the numerical
approximation of the electric field.
Considering the shapes  studied in Section \ref{sec:shape}, the
predictions of unknown $k_i$ are comparable, except that now $k_i$
tends to be underestimated, not overestimated. The corrections
of the shapes follow similar trends. 



\section{Conclusions}
\label{sec:conclusions}

We have adapted topological field based imaging to work
with the experimentally measurable holographic intensity.
Assuming pure polarization of the incident light and
neglecting the non polarized components of the electric field,
we can reconstruct objects recorded in an in-line hologram
with only the wavelength of the single incident wave and the
ambient permittivity as prior information.
The method we present here works in the scalar field
approximation. This should be valid for many experiments,
but future work can extend the method to the full vector
Maxwell equations.

By tracking peaks of the topological
fields, we can detect multiple objects, convex and non
convex shapes, ranging from sizes comparable to the
employed wavelength to sizes of a few nanometers, with
nanometer precision.
Our method does not require any specific object parametrization. 
Initial reconstructions show a different resolution in the incidence 
direction of the light and on planes orthogonal to it.
Whereas shapes, sizes and positions are reasonably approximated
on $xy$ planes, objects are displaced and elongated in the
direction $z$. This later feature undergoes a transition as
the size of the object grows above the wavelength, which
facilitates its location in the $z$ direction between peaks of the
topological fields. 
Here, we consider sizes below this transition. We show that
the information on the number of objects and 
their positions provided by the initial reconstructions 
may be corrected by iteration.
To this purpose, we propose strategies based on two different functionals.
The first one quantifies the deviation  between the hologram created 
by the true object and the hologram associated to the proposed 
shape, generated numerically. The second one replaces the hologram
by an approximation of the full electric wave field obtained numerically 
from it. In both cases, their topological derivatives and topological
energies produce similar initial reconstructions, which illustrates the
robustness of these methods to noise.  Then, topological derivative
based iterations allow us to remove the offset and detect scatterers
that might have be missed in the initial reconstruction, as a result of 
the presence of more prominent ones (due to their location or to
higher permittivity contrast). Moreover, when the optical properties 
of objects are unknown, they can be inferred in an additional optimization step.
The results obtained with the first  functional are more straightforward.
However, the second functional might have the potential to allow 
for better results, since
reconstructions using synthetic electric fields as data have been
shown to have the ability to the remove the elongation \cite{siims}
by iteration. Currently,
this possibility is limited by the quality of our numerical predictions
of the electric field, obtained here by a combination of topological 
methods, gradient techniques and gaussian filters. Notice that 
the light field scattered by objects $E_{sc}$ is not a measurable magnitude
in practice. Only interference patterns such as holograms $|E_{sc}+E_{inc}|^2$ are
measurable. We succeed in obtaining reasonable numerical 
predictions of the electric field at the detectors though. 
This might pave the way to adapt to this framework techniques 
developed for larger wavelengths based on the knowledge of the 
full electric field at the  detectors 
\cite{coltonsampling,capatano,chaumet,eyraud,music,li,yu,zaeytijd}.

We have focused here on methods that need no a priori
information, but may use it if available, as shown in Figure \ref{fig9}.
This technique could serve to determine priors for bayesian methods \cite{bayesian,eyraud}.
Complementary shape or parameter optimization techniques might be incorporated to refine the approximation as well \cite{caubetfluid,oliver1,santosa}, using this holography functional and adjoint. In case the exact number  of objects is determined, shape optimization based on deforming contours may  improve the descriptions of shapes \cite{caubetfluid}. 
When a good approximation to permittivity variations is available, Newton type algorithms \cite{capatano,chaumet,li,yu,zaeytijd} may be used to refine it. 
Restrictions in the incidence directions and
the extent of the recording region increase the possibility of converging to spurious minima in the optimization reformulations of the inverse problems unless good enough initial guesses of objects and permittivities are encountered.  This type of inverse problems being ill posed, the quality of the final reconstruction is limited by the choice of incident directions (only one in these microscopy set-ups), the spread of the region where data  are recorded (a flat screen behind the objects) and the distance to it.

\appendix

\section{Derivatives of the holography cost functional}
\label{sec:derivatives}

As a previous step to differentiate functional (\ref{costH}),
we rewrite the constraint (\ref{forward}) in variational form.
First, we replace the transmission problem (\ref{forward})  by an
equivalent problem set in a bounded region
$B_R$ containing the objects $\Omega$ and the detectors
$\mathbf x_j$, $j=1,\ldots,N$. To do so, we take $B_R$ to be
a ball of large enough radius $R$ with boundary  $\Gamma_{R}
=\partial B_R$. The Dirichlet--to--Neumann (also called
Steklov--Poincar{\'e}) operator associates to any Dirichlet data on
$\Gamma_{R}$ the normal derivative of the solution of the exterior
Dirichlet problem:
\[
\begin{array}{rcl}
L_{k_e}:H^{1/2}(\Gamma_R)&\longrightarrow &H^{-1/2}(\Gamma_R)\\
f&\longmapsto&\partial_{\bf n} w
\end{array}
\]
where $w\in H^{1}_{loc}(\mathbb{R}^{3} \setminus \overline{B}_R), $
 is the unique solution of
\[
\left\{\begin{array}{ll}  \Delta w+  k_e^2 w=0,
&\qquad \mbox{in $\mathbb{R}^{3} \setminus \overline{B}_R$},\\
w=f,&\qquad\mbox{on $\Gamma_R$},\\
\displaystyle  \lim_{r\to\infty}r(\partial_r w-\imath k_e w)=0.
\end{array}\right.
\]
$H^{1}_{loc}(\mathbb{R}^{3} \setminus \overline{B}_R)$ denotes the
Sobolev space of functions that are locally in $H^{1}$,
whereas $H^{1/2}(\Gamma_R)$ and $H^{-1/2}(\Gamma_R)$
are the  trace spaces \cite{gilbart,nedelec}. 
The vector $\mathbf n$ is the unit outer normal. 
This operator allows us  to replace the radiation condition at infinity by 
a non--reflecting boundary condition on $\Gamma_R$ \cite{kellerGivoli}:
$ \partial_{\bf n}(E-E_{inc})=L_{k_e}(E-E_{inc}).$
The variational formulation of problem (\ref{forward}) becomes: Find
$E \in H^1(B_R)$ such that
\begin{equation}\label{Variational}
 b(\Omega; E, v)= \ell(v),\qquad\forall v\in
H^1(B_R),
\end{equation}
\[
\begin{array}{rcl}
{\rm with \hskip 1.4cm}
\displaystyle  b(\Omega;E,v) &:=& \displaystyle  \int_{B_R \setminus\overline{\Omega}} (  \nabla E\nabla {\overline v}- k_e^2 E{\overline v}) d{\bf z}
+\int_{\Omega} (\beta \nabla E\nabla {\overline v}-
\beta k_i^2 E{\overline v}) d{\bf z}
\\ & &\displaystyle  -\int_{\Gamma_R}  L_{k_e}(E)\,\overline{v} \, dS,\qquad\forall
E,\,v\in H^1(B_R),\\ 
\ell(v) &:=& \displaystyle  \int_{\Gamma_R} (\partial_{\bf
n}E_{inc}-L_{k_e}(E_{inc}))\,\overline{v}\, dS, \qquad\forall v\in H^1(B_R).
\end{array}
\]
Assuming that $k_e$ is constant near $\Gamma_R$ and outside $B_R$, 
existence of a unique solution is guaranteed for bounded $k_e(\mathbf x)$ and $k_i(\mathbf x)$, and Lipschitz domains $\Omega$ \cite{sphericalkirsch,nedelec,selgas}.
Elliptic regularity for $-\Delta$ implies then that the solution  belongs to 
$H^2(\Omega')$ for any smooth $\Omega'  \subset \Omega$ or 
$\Omega' \subset B_R\setminus \overline{\Omega}$ \cite{gilbart,grisvard}. 
Sobolev's embeddings ensure continuity in $\Omega'$, and continuity of 
derivatives when $k_e$,$k_i$ are differentiable ($W^{1,\infty}$ suffices) 
and bounded. When $\Omega$ is a $C^2$ domain \cite{gilbart}, we may 
take $\Omega' =\Omega, B_R\setminus \overline{\Omega}$,
and, in fact, the solution is continuous in $\mathbb R^3$.
If $k_e$ and $k_i$ are constant, the solution admits  integral expressions
both in $\Omega$ and $\mathbb R^3\setminus \overline{\Omega}$
in terms of the Green functions of Helmholtz equations 
\cite{coltonkress,sphericalkirsch,nedelec}. 
In absence of $\Omega$, $E$ is as smooth as $E_{inc}$ permits.

\subsection{Topological derivative}
\label{sec:td}

Given a region ${\cal R} \subset \mathbb R^3$, the topological 
derivative is a scalar field defined for each $\mathbf x \in {\cal R}$, 
which measures the variation of the functional when removing
from ${\cal R}$ small balls $B_\varepsilon=B({\mathbf x}, \varepsilon)$
centered at $\mathbf x$ 
\cite{Sokowloski}:
\begin{eqnarray} \label{topderdef} \hskip 4mm
D_T({\mathbf{x}},{\cal R}) := \lim_{\varepsilon\to 0}
{J({\cal R}_\varepsilon) - J({\cal R}) \over {\cal V}(\varepsilon)},
\end{eqnarray}
where ${\cal V}(\varepsilon)= {4\over 3} \pi \varepsilon^3$ is the volume 
of a three dimensional ball  of radius $\varepsilon$.

{\bf Theorem 1.} {\it When   $k_e$, $k_i$, and $\beta$ are constant
the topological derivative of functional (\ref{costH}) with 
$\Omega=\emptyset$ is given by
\begin{eqnarray} \hskip 3mm
D_T(\mathbf{x},\mathbb R^3) =
{\rm Re} \left[ 3{1 - \beta \over 2 + \beta}
\nabla E(\mathbf{x})\! \cdot \! \nabla
 \overline{P}(\mathbf{x})
\!+\! (\beta k_i^2  - k_e^2)  E(\mathbf{x})   \overline{P}(\mathbf{x})
\right], \; \mathbf x \in \mathbb R^3,
\label{dtemptyap}
\end{eqnarray}
where $E$ is the solution of the forward problem: 
\begin{eqnarray} \label{forwardemptyap}
\left\{
\begin{array}{ll}
\Delta E + {k}_e^2 E=  0 &\quad\mbox{in ${\mathbb R}^3$}, \\
\displaystyle\lim_{r\to \infty} r\big(\partial_r (E-E_{inc}) 
-\imath {k}_e (E-E_{inc}) \big)=0,
\end{array}
\right.
\end{eqnarray}
and $\overline{P}$ is the solution of the conjugate adjoint problem:
\begin{eqnarray} \label{adjointemptyap}
\left\{
\begin{array}{ll}
\Delta \overline{P} + {k}_e^2 \overline{P}= 2 \sum_{j=1}^N
 ({\mathcal I}-|E|^2) \overline{E} \, \delta_{{\mathbf x}_j}
&\quad\mbox{in ${\mathbb R}^3$},
\\
\displaystyle\lim_{r\to \infty} r\big(\partial_r \overline{P} 
-\imath {k}_e\overline{P} \big)=0,
\end{array}
\right.
\end{eqnarray}
$\delta_{\mathbf x_j}$ being Dirac masses supported at the
receptors. 
For a plane wave, $E=E_{inc}$ and $\overline{P}$ is given by:
\begin{eqnarray}\label{adjointemptyHap}
\overline{P}({\mathbf x})
=- \sum_{j=1}^N {e^{\imath {k}_e |{\mathbf x}-{\mathbf x_j}|}
\over 4 \pi |{\mathbf x}-{\mathbf x_j}|}
[2 ({\mathcal I}({\mathbf x_j})-|E({\mathbf x_j})|^2 ) 
\overline{E}(\mathbf x_j) ].
\end{eqnarray}
}

{\bf Proof.}
To prove this expression, we set  ${\cal R}= \mathbb R^3$ and
exploit the relation with shape derivatives established in  \cite{feijoo}
\footnote{Reference \cite{feijoo} takes
${\cal V}(\varepsilon)$ to be negative. This does not affect
the validity of formula (\ref{topder}) but changes the sign of
$D_T({\mathbf{x}},{\cal R})$ at each point.}: 
\begin{eqnarray}\label{topder} \hskip 4mm
D_T({\mathbf{x}},{\cal R}) 
= \lim_{\varepsilon\to 0} \frac { <D{J}({\cal R}_\varepsilon),
\mathbf V>}{{\cal V}'(\varepsilon)}  
= \lim_{\varepsilon\to 0} \frac { 1}
{{\cal V}'(\varepsilon)} \,\frac{d}{d\tau} {J}(\varphi_\tau(
{\cal R}_\varepsilon))\Big|_{\tau=0},
\end{eqnarray}
where ${\cal V}'(\varepsilon)= 4 \pi \varepsilon^2$
and ${\cal R}_\varepsilon= {\cal R} \setminus \overline{B_\varepsilon} $.
The vector field $\mathbf  V$ is an extension to  $\mathbb{R}^3$ of
$\mathbf  V= -\mathbf  n({\mathbf z}),\, {\mathbf z}\in
\Gamma_{\varepsilon}=\partial B_\varepsilon({\mathbf{x}}),$ where 
the normal $\mathbf n(\mathbf  z): ={ \mathbf x- \mathbf  z \over 
|\mathbf  x - \mathbf  z| }$
points inside the ball, and vanishes away from a narrow 
neighborhood of $\partial B_\varepsilon$.
The shape derivative along the vector field ${\bf V}$ is defined 
as
\begin{eqnarray}\label{shapeder}
<D{J}({\cal R}_\varepsilon), \mathbf V> :=
\frac{d}{d\tau}\,{J}(\varphi_\tau({\cal R}_{\varepsilon}))
\Big|_{\tau=0},
\end{eqnarray}
for the family of deformations  
$\varphi_\tau({\mathbf{z}}):={\mathbf{z}}
+\tau{\bf V}({\mathbf{z}}),\, {\mathbf{z}}\in
\mathbb{R}^3,$ $\tau>0$. For any region ${\cal R}$, 
the deformed domain
$\varphi_\tau({\cal R})$ is the image of ${\cal R}$ by the
deformation: $\varphi_\tau({\cal R})=\{ \mathbf  z + \tau
\mathbf  V(\mathbf  z) \;|\; \mathbf z \in {\cal R} \}.$
Evaluating ${ J}$ on the deformed regions, we define a scalar function 
${J}(\varphi_\tau({\cal R}))$ of the deformation parameter $\tau$ which 
can be differentiated with respect to it.

{\it Step 1: Computation of the shape derivative.}
We evaluate (\ref{costH}) in the deformed domains 
${\cal R}_{\varepsilon,\tau}= \varphi_\tau({\cal R}_{\varepsilon})$, 
and denote by $E_{\varepsilon,\tau}\in H^1(B_R)$ the solution of 
(\ref{Variational}) with object $  B_{\varepsilon,\tau}$. 
Notice that $\mathbf V$ vanishes on $\Gamma_R$,  
and at $\mathbf x_j$, $j=1,\ldots,N$.
Differentiating with respect to $\tau$ we obtain
\begin{equation}\label{derivadatau}
\frac{d}{d\tau}\,J({\cal R}_{\varepsilon,\tau})
\bigg|_{\tau=0}=2\, {\rm Re} \bigg[ \sum_{j=1}^N 
(|E_{\varepsilon}(\mathbf x_j)|^2-{\cal I}(\mathbf x_j)) 
\overline{E_{\varepsilon}(\mathbf x_j)}
\dot E_{\varepsilon}(\mathbf x_j)\bigg],
\end{equation}
where $E_{\varepsilon}=E_{\varepsilon,0}$ and
$\dot{E}_{\varepsilon}=\frac{d}{d\tau}\, E_{\varepsilon,\tau}
\big|_{\tau=0}$.
We follow \cite{feijooOberai,feijoo} and evaluate this derivative avoiding 
the computation of $\dot{E}_{\varepsilon}$ by introducing the Lagrangian 
functional
\[
{\cal L}({\cal R}_{\varepsilon,\tau};v,p) := J({\cal R}_{\varepsilon,\tau})
- \mbox{Re}[b(B_{\varepsilon,\tau};v,p)-\ell(p)],\qquad\forall
v,\, p \in H^{1}(B_R).
\]
Using ${\cal L}({\cal R}_{\varepsilon,\tau};E_{\varepsilon,\tau},p)
= J({\cal R}_{\varepsilon,\tau})$, (\ref{shapeder})  
and (\ref{derivadatau}) we obtain for any 
$p \in H^{1}(B_R)$
\begin{eqnarray}
<DJ({\cal R}_{\varepsilon}), \mathbf V>&=& \frac{d}{d\tau}
{\cal L}({\cal R}_{\varepsilon,\tau};E_{\varepsilon,\tau},p) \bigg|_{\tau=0} =  
- \mbox{Re}\left[\frac{d}{d\tau}b(B_{\varepsilon,\tau};E_{\varepsilon},p)
\bigg|_{\tau=0} \right]\nonumber \\
[-1ex]
&- &\mbox{Re}\bigg[b(B_{\varepsilon};\dot{E}_{\varepsilon},p)\bigg] +
2\, {\rm Re} \bigg[ \sum_{j=1}^N 
(|E_{\varepsilon}(\mathbf x_j)|^2-{\cal I}(\mathbf x_j)) 
\overline{E_{\varepsilon}(\mathbf x_j)}\dot E_{\varepsilon}(\mathbf x_j)\bigg].
\label{shapelagrangian}
\end{eqnarray}
Choosing $ p=P_{\varepsilon}$, where $\overline P_\varepsilon$ is the 
solution of 
\begin{eqnarray} \label{adjointepsilon}
\left\{
\begin{array}{ll}
\Delta \overline P_{\varepsilon} + {k}_e^2 \overline P_{\varepsilon}= 
2 \sum_{j=1}^N  ({\mathcal I}-|E_\varepsilon|^2) \overline E_\varepsilon \, \delta_{{\mathbf x}_j}
&\quad\mbox{in ${\mathbb R}^3\setminus\overline{B_\varepsilon}$},
\\
\Delta \overline P_{\varepsilon} + {k}_i^2 \overline P_{\varepsilon} =0,
&\quad\mbox{in $ B_\varepsilon $},\\
\overline P_{\varepsilon}^-= \overline P_{\varepsilon}^+,&\quad 
\mbox{on $\partial B_\varepsilon$},\\
\beta \partial_{\mathbf{n}}  \overline P_{\varepsilon}^-= 
\partial_{\mathbf{n}} \overline P_{\varepsilon}^+, &\quad 
\mbox{on $\partial B_\varepsilon$},\\
\displaystyle\lim_{r\to \infty} r\big(\partial_r \overline P_{\varepsilon} 
-\imath {k}_e \overline P_{\varepsilon} \big)=0,
\end{array}
\right.
\end{eqnarray}
the two terms  involving $\dot E_{\varepsilon}$  in (\ref{shapelagrangian})  cancel. The shape derivative is then given by
$-{\rm Re}[\frac{d}{d\tau}
b(B_{\varepsilon,\tau};E_{\varepsilon},P_\varepsilon)
\big|_{\tau=0}]$.
That term has been computed in detail in \cite{cime2008} (pages
117-118) for a different adjoint field and in two dimensions. However, the
final result does not depend  neither on the dimension nor on the source
for the adjoint in the detector region. Reproducing those calculations
we find,
\begin{eqnarray}
<DJ({\cal R}_{\varepsilon}), \mathbf  V>&=&  \mbox{\rm Re} \bigg[
\int_{\partial B_\varepsilon} (-\beta \nabla
E_{\varepsilon}^- \cdot  \nabla \overline P_{\varepsilon}^- +  
k_i^2 E_{\varepsilon}^- \overline P_{\varepsilon}^- +
2 \beta  \partial_{\bf n}E_\varepsilon^-\partial_{\bf n} \overline P_{\varepsilon}^-)  
\, dS
\nonumber \\  [-1ex]
&&\qquad \, -\int_{\partial B_\varepsilon} 
(- \nabla E_{\varepsilon}^+ \cdot  \nabla \overline P_{\varepsilon}^+ +
k_e^2 E_{\varepsilon}^+ \overline P_{\varepsilon}^+ +2
\partial_{\bf n} E_{\varepsilon}^+\partial_{\bf n} \overline P_{\varepsilon}^+)   
\, dS
\bigg]. \label{sphapeintermediate}
\end{eqnarray}
We use the transmission boundary conditions at the interface to rewrite this expression in terms of the inner values. For any function $u$ defined
on  the sphere we have 
$\nabla u = (\nabla u \cdot \mathbf n) \mathbf n+ \nabla_S u$,
where $\nabla_S u$ is the surface gradient. In spherical
coordinates $\nabla_S u = {1\over r} {\partial u \over \partial \theta} 
\hat{\mathbf e}_\theta
+ {1\over r \sin(\theta)} {\partial u \over \partial \phi} \hat{\mathbf e}_\phi$.
Continuity of a function across the surface, $u^+=u^-$, implies continuity of 
the surface gradients: $\nabla_S u^- = \nabla_S u^+$.
Therefore, (\ref{sphapeintermediate}) becomes
\begin{eqnarray} 
<DJ({\cal R}_{\varepsilon}), \mathbf  V>&=&  \mbox{\rm Re} \bigg[
\int_{\partial B_\varepsilon} \Big( (1-\beta) \nabla_S
E_{\varepsilon}^-  \cdot  \nabla_S \overline P_{\varepsilon}^-  \nonumber \\
[-1ex]
&& \qquad +  (k_i^2-k_e^2) E_{\varepsilon}^- \overline P_{\varepsilon}^- +
 \beta (1-\beta)  \partial_{\bf n}E_\varepsilon^-\partial_{\bf n} 
 \overline P_{\varepsilon}^-  \Big) \, dS
\bigg]. \label{shapefinal}
\end{eqnarray}

%
%

{\it Step 2: Passage to the limit.} To calculate the limit (\ref{topder})
we need to investigate the asymptotic behavior of 
$E_{\varepsilon}, \overline P_{\varepsilon}$ and their gradients.
The fields $E_\varepsilon$ and  $\overline P_{\varepsilon}$
are given by the series expansions detailed in Appendix 
\ref{sec:explicitforwardadjoint}. When  $\varepsilon \rightarrow 0$, 
we have
\begin{eqnarray}
E_{\varepsilon}^-(\mathbf x + \varepsilon \boldsymbol \xi) 
\rightarrow  E(\mathbf x),  \qquad
\overline P_{\varepsilon}^-(\mathbf x + \varepsilon \boldsymbol \xi) 
\rightarrow  \overline P(\mathbf x),  
\label{limitE} \\
\nabla E_{\varepsilon}^-(\mathbf x + \varepsilon \boldsymbol \xi) 
\rightarrow  {3\over 2+ \beta} \nabla E (\mathbf x),   \qquad
\nabla \overline P_{\varepsilon}^-(\mathbf x + \varepsilon \boldsymbol \xi)   \rightarrow  {3\over 2+\beta } \nabla \overline P (\mathbf x), 
\label{limitdE}
\end{eqnarray}
uniformly for $|\boldsymbol \xi| = 1$.
Here, $E=E_{inc}$ and $\overline P$  given by (\ref{adjointemptyHap})
are the solutions of the forward and adjoint problems 
(\ref{forwardemptyap})-(\ref{adjointemptyap}).
To justify this, we observe that the coefficients
of the series expansion for $E_\varepsilon$ inside the sphere 
$B_\varepsilon$ take the form:
\begin{eqnarray*}
b_{n,m} =  \delta_n(\varepsilon) \tilde b_{n,m}, \quad
\delta_n(\varepsilon) = {k_e  j_n'(k_e \varepsilon) h_n^{(1)}(k_e \varepsilon) 
\!-\! k_e  j_n(k_e \varepsilon) (h_n^{(1)})'(k_e \varepsilon) \over
\beta k_i j_n'(k_i \varepsilon) h_n^{(1)}(k_e \varepsilon) 
\!-\! k_e j_n(k_i \varepsilon) h_n^{(1)'}(k_e \varepsilon)}, 
\end{eqnarray*}
where $\tilde b_{n,m}$ are the coefficients for $E.$
The spherical Bessel functions $j_n$ and $h_n^{(1)}=j_n+ \imath y_n$ 
have the following  asymptotic behavior:
$j_n(k \varepsilon) \sim  (k \varepsilon)^n {2^n n! \over (2n+1)!}$,
$y_n(k \varepsilon) \sim -(k \varepsilon)^{-(n+1)} {(2n-1)! \over
2^{n-1} (n-1)!}$ as $\varepsilon \rightarrow 0$.
Moreover, $z_n(k \varepsilon)' = -z_{n+1}(k \varepsilon) 
+ n (k \varepsilon)^{-1} z_n(k \varepsilon)$ for $z_n=j_n,h_n^{(1)}$
\cite{sphericalkirsch}. Thanks to this, we find that the amplification
factors for the coefficients of the spherical harmonics when
comparing the series for $E_\varepsilon$ and $E$ are:
\begin{eqnarray*} \label{amplificationE}
a_n(\varepsilon) = {j_n(k_i \varepsilon) \over j_n(k_e \varepsilon)} 
\delta_n(\varepsilon) \rightarrow  {2n + 1
\over \beta n + (n+1)}, \qquad {\rm as } \, \varepsilon \rightarrow 0.
\end{eqnarray*}
For the expansions of 
$E_\varepsilon(\mathbf x + \varepsilon \boldsymbol \xi)$ 
and $E(\mathbf x + \varepsilon \boldsymbol \xi)$ the relevant 
terms as $\varepsilon \rightarrow 0$ correspond to $n=0$, with 
$a_0(\varepsilon) \rightarrow 1.$ This implies (\ref{limitE}).
Since $P_0^0$ is constant, the series for the derivatives with
respect to $\theta$ and $\phi$ start at $n=1$, which provides the
leading terms with an amplification factor  
$a_1(\varepsilon) \rightarrow {3 \over \beta + 2}.$ 
The leading term in the series expansion for the derivatives in the 
radial direction is found for $n=1$, with the same amplification factor.
This implies (\ref{limitdE}). Similar arguments work for 
$\overline P_\varepsilon$ and $\overline P.$
\\
To calculate  the limit (\ref{topder}), we write 
$\nabla_S E_\varepsilon^- \cdot \nabla_S \overline P_\varepsilon^-
= \nabla E_\varepsilon^- \cdot \nabla \overline P_\varepsilon^- 
- \partial_{\mathbf n} E_\varepsilon^- \partial_{\mathbf n} P_\varepsilon^-$
in (\ref{shapefinal}), use (\ref{limitE})-(\ref{limitdE}) and 
$\int_{\partial B_\varepsilon} (\mathbf w \cdot \mathbf n)  
(\mathbf v \cdot \mathbf n) d S = {4\over 3} \pi \varepsilon^2 \mathbf w 
\cdot \mathbf v.$
$\square$

Expression (\ref{dtemptyap}) was obtained by formal asymptotic expansions
of the integral equations for a different cost functional and a different
adjoint in \cite{guzinaacoustic}. When $\Omega \neq \emptyset$, we set
${\cal R} = \mathbb R^3 \setminus \overline \Omega$. Then, (\ref{topderdef})
defines the topological derivative for points $\mathbf x \notin \Omega$.
The definition was extended to $\mathbf x \in \Omega$ in \cite{ip2008}
\footnote{For simplicity we chose the sign so that the resulting expression
is globally continuous in $\mathbb R^3$ when $\beta=1$.}:
\begin{eqnarray} \label{topderdefextended} \hskip 4mm
D_T({\mathbf{x}},\mathbb R^3 \setminus \overline \Omega) 
:= \lim_{\varepsilon\to 0}
{J( \mathbb R^3 \setminus \overline \Omega) -
J( (\mathbb R^3 \setminus \overline \Omega) \cup B_\varepsilon) 
\over {\cal V}(\varepsilon) }.
\end{eqnarray} 
This limit measures the variation of the functional when adding  points
to ${\cal R} = \mathbb R^3 \setminus \overline \Omega$, that is, when
removing them from $\Omega$.

{\bf Theorem 2.} {\it When  $k_e$ and $k_i$ are constant and $\beta=1$,
the topological derivative of functional (\ref{costH}) in presence of
an object $\Omega$ is given by
\begin{eqnarray} \hskip 3mm
D_T(\mathbf{x},\mathbb R^3 \setminus \overline \Omega) =
{\rm Re} \left[( k_i^2  - k_e^2)  E(\mathbf{x})   \overline{P}(\mathbf{x})
\right], \qquad \mathbf x \in \mathbb R^3,
\label{dtfullap}
\end{eqnarray}
where $E$ is the solution of the forward problem (\ref{forward})
and $\overline{P}$ is the solution of the conjugate adjoint problem
\begin{eqnarray} \label{adjointobjectap}
\left\{
\begin{array}{ll}
\Delta \overline{P} + {k}_e^2 \overline{P}= 2 \sum_{j=1}^N
({\mathcal I}-|E|^2) \overline{E} \delta_{{\mathbf x}_j}
&\quad\mbox{in ${\mathbb R}^3\setminus\overline{\Omega}$}, \\
\Delta \overline{P} + {k}_i^2 \overline{P} =0,&\quad\mbox{in $\Omega$},\\
\overline{P}^-= \overline{P}^+,&\quad \mbox{on $\partial \Omega$},\\
\beta \partial_{\mathbf{n}}  \overline{P}^-= 
\partial_{\mathbf{n}} \overline{P}^+, &\quad 
\mbox{on $\partial  \Omega$},\\
\displaystyle\lim_{r\to \infty} r\big(\partial_r \overline{P}
-\imath {k}_e \overline{P} \big)=0.
\end{array}
\right.
\end{eqnarray} 
Formula (\ref{dtfullap}) still holds for differentiable  $k_i,k_e$
replacing $k_i,k_e$ by $k_i(\mathbf x),k_e(\mathbf x)$.}

{\bf Proof.} Let us consider first points
$\mathbf x \in {\cal R}= \mathbb R^3 \setminus \Omega$. Since
$\mathbf V$ vanishes on $\partial \Omega$, the only difference
with the proof of Theorem 1 arises when justifying the limit (\ref{limitE}). 
Functions $E_\varepsilon$ and $\overline{P}_{\varepsilon}$ are
solutions of problems of the form (\ref{forward}) and (\ref{adjointobjectap})
with objects $\Omega \cup B_\varepsilon$ instead of $\Omega$. 
To evaluate the order of the correction due to $B_\varepsilon$ in 
$E$ and $\overline{P}$ we compute the operator Dirichlet-to-Neuman 
$N_{k_i}$ which associates to any Dirichlet data on $\partial B_\varepsilon$ 
the normal derivative on $\partial B_\varepsilon$ of the solution of 
$\Delta u + k_i^2 u=0$ in $B_\varepsilon$ with the given trace on
$\partial B_\varepsilon$.  The transmission conditions
for $E_\varepsilon$ and $\overline{P}_\varepsilon$ on $\partial B_\varepsilon$ may be rewritten in the form $
N_{k_i}(u) = \partial_{\mathbf n} u$
for the exterior problem. 
Using the expansions in  Appendix \ref{sec:explicitforwardadjoint} to evaluate this condition,  we see that the corrections due to the ball $B_\varepsilon$ do not appear to zero order in $\varepsilon$ and (\ref{limitE}) holds. 

When $\mathbf x \in \Omega$, the roles of $k_i$ and $k_e$ are exchanged in Step 1 of the proof of Theorem 1. Therefore the shape derivative has the opposite sign.  The sign choice in definition (\ref{topderdefextended})
cancels this effect. The procedure to pass to the limit is the same.

If we allow $k_e$ and $k_i$ to vary in space, Step 1 remains unchanged.
For Step 2, we  expand $k(\mathbf x + \varepsilon \boldsymbol \xi)
=k(\mathbf x) + \varepsilon r(\mathbf x) $, $r$ being a bounded function
and take limits as before. $\square$


We consider now variations of the functional with
respect to the coefficients.

{\bf Theorem 3.} {\it Given differentiable functions $k_i$ and $\psi$, the 
derivative of 
\begin{eqnarray}
J(\delta) = J( k_{i} +\delta \psi) = {1\over 2}  \sum_{j=1}^N
| |E_\delta(\mathbf{x}_j)|^2- {\mathcal I}(\mathbf{x}_j)|^2, \label{costk}
\end{eqnarray}
where $E_\delta$ is the solution of the forward problem (\ref{forward})
with coefficient $k_{i}+\delta \psi$, $\delta >0$, is
\begin{eqnarray}
{d J \over d \delta} \bigg|_{\delta=0}=   2 \, \mbox{\rm
Re}\left[\int_{\Omega}  \psi \beta {k}_{i}\, 
E  \overline{P}    \,d{\mathbf{z}}\right],
\label{derk}
\end{eqnarray}
$E$ and $\overline P$ being solutions of the forward and 
adjoint problems (\ref{forward}) and (\ref{adjointobjectap}) with
coefficient $k_i$.}

{\bf Proof.}
The derivative with respect to $\delta$ is 
\begin{equation}\label{derivadadelta}
\frac{d}{d\delta}\,J\bigg|_{\delta=0} = 2\, {\rm Re} \bigg[ \sum_{j=1}^N 
(|E(\mathbf x_j)|^2-{\cal I}(\mathbf x_j)) 
\overline{E(\mathbf x_j)} \dot E(\mathbf x_j)\bigg],
\end{equation}
where $\dot{E}=\frac{d}{d\delta}\, E_{\delta}
\big|_{\delta=0}$. We avoid computing
$\dot{E}$ by introducing a modified functional ${\cal L}$ and an
adjoint problem. Recall that $E_{\delta} \in H^1(B_R)$ is a solution of
\begin{eqnarray*}
\displaystyle  b(\delta; E_{\delta}, v)= \ell(v),\qquad \forall v\in
H^1(B_R),
\end{eqnarray*}
where
\[
\begin{array}{rcl}  
\displaystyle  b(\delta;u,v) \!\!\!&:=&\!\!\! \displaystyle  \int_{B_R \setminus
\overline \Omega}
(\nabla u\nabla {\overline v}- k_e^2 u {\overline v}) d{\mathbf{z}}
+\int_{\Omega} \left( \beta  \nabla u \nabla {\overline v}-
\beta (k_i + \delta \psi)^2 u {\overline v} \right) d{\mathbf{z}}
\\ \hskip -2.5cm  &&\displaystyle  -\int_{\Gamma_R}  L_{k_e}(u)\,\overline{v} \,
dS,\qquad \forall u,\,v\in H^1(B_R),
\end{array}
\]
and $\ell$ is defined in (\ref{Variational}).
Then,
\[
J(\delta)={\cal L}(\delta;E_{\delta},p):=J(\delta)
-\mbox{Re}[b(\delta;E_{\delta},p)-\ell(p)],\qquad\forall p\in
H^{1}(B_R),
\]
and
\begin{eqnarray*}
\frac{d}{d\delta}\,J\big|_{\delta=0} &=& \frac{d}
{d\delta} {\cal L}(\delta; E_\delta,p) \big|_{\delta=0}=
-\mbox{Re}\left[\frac{d}{d\delta}b(\delta; E,p)\big|_{\delta=0}
 \right] - \mbox{Re}\bigg[b(0;\dot{E},p)\bigg]
 \nonumber \\  [-1ex]
&& \hspace*{3cm}
+ 2\, {\rm Re} \bigg[ \sum_{j=1}^N 
(|E(\mathbf x_j)|^2-{\cal I}(\mathbf x_j)) 
\overline{E(\mathbf x_j)} \dot E(\mathbf x_j)\bigg],
\end{eqnarray*}
thanks to (\ref{derivadadelta}).  If $p=P$ 
is a solution of (\ref{adjointobjectap}),
the terms involving $\dot{E}$ vanish.  Finally,
\[
\frac{d}{d\delta}\,J\big|_{\delta=0} 
= -\mbox{Re}\left[\frac{d}{d\delta}b(\delta; E,P)\big|_{\delta=0} \right] 
= \mbox{Re} \left[ \int_{\Omega}
 2\psi \beta k_{i}  \, E  \overline P   \,  d \mathbf z  \right]. \square
\]

\section{Explicit forward and adjoint fields for a sphere}
\label{sec:explicitforwardadjoint}

We consider here the forward  and adjoint problems when $\Omega=B_R$
is a sphere centered at $(0,0,0)$ with radius $R$ and the coefficients
$k_e$, $k_i$ are constant. Both can rewritten as
transmission problems of the form
\begin{eqnarray} \label{sphereauxiliar}
\left\{
\begin{array}{ll}
\Delta u +  {k}_e^2 u= 0, 
&\quad\mbox{in ${\mathbb R}^3\setminus\overline{B_R}$},
\\
\Delta u +  {k}_i^2 u =0, &\quad\mbox{in $B_R$},\\
u^-= u^+ + U ,&\quad \mbox{on $\partial B_R$},\\
\beta \partial_{\mathbf{n}}  u^- = \partial_{\mathbf{n}} u^+
+ \partial_{\mathbf{n}} U, &\quad \mbox{on $\partial B_R$},\\
\displaystyle\lim_{r\to \infty} r\big(\partial_r u
- \imath  {k}_e u\big)=0,
\end{array}
\right.
\end{eqnarray}
for adequate choices of $U$. These solutions admit series
expansions \cite{coltonkress}:
\begin{eqnarray}
u({\mathbf x}) &=& \sum_{n=0}^{\infty}\sum_{m=-n}^n 
a_{nm} h^{(1)}_n(k_e |{\mathbf x}|) Y^m_n(\hat{\mathbf x}),
 \quad |{\mathbf x}| \geq  R,  
\label{uscattered} \\[-1ex]
u({\mathbf x}) &=& \sum_{n=0}^{\infty}\sum_{m=-n}^n 
b_{nm} j_n(k_i |{\mathbf x}|) Y^m_n(\hat{\mathbf x}),
 \quad |{\mathbf x}| \leq  R,
\label{utransmitted}
\end{eqnarray}
where ${\mathbf x} = |{\mathbf x} | \hat{\mathbf x}$,
$j_n$ are the spherical Bessel functions of the first kind,
$h^{(1)}_n$ are the spherical Hankel functions and
$Y^m_n$ are the standard spherical harmonics,
\[
Y^m_n(\theta,\phi) = \sqrt{ {2n+1 \over 4 \pi}
{(n-|m|)!  \over (n+|m|)!} } P^{|m|}_n(\cos(\theta))e^{\imath m \phi}, 
\]
for associated Legendre polynomials $P^{|m|}_n$.
More precisely, if $U$  can be expanded as
\[
U({\mathbf x}) = \sum_{n=0}^{\infty}\sum_{m=-n}^n 
u_{nm} j_n(k_e |{\mathbf x}|) Y^m_n(\hat{\mathbf x})
\]
in a ball containing $B_R$, the coefficients in 
(\ref{uscattered})-(\ref{utransmitted}) are computed as follows.

On the boundary of the sphere $|{\mathbf x}|=R$, the
transmission conditions hold. We impose these relations on
the inner and outer series expansions and equate
the coefficients of $Y^m_n(\hat{\mathbf x})$
since the spherical harmonics form a basis in
$L^2(\partial B_1)$ \cite{coltonkress}. This yields the relations:
\begin{eqnarray}
u_{nm} j_n(k_e R) 
+ a_{nm} h^{(1)}_n(k_e R) - b_{nm} j_n( k_i R) = 0, \nonumber  \\
u_{nm} k_e  j_n'(k_e R) 
+ a_{nm} k_e h^{(1)'}_n(k_e R) - \beta b_{nm} k_i j_n'( k_i R) = 0. 
\nonumber
\end{eqnarray}
Solving the system we obtain the value of the
coefficients:
\begin{eqnarray}
a_{nm} = u_{nm} a_n(R) = u_{nm}
{k_e j_n(k_i R)  j_n'(k_e R) - \beta k_i j_n'(k_i R) j_n(k_e R)
\over
\beta k_i j_n'(k_i R) h_n^{(1)}(k_e R) - k_e j_n(k_i R) h_n^{(1)'}(k_e R)},
\label{anm} \\  
b_{n,m} =  u_{nm} b_n(R) = u_{mn} 
{k_e  j_n'(k_e R) h_n^{(1)}(k_e R) 
 - k_e  j_n(k_e R) (h_n^{(1)})'(k_e R) \over
\beta k_i j_n'(k_i R) h_n^{(1)}(k_e R) 
-  k_e j_n(k_i R) h_n^{(1)'}(k_e R)}.
\label{bnm}
\end{eqnarray}
To calculate these coefficients, notice that the spherical Bessel function 
is related to the Bessel functions of the first kind by 
$j_n(s)= \sqrt{\pi \over 2 s} J_{n+1/2}(s)$ \cite{sphericalkirsch}. 
The spherical Hankel function is related to the Hankel 
functions of the first kind by $h_n^{(1)}(s)= \sqrt{\pi \over 2 s}
H_{n+1/2}(s)$. Their derivatives are evaluated using the
formula $f_n'(s) = {n\over s} f_n(s)- f_{n+1}(s),$ which holds
for both $j_n$ and $h_n^{(1)}.$

\subsection{Forward field for a sphere}
\label{sec:forwardexplicit}

Setting the total field $E=E_{sc}+E_{inc}$ outside the object and
$E=E_{tr}$ inside, the forward problem (\ref{forward})
can be rewritten as
\begin{eqnarray} \label{forwardincident}
\left\{
\begin{array}{ll}
\Delta E_{sc} +  {k}_e^2 E_{sc}= 0, 
&\quad\mbox{in ${\mathbb R}^3\setminus\overline{B_R}$}, \\
\Delta E_{tr} +  {k}_i^2 E_{tr}=0, &\quad\mbox{in $B_R$},\\
E_{tr}= E_{sc}+ E_{inc} ,&\quad \mbox{on $\partial B_R$},\\
\beta \partial_{\mathbf{n}}  E_{tr}= \partial_{\mathbf{n}} E_{sc}
+ \partial_{\mathbf{n}} E_{inc}, &\quad \mbox{on $\partial B_R$},\\
\displaystyle\lim_{r\to \infty} r\big(\partial_r E_{sc}
- \imath  {k}_e E_{sc}\big)=0.
\end{array}
\right.
\end{eqnarray}
The incident wave 
$E_{inc}= e^{\imath k_e \hat{\mathbf d} \cdot {\mathbf x}}$ admits
the Jacobi-Anger expansion \cite{coltonkress,sphericalkirsch}:
\begin{eqnarray}
E_{inc}({\mathbf x}) = 4 \pi \sum_{n=0}^{\infty}\sum_{m=-n}^n 
\imath^n j_n(k_e |{\mathbf x}|) Y^m_n(\hat{\mathbf x}) 
\overline{Y^m_n(\hat{\mathbf d})}.
\label{expincident}
\end{eqnarray}
Using the series expansions (\ref{uscattered})-(\ref{utransmitted}) and  (\ref{expincident}), the expression of the coefficients (\ref{anm})-(\ref{bnm}),
together with the identity
\begin{equation}
\sum_{m=-n}^n \overline{Y^m_n(\hat{\mathbf d})} Y^m_n(\hat{\mathbf x})
= {2n +1 \over 4 \pi} P_n( \hat{\mathbf x} \cdot \hat{\mathbf d}),
\label{legendre}
\end{equation}
where $P_n$ stands for the Legendre polynomial, the scattered and 
transmitted fields take the form:
\begin{eqnarray}
E_{sc}({\mathbf x}) &=& \sum_{n=0}^\infty  \imath^n (2n+1) a_n(R)
 h_n^{(1)}(k_e |{\mathbf x}|) P_n( \hat{\mathbf x} \cdot \hat{\mathbf d}),
\label{series} \\ [-1ex]
E_{tr}({\mathbf x}) &=& \sum_{n=0}^\infty  \imath^n (2n+1) b_n(R) 
j_n(k_i |\mathbf x|) P_n( \hat{\mathbf x} \cdot \hat{\mathbf d}).
\label{seriestr} 
\end{eqnarray}

When the ball is centered about a point ${\mathbf x_0},$ we make
a change of variables to locate  $\mathbf{x}_0$ at $(0,0,0)$ and be 
able to use this expansion. Then, identity (\ref{expincident}) is multiplied
by a complex exponential  factor, and so is the coefficient $u_{nm}$
in the right hand side of the definitions (\ref{anm})-(\ref{bnm}) of $a_n$
and $b_n$. For instance, when
$\hat{\mathbf d}=(0,0,1)$ and the ball is located at
$(0,0,z_i)$, the right hand side in series expansions (\ref{series})
and (\ref{seriestr}) is multiplied by $e^{\imath k_e z_i}$.

The terms in the series decrease fast after oscillating up to about 
$n=k_eR$. Therefore, truncating to $2k_eR$ terms should be enough
for large enough spheresÊ \cite{sphericalturley}.

\subsection{Adjoint field for a sphere}
\label{sec:forwardadjoint}

The conjugate of the adjoint field defined by (\ref{adjointobjectap}) 
satisfies:
\begin{eqnarray} \label{adjointconjugate}
\left\{
\begin{array}{ll}
\Delta Q + {k}_e^2 Q= \sum_{j=1}^N
d_j \delta_{{\mathbf x}_j}&
\quad\mbox{in ${\mathbb R}^3\setminus\overline{B_R}$},
\\
\Delta Q+ {k}_i^2 Q=0,&\quad\mbox{in $B_R$},\\
Q^-= Q^+,&\quad \mbox{on $\partial B_R$},\\
\beta \partial_{\mathbf{n}}  Q^-= \partial_{\mathbf{n}} Q^+, 
&\quad \mbox{on $\partial B_R$},\\
\displaystyle\lim_{r\to \infty} r\big(\partial_r Q
- \imath {k}_e Q\big)=0,
\end{array}
\right.
\end{eqnarray}
where  $d_j = 2 ({\cal I}(\mathbf{x}_j)- |E(\mathbf{x}_j)|^2) 
\overline{E(\mathbf{x}_j)}$ or 
$d_j=\overline{E_{approx,j}-E(\mathbf{x}_j)}$. 
The function $Q_0({\mathbf x})
=- \sum_{j=1}^N {e^{\imath k_e |{\mathbf x}-{\mathbf x_j}|}
\over 4 \pi |{\mathbf x}-{\mathbf x_j}|} d_j$ is a solution
of this problem in absence of objects. Setting 
$Q=Q_0+\hat Q$ outside $B_R$ and keeping $Q = \hat Q$,
inside, the function $\hat Q$ is the solution of a problem
with the structure (\ref{forwardincident}), where $E_{inc}$ is 
replaced by $Q_0$. To expand $Q_0$ in series of spherical
harmonics we use the addition theorem for fundamental
solutions of Helmholtz operators \cite{coltonkress}:
\begin{eqnarray*}
\label{greenexpansion}
G(\mathbf x, \mathbf x_0)= {e^{\imath k_e |{\mathbf x}-{\mathbf x_0}|}
\over 4 \pi |{\mathbf x}-{\mathbf x_0}|} = \\
\left\{
\begin{array}{l}
\imath k_e \sum_{n=0}^\infty \sum_{m=-n}^m h_n^{(1)}(k_e |\mathbf x|)
Y_n^m(\hat{\mathbf x}) j_n(k_e |\mathbf x_0|) 
\overline{Y_n^m(\hat{\mathbf x}_0)}, \quad |\mathbf x | > |\mathbf  x_0 |\\
\imath k_e \sum_{n=0}^\infty \sum_{m=-n}^m h_n^{(1)}(k_e |\mathbf x_0 |)
Y_n^m(\hat{\mathbf x}_0) j_n(k_e |\mathbf x|) 
\overline{Y_n^m(\hat{\mathbf x})}, \quad |\mathbf x| < |\mathbf x_0|
\end{array} \right.
\end{eqnarray*}
Our choice of spherical harmonics \cite{coltonkress} satisfies
$\overline{Y_n^m(\hat{\mathbf x})}=Y_n^{-m}(\hat{\mathbf x}),$
which yields the following expansion for $Q_0$ at the interface
$|\mathbf x|=R< |\mathbf x_j|$, $j=1,\ldots,N$:
\begin{eqnarray}
\label{expansionQ0}
Q_0({\mathbf x}) =- \imath k_e \sum_{n=0}^\infty \sum_{m=-n}^m  
\left[ \sum_{j=1}^N d_j  h_n^{(1)}(k_e |\mathbf x_j|) \overline{Y_n^m(\hat{\mathbf x}_j)} \right] j_n(k_e |\mathbf x|)  Y_n^m(\hat{\mathbf x}).
\end{eqnarray}
Using the series expansions (\ref{uscattered})-(\ref{utransmitted}) and  (\ref{expansionQ0}), the expression of the coefficients (\ref{anm})-(\ref{bnm}),
as well as the identity (\ref{legendre}) we find:
\begin{eqnarray}
Q&=& Q_0 \!-\! \imath k_e \sum_{n=0}^{\infty}
{2n \!+\!1 \over 4 \pi} a_n(R)  h^{(1)}_n(k_e |{\mathbf x}|)  
s_{n}(\hat{\mathbf x}),
\quad |{\mathbf x}| \geq R, 
\label{expqscattered} \\ [-1ex]
Q &=&  \!-\! \imath k_e \sum_{n=0}^{\infty}
{2n \!+\!1 \over 4 \pi} b_n(R)  j_n(k_i |{\mathbf x}|)  
s_{n}(\hat{\mathbf x}),
\quad |{\mathbf x}| \leq R,
\label{expqtransmitted}
\end{eqnarray}
setting
$
s_{n}(\hat{\mathbf x})=\sum_{j=1}^N d_j h_n^{(1)}(k_e |\mathbf x_j|)  
P_n( \hat{\mathbf x} \cdot \hat{\mathbf x}_j).
$

\section*{Acknowledgments} 
T.G. Dimiduk and A. Carpio thank V.N. Manoharan for
discussions of holography and the Kavli Institute Seminars at
Harvard for the interdisciplinary communication environment
that initiated this work. A. Carpio thanks M.P. Brenner for 
hospitality while visiting Harvard  University.

\end{document}